\documentclass[superscriptaddress,twocolumn,prc]{revtex4} 
\usepackage{graphicx}

\newcommand{\be}{\begin{eqnarray}}
\newcommand{\ee}{\end{eqnarray}}

\begin{document}

\title {
A Hydrodynamic Description of Heavy Ion Collisions at the SPS and RHIC 
}
\author{D. Teaney} 
\affiliation{
 Department of Physics and Astronomy, State University of New York,
     Stony Brook, NY 11794
}
\affiliation{
 Department of Physics, Brookhaven National Laboratory 
 Upton, NY 11973
}
\author{J. Lauret}
\affiliation{
 Department of Physics, Brookhaven National Laboratory 
 Upton, NY 11973
}
\author{E.V. Shuryak}
\affiliation{
 Department of Physics and Astronomy, State University of New York,
     Stony Brook, NY 11794
}

\date{\today}
\begin{abstract}

A hydrodynamic + cascade model of relativistic
heavy ion collisions is presented and compared 
to available hadronic data from the  SPS to RHIC.  
The model consistently reproduces the 
radial and elliptic flow data for  different particles, 
collision energies, and impact parameters.
Three ingredients are essential to the success: 
(a) a reasonable EOS exhibiting the hard and soft
features of the QCD  phase transition, (b) thermal hadronization at the
phase boundary, and (c) subsequent hadronic rescattering.
Some  features of the RHIC data are readily explained:
(i) the observed elliptic flow and its dependence  on $p_T$ and mass,
(ii) the anomalous $\bar{p}/\pi^{-}$ ratio for $p_T \approx 2.0\,\mbox{GeV}$, 
(iii) the difference in the slope parameters measured 
by the STAR and PHENIX collaborations, and (iv)
the respectively strong and weak impact parameter
dependence of the $\bar{p}$
and $\phi$ slope parameters.  For an EOS without
the hard and soft features of the QCD phase transition,
the broad consistency with the data is lost.


\end{abstract}
\maketitle
\vspace{0.1in}
\newpage

\section{Introduction}\label{intro}
\subsection{Reaching the Macroscopic Limit}

Excited nuclear matter has been created by 
colliding Pb and Au ions at the SPS  
($\sqrt{s}=17\,\mbox{GeV per nucleon}$) 
and RHIC ($\sqrt{s}=130\,\mbox{GeV per nucleon}$)
accelerators \cite{QM97,QuarkMatter99,QM2001}.
For infinitely large nuclei, the excited matter can be
characterized by macroscopic quantities
-- pressure, temperature, viscosity, etc.
Lattice QCD simulations indicate that for 
temperatures larger than $T_{c} \approx 160~\mbox{MeV}$, 
confined nuclear matter 
morphs into a phase of deconfined quarks and gluons 
-- the Quark Gluon Plasma (QGP) \cite{ShuryakQGP,Lattice-Eos}. The
possibility of observing the QGP in
real nuclei  
has motivated the heavy ion experimental program.

If the system is macroscopic then thermodynamics 
describes the static properties of the
matter and hydrodynamics describes the dynamic properties
of the matter.
In fact, the observed particle
ratios are remarkably close to
the particle ratios in an ideal gas of hadrons at
a temperature, $T\approx165\,\mbox{MeV}$
 \cite{Stachel-Thermal,Becattini-Thermal,Sollfrank-Thermal,Redlich-Thermal}.
This suggests
that the system evolved from a state close to
thermal equilibrium at the phase transition boundary. However,
the same thermal description
reproduces the hadron ratios
in proton-proton and $e^{+}e^{-}$ collisions, where 
the system size is small  and equilibration seems impossible.
The success of thermodynamics seems to reflect phase space 
rather than the equilibration of macroscopic system.

Therefore, a thermodynamic (static) description, divorced
from a hydrodynamic (dynamic) description, can not
unambiguously signal a macroscopic
state.
It is important that  
elementary proton-proton and  $e^+e^-$ collisions do not 
exhibit hydrodynamic behavior.  
An analysis of hadronic spectra \cite{SZ-FlowProfile} shows
little sign of the transverse expansion 
predicted by hydrodynamics.
Thus, the excited systems produced in these elementary collisions are $not$ macroscopic.

In contrast, experiments with heavy ions do 
show evidence for a hydrodynamic expansion. 
Momentum correlations, colloquially known as $flow$,
are observed at the SPS and RHIC.
In PbPb collisions, the particles emerge from the
collision with
a collective transverse velocity of approximately
$1/2\,c$ \cite{RadialFlow-Review}.
This radial flow is firmly established from a
combined
analysis of particle spectra, HBT correlations, and
deuteron coalescence \cite{Uli-Expand}. 
In non-central collisions, the particles emerge with 
an elliptic flow (see for example\cite{QuarkMatter99,QM2001}). Elliptic flow is quantified by 
$v_2$, the asymmetry  of the angular distribution 
\begin{eqnarray}
v_2 &=& \left\langle \cos(2\,\phi) 
          \right\rangle  
\end{eqnarray} 
where $\phi$ is measured around the beam axis with
respect to the impact parameter.
Radial and elliptic flow data are measured as 
a function of transverse momenta, particle type,
impact parameter,
 and collision energy. This wealth of momentum
correlations severely constrains viable models of the heavy ion
collision. 

Several microscopic models  have been used 
to explain the available heavy-ion data.  
The first is a dilute parton model, which
is quantified with the HIJING event generator \cite{HIJING}.
Dilute parton models are based 
upon the extrapolation of perturbative QCD from high $p_T$ down 
to a scale of $\sim 1\,\mbox{GeV}$. 
For central AuAu collisions at RHIC, HIJING 
predicted a mini-jet multiplicity of $\frac{dN^g}{dy}\sim 200$, 
which  is insufficient to generate
the strong hydrodynamic response observed at RHIC \cite{Molnar-Elliptic}.
The second is a string model, 
which is quantified with the UrQMD event generator. 
In string-models non-interacting strings decay into
hadrons which subsequently interact. 
Due to the small transverse pressure at early
times,  UrQMD predicted a decrease in elliptic flow from the 
SPS to RHIC \cite{UrQMD-Elliptic}. 
A $\approx 50\%$ increase was observed. 
In contrast to these microscopic models, 
hydrodynamic calculations at the SPS and RHIC, give a good description
of the observed radial and elliptic flows 
\cite{Sollfrank-BigHydro,Schlei-BigHydro,Kolb-UU,Kolb-LowDensity,Kolb-Radial,Htoh}, but
offer no insight into the microscopic mechanism of equilibration. 

Accepting the macroscopic approach and its limitations, the
phase transition to the QGP influences both
the radial and elliptic flows. Lattice simulations
indicate \cite{Lattice-Eos} that 
over a wide range  of 
energy densities 
$e=0.5-1.3\,\mbox{GeV/fm}^{3}$, 
the
temperature and pressure are nearly constant and the speed of
sound is approximately zero, $\frac{dp}{de} \approx 0$. 
Because the speed of sound is small in this range, 
the pressure can not effectively accelerate the matter \cite{HS-soft,Rischke-Log}.
However, when the
initial energy density is well above the 
transition region, the matter enters the hard QGP phase.
The speed of  sound approaches $\sqrt{1/3}$ and the 
pressure drives collective motion.
At a time of $\sim 1~\mbox{fm/c}$, 
the energy density at the SPS and
RHIC are very approximately 
$4$ and $8~\mbox{GeV/fm}^{3}$ \cite{NA49-EnergyDensity,Phobos-Multiplicity}. 
Based on these experimental estimates, the hard QGP
phase
is expected to live significantly longer at RHIC than
at the 
SPS. The final radial and elliptic flows of the produced particles should
reflect this 
difference \cite{VanHove-T,Ollitrault-MixedPhase,Kataja-MixedPhase,Ollitrault-Elliptic,Kolb-UU,Rischke-Px}.

Certainly, hydrodynamics is not applicable when the
particles decouple from the collision and this
``freezeout''  must be  modeled in order to
compare the observed hadron spectra to hydrodynamic
calculations.  Usually, a naive freezeout prescription is 
taken:
A ``freezeout Temperature'' $T_f$, is specified; 
 thermal and chemical equilibrium are assumed for
$T<T_f$;  
the spectrum of particles passing 
through the $T_f$ isotherm is calculated;
 $T_f$ is  finally adjusted to
match the single particle spectrum of pions and
nucleons. 
Of course, this 
prescription is unrealistic and takes away from 
the predictive power of hydrodynamics. Nevertheless,
the approach
successfully describes many radial and elliptic
observables 
from the SPS to 
RHIC \cite{Sollfrank-BigHydro,Schlei-BigHydro,Kolb-LowDensity,Kolb-Radial}.

However, the naive freezeout prescription
 fails in a number of respects.
First, on the time scale of the collision $\sim 10~\mbox{fm/c}$, 
hadronic reactions do not alter the
hadron composition and chemical equilibrium is not
maintained (see for example 
\cite{Uli-Chemical,HydroUrqmd,UrQMD-Hagedorn}).  
Therefore in the late hadronic stages,  
chemical freezeout must be modeled in order to
describe the resonance contribution and 
the particle ratios.  
Second,
different particle types freezeout at different times
and with different transverse velocities. With a universal
freezeout temperature, the transverse flow of the strange
particles
$\Lambda,\,\Xi,\,\Omega$, is never reproduced \cite{Sorge-Strange}.
Third, at the SPS and RHIC, integrated elliptic flow 
is a strong function of the freezeout temperature.
When the
universal freezeout temperature is adjusted to match
the nucleon spectrum, the integrated pion elliptic flow is too 
large \cite{Kolb-Flow}. In
reality, pions and nucleons freezeout at different times and
temperatures.

To model freezeout, Bass and Dumitru \cite{HydroUrqmd} replaced the
hadronic phase of the hydrodynamics with a hadronic transport 
model, UrQMD. In this approach, the switch from hydro
to cascade is 
made at a switching temperature, $T_{switch}\approx
T_{c}$. The
spectrum of particles leaving the surface is taken 
as the input to the cascade and the attendant
theoretical problems
are ignored. The approach worked. 
Chemical freezeout was 
incorporated into a comprehensive dynamical picture. 
The flow of the multi-strange baryons was reproduced. 
When a similar hydro+cascade model \cite{Htoh} 
was applied to non-central collisions, elliptic flow
was also reproduced at the 20\% level. 
Furthermore, with the $T_{f}$ indeterminacy removed, 
these ``simple'' boost invariant hydro+cascade models
were rather predictive -- only  $\frac{dN}{dy}$ and 
the $\bar{p}/p$ ratio have to be specified.  

\subsection{Brief Summary}

In this work, we compare the hydro+cascade model 
of \cite{Htoh} to the flow systematics at the SPS and RHIC. 
The model uses hydrodynamics to model the initial 
stage of the collision, and the hadronic cascade, Relativistic Quantum
Molecular Dynamics (RQMD v2.4), to model the final
stages of the collision \cite{Sorge-RQMD}.

The Equation of State (EOS)
is varied systematically and results are compared 
to the whole body of flow data from the SPS to RHIC.
A family of EOS, labeled by the  
value of the Latent Heat (LH) is constructed;  
LH4, LH8, LH16$...$ denote increasingly soft 
EOS with latent heats $0.4,0.8,1.6...\,\mbox{GeV/fm}^{3}$ respectively. 
As a limiting case, the latent heat is made very large forming 
LH$\infty$. A Resonance Gas (RG) EOS is also studied.  
For an EOS with both the hard and soft features of the QCD
phase transition, the model is broadly consistent with the
body of available data. The best overall consistency 
with the data is found with LH8.
For an EOS with only hard (e.g. RG) or only 
soft features (e.g. LH$\infty$), 
the broad consistency with the data is lost. 

The model consists of three distinct components. The first 
component solves the equations of 
relativistic hydrodynamics in the transverse plane, assuming 
Bjorken scaling \cite{Bjorken-83}. The switching surface, or the
isotherm where $T_{switch}=160\,\mbox{MeV}$, is found. The sensitivity
of the model results to $T_{switch}$ will be discussed in a separate
publication where chemical freezeout is also addressed \cite{Teaney-Chemical}. 
The second component converts the macroscopic hydrodynamic
variables on the switching surface into
hadrons according to the Cooper-Frye prescription augmented with
a theta function rejecting backward moving 
particles \cite{CooperFrye}. 
Finally, the
third component, the hadronic cascade RQMD  
sequentially rescatters the generated hadrons and models
the hadronic freezeout stage of the collision. 
Throughout the analysis the role of hadronic
rescattering is assessed. In all figures, the  
``Hydro+RQMD'' curves incorporate hadronic rescattering and 
resonance decays while the  
``Hydro Only'' curves only incorporate resonance decays.

As outlined in the abstract, several features of 
the first RHIC data are readily explained in the
course of this analysis.

\section{Model Description and the EOS}
\label{FLModel}

\subsection{Hydrodynamics}
\label{Hydrodynamics}

Relativistic hydrodynamics is a set of conservation laws for the
 stress tensor ($T^{\mu\nu}$) 
and for the conserved currents ($J_{i}^{\mu}$), 
$\partial_{\mu}T^{\mu \nu}=0$ and $\partial_{\mu}J_{i}^{\mu}=0$.  
In equilibrium, $T^{\mu\nu}$ and $J_{i}^{\mu}$
are related to the bulk properties of the fluid by
the relations,
$T^{\mu \nu} = (\epsilon + p) U^{\mu} U^{\nu} - p g^{\mu \nu}$ and 
$J_{i}^{\mu} = n_{i} U^{\mu} $ \cite{LL-Hydro}. 
Here $\epsilon$ is the energy density, $p$ is the pressure,
$n_i$ is the number density of the corresponding current, and 
$U^{\mu}=\gamma(1, \bf{v})$ is the proper velocity of the fluid. 
In strong interactions, the conserved currents are isospin ($J_{I}^{\mu}$),
strangeness ($J_{S}^{\mu}$), and baryon number ($J_{B}^{\mu}$). 
For the hydrodynamic evolution, 
isospin symmetry is assumed and  the net strangeness is set to zero;
 therefore only the baryon current $J_{B}$ is considered below.

The equations of motion may be expressed in terms of the variables 
$\tau=\sqrt{t^{2}-z^{2}}$ and $\eta=\frac{1}{2} \log (\frac{t + z}{t-z})$,
which are respectively referred to as the Bjorken proper time and the 
spatial rapidity.  
Boost invariance assumes that the solution for any value of $\eta$ 
may be found by boosting the solution at $\eta=0$ to a frame moving
with velocity $v=\tanh (\eta)$ in the  negative z-direction.  
With this assumption, the equations of motion become two dimensional
\cite{Bjorken-83,Ollitrault-Elliptic} and are given at $\eta=0$ by
\begin{eqnarray}
\label{EOM}
  \partial_{\tau}(\tau T^{00}) + \partial_{x}(\tau T^{0x}) + 
  \partial_{y}(\tau T^{0y}) &=&  -p \\
  \partial_{\tau}(\tau T^{0x}) + \partial_{x}(\tau T^{xx}) + 
  \partial_{y}(\tau T^{xy}) &=&  0  \nonumber \\ 
  \partial_{\tau}(\tau T^{0y}) + \partial_{x}(\tau T^{xy}) + 
  \partial_{y}(\tau T^{yy}) &=&  0   \nonumber \\
  \partial_{\tau}(\tau J_{B}^{0}) + \partial_{x}(\tau J_{i}^{B}) + 
  \partial_{y}(\tau J_{B}^{y}) &=&  0 \,.  \nonumber
\end{eqnarray}
Integrating over the transverse plane, one finds that
net baryon number per unit spatial rapidity, 
 $\int dxdy \,(\tau J_{B}^{0})$, and 
the transverse 
momentum per unit rapidity, $\int dxdy \,(\tau T^{0x})$ 
as well as $\int dxdy\, (\tau T^{0y})$, are conserved.
The energy per unit rapidity, $\int dxdy\,(\tau T^{00})$,
decreases due to the work done per unit time \cite{MG84-Work} by the pressure in the 
longitudinal direction, $\int dxdy \,p$. 

For an ideal fluid, entropy conservation
can be derived \cite{LL-Hydro}, $\partial_{\mu}(S^{\mu})=0$.
The entropy current is defined as $S^{\mu}\equiv s\,U^{\mu}$, 
where $s$ is the entropy density and $U^{\mu}$ is the
fluid 4-velocity. 
For a Bjorken expansion entropy conservation  
becomes
\begin{eqnarray}
\label{entropy_conserve}
  \partial_{\tau}(\tau S^{0}) + \partial_{x}(\tau S^{x}) + 
  \partial_{y}(\tau S^{y}) &=&  0  \,. 
\end{eqnarray}
Integrating over the transverse plane, we find that 
\begin{eqnarray}
\label{total_entropy}
   \int dx\,dy\,\tau\,s \gamma & =&  \frac{dS_{tot}}{d\eta}   
\end{eqnarray}
is a constant of the motion.
This relation is monitored to test the accuracy of the solution.

These equations are solved numerically 
with a Gudunov method \cite{Hydro-Leveque}.
Using second order operator splitting \cite{Hydro-Leveque},
a single time step separately updates the x-direction, the 
y-direction, and  the loss terms on the  r.h.s. of Eq.
\ref{EOM}. Different splittings gave only 
negligibly different results.
The simple RHLLE Riemann solver 
was used  
for the updates in the x and y directions \cite{Rischke-RHLLE,Rischke-PlasmaTest}. 
A second order (in $\tau$) Runge-Kutta
stepper was used for the r.h.s. update.
 
\subsection{Initial Conditions}
\label{FLModel-InitCond}

To model the initial conditions, 
the entropy and baryon distributions at a Bjorken time of
$\tau_{0}=1 ~\mbox{fm/c}$, 
are made proportional to the distribution of participating nucleons
in the transverse plane. 
Since entropy and baryon number are conserved per unit rapidity, 
the final yields of pions and nucleons  
are then proportional to the number of participants.
The initial conditions are similar to sWN (entropy per Wounded Nucleon) 
initial conditions in \cite{Kolb-Centrality} and to the initial conditions
of  \cite{Ollitrault-Elliptic}. 

For all subsequent discussions, we consider 
two identical (for simplicity) nuclei with  atomic number A and B, and
nucleon distributions $\rho_{A}(\vec{r})$ and 
$\rho_{B}(\vec{r})$,
collide along the z-axis with impact parameter $\vec{b}$, 
pointing in the x-direction from the center of nucleus A,
 $(x_{A},y_{B})\equiv(-b/2,0)$, to the center of nucleus B, 
 $(x_{B},y_{B})\equiv(+b/2,0)$.  
The nucleon distribution $\rho_{A}$ is 
parameterized as a Woods-Saxon distribution, $\rho(\vec{r}) \propto 
\frac{1}{e^{(r - R_A)/\delta}+1}$,  and is normalized to the 
atomic number A. The parameters are 
$\delta =0.55\,\mbox{fm}$, $R_{A} = 1.08\,A^{1/3}$. 
The $R_{A}$ used is 4\% below the value used by the STAR collaboration
\cite{SN402}.
The number of participating nucleons per unit area, $\frac{dN_{p}}{dx\,dy}$,
at a position $\vec{x}_{T}=(x,y)$ in the transverse plane is given by
\begin{widetext}
\begin{eqnarray}
\label{glauber}
   \frac{dN_{p}}{dx\,dy} &=& T_A (\vec{b}/2+\vec{x}_{T})  
          \left\{ 1 - \left[1 -\frac {\sigma_{NN}T_B(-\vec{b}/2 + 
          \vec{x}_{T})}{B}\right]^{B}\right\}  
          +  T_B (-\vec{b}/2 + \vec{x}_{T})
          \left\{ 1 - 
          \left[1 -\frac {\sigma_{NN}T_A(\vec{b}/2 + 
          \vec{x}_{T})}{A}\right]^{A} \right\}.\nonumber \\ 
\end{eqnarray}
\end{widetext}
Here, $T_A(\vec{x}_{T})=\int \rho_{A} (x,y,z) dz$ 
is the thickness of a nucleus at position (x,y) and  $\sigma_{NN}$ is
the inelastic nucleon-nucleon cross section. For the sake of 
comparison, $\sigma_{NN}$ 
is taken as $33\,\mbox{mb}$ both at the SPS and
RHIC.
For large $A$, $[1 -\frac {\sigma_{NN}T_A(\vec{x}_{T})} {A}]^{A}
\approx \exp(-\sigma_{NN}T_A(\vec{x}_{T}))$, and 
often Eq.~\ref{glauber} is re-written in terms of exponents.

With the number of participants specified, the  
initial entropy and (net) baryon densities at time
$\tau_{0}=1\,\mbox{fm/c}$, are then fixed with two constants 
$C_{s}$ and $C_{n_{B}}$ with
\begin{eqnarray}
\label{initcond}
   s    (x,y,\tau_{0}) &=& \frac{C_{s}}     {\tau_{0}} \frac{dN_{p}}{dx\,dy} \\
   n_{B}(x,y,\tau_{0}) &=& \frac{C_{n_{B}}} {\tau_{0}} \frac{dN_{p}}{dx\,dy} \, .  
\end{eqnarray}
The two dimensionless constants $C_s$ and $C_{n_{B}}$ 
are the entropy and net baryon number produced 
per unit spatial rapidity  per participant.  At the
SPS (see Sect. \ref{FLRadialFlow-SPS}), $C_{s}$ and $C_{n_{B}}$ were 
 adjusted to fit the 
 total yield of charged particles and the net yield of protons, respectively.
  At RHIC, $C_{s}$ was adjusted
to match the PHOBOS multiplicity $\frac{dN}{d\eta} = 555\pm12\mbox{(stat)}\pm35\mbox{(syst)}$ \cite{Phobos-Multiplicity}. At the,
time the $\bar{p}/p$ ratio was not known and 
$s/n_{B}=C_{s}/C_{n_{B}}$ was estimated from UrQMD simulations to
be $\approx 150$. This gives  the ratio $\bar{p}/p = 0.45$. Later, the
STAR and PHOBOS collaborations measured the ratios,  
$\bar{p}/p =0.65\pm.01\mbox{(stat)}\pm.07\mbox{(syst)}$ and 
$\bar{p}/p=0.60\pm.04\mbox{(stat)}\pm.06\mbox{(syst)}$  respectively
\cite{STAR-ppbar,Phobos-ppbar}.  
Since the measured ratio is close to the ratio initially used, 
 and since a full simulation takes several CPU days, 
the UrQMD-based estimate $s/n_{B}\approx150$ was
used throughout this work. This makes the model $\bar{p}$ yield approximately
15\% too low and the model proton yield approximately 
15\% too high. This correction
will be accounted for in future works.
A summary of the parameters is given in Table \ref{snbtable}.
\begin{table}[!tbp]
\begin{tabular}{|l|r|r|} \hline
 Parameter/Value    &PbPb SPS & AuAu RHIC  \\ \hline\hline
$C_s$                  & 8.06 & 14.42  \\ \hline
$C_{n_B}$              &  0.191 &  0.096  \\ \hline
$\tau_{0}$ (fm)        &  1.0 &  1.0  \\ \hline
$\sigma_{NN}$ (mb)       &  33  &  33    \\ \hline\hline
$s/n_{B}=C_s/C_{n_B}$              &  42  &  150  \\ \hline
$e_0 \, (\mbox{GeV/fm}^{3}) - LH8$ &8.2 & 16.7 \\ \hline
$e_0 \,(\mbox{GeV/fm}^{3}) - LH\infty$ &6.4 & 11.2 \\ \hline
$\langle e \rangle \,(\mbox{GeV/fm}^{3}) - LH8$ &5.4 & 11.0 \\ \hline
$\langle e \rangle \,(\mbox{GeV/fm}^{3}) - LH\infty$ &4.5 & 7.9 \\ \hline
\end{tabular}
\caption[Table of parameters characterizing the initial conditions]{
\label{snbtable}
A  summary of  the input parameters to the model. $C_s$ and
$C_{n_B}$ are respectively 
the entropy and baryon number per participant
per unit rapidity. The
values above the double line are the input parameters. The values
below the double line are derived from the input parameters. 
The initial energy density 
depends on the EOS and impact parameter. 
For central collisions and for two 
EOS spanning the gamut, we quote 
the initial energy 
density in the center of the collision ($e_{0}$) 
and the initial energy density 
averaged over the transverse plane with respect to the number
of participants ($\langle e \rangle$). }
\end{table}

\begin{figure}
\begin{center}
  \includegraphics[height=3.0in,width=3.0in]{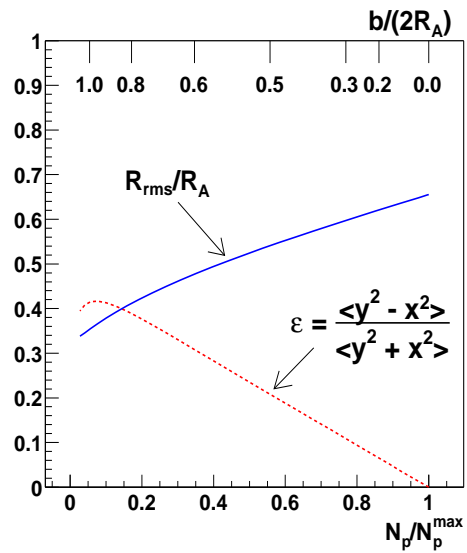}
\end{center}
\caption[Parameters characterizing the initial Glauber distribution
as a function of impact parameter] {
   The derived quantities  $\epsilon$ and $R_{rms}/R_{A}$, defined
   by Eq.~\ref{GlauberEquRms} and \ref{GlauberEquEps},
	as a function of the number
   of participants relative to the maximum number. The axis on
   top of the graph shows the impact parameter b relative
   to $2 R_{A}$. The curves are drawn for  
   PbPb collisions at the SPS, but depend only slightly on the
   colliding system and energy.
}
\label{GlauberFig}
\end{figure}

Two quantities, which 
will be used extensively in the analysis in Sect.
\ref{FLSpaceTime} and Sect. \ref{FLEllipticFlow},  are defined as 
\begin{eqnarray}
\label{GlauberEquRms}
   R_{rms} &\equiv& \sqrt{ \langle x^2 + y^2 \rangle } \\
\label{GlauberEquEps}
   \epsilon &\equiv&  \frac{ \langle y^2 - x^2 \rangle }
                {\langle y^2 + x^2 \rangle} \,,
\end{eqnarray}
where the average is taken over the initial entropy distribution
of Eq.\,\ref{initcond}. 
These quantities are plotted as a function 
of the number of participants  
relative to central collisions in Fig.~\ref{GlauberFig}. 
$\epsilon$ measures the initial 
elliptic deformation of the overlap region and grows approximately
linearly with $N_p$.

For the calculation presented, the entropy and therefore  the number of
charged particles scales as the number of participants. Recently,
the experiments have reported that the charged
particle multiplicity grows slightly faster than 
the number of participants \cite{PHENIX-Centrality,Phobos-Centrality}. 
This slight growth can be incorporated into 
hydrodynamics \cite{Kolb-Centrality}, but instead 
 the experimental $dN_{ch}/dy$ is compared directly
to the model $dN_{ch}/dy$. This makes the model impact parameter 
slightly larger than the impact parameter determined by 
the experimental collaborations.

\subsection{Equation of State}
\label{FLModel-EOS}

To solve the equations of motion, we need an Equation of
State (EOS), or a relation between the pressure ($p$) and  the 
energy and baryon densities ($e$ and $n_B$ respectively). 
In many previous
hydrodynamic calculations, a bag model EOS is 
used \cite{Sollfrank-BigHydro,HydroUrqmd,HydroUrqmdHBT,Raju-ResonanceGas}.
This has some advantages, since the degrees of
freedom are explicit in both phases. 
However a typical bag model results
in an EOS with a large latent heat,  LH=$1-1.5\,\mbox{GeV/fm}^{3}$.
Furthermore it is difficult to adjust the latent heat independently of
$T_{c}$ in such models of the phase transition.

We have taken a more pragmatic approach and have constructed
a thermodynamically consistent  EOS with a variable latent
heat in the $e$ and $n_B$ plane.  First, note the
following two derivatives which apply along the
path where $n_{B}/s$ is constant
\begin{eqnarray}
\label{thermal-eoscs2}
 \left(\frac{dp}{de}\right)_{n_{B}/s} &\equiv& \,c_{s}^{2} \\
\label{thermal-eosds}
 \left(\frac{ds}{de}\right)_{n_{B}/s} &=& \frac{s}{p+e} \,.
\end{eqnarray}
The first of these is simply the definition of the speed 
of sound. The second relation  is surprising:  
it does not contain the chemical potential $\mu_B$ explicitly. 
(It follows by noting that
$\left(\frac{ds}{de}\right)_{n_B/s} = 
\left( \frac{ds}{de} \right)_{n_B} + 
\left( \frac{ds}{dn_B}\right)_{e} \frac{n_B}{s} 
\left(\frac{ds}{de}\right)_{n_B/s}$ and solving for 
$\left(\frac{ds}{de}\right)_{n_B/s}$, by using thermodynamic
identities). Given the speed of sound everywhere and the 
entropy on a single arc in the 
$e,n_B$ plane, these derivatives may be integrated to
determine the entropy, s($e$,$n_B$). From 
the entropy, all other thermodynamic 
functions\, (e.g., T and $\mu_B$) may be determined. Below, 
only the speed of sound is specified.

For smooth flows, entropy and baryon number are separately conserved.
If at some initial time $n_B/s$ is constant everywhere in 
space, the two conservation laws imply that $n_B/s$ is constant 
everywhere in space $and$ time \cite{LL-Hydro}. 
For the initial
conditions specified in Sect. \ref{FLModel-InitCond},\, $n_B/s = 
C_{n_B}/C_{s}$ is
constant in space and remains constant as the system evolves.
Therefore, the pressure is needed
only along the path $n_{B}/s = C_{n_B}/C_{s}$.
This may be directly verified by 
fully differentiating $\partial_{\mu} T^{\mu\nu}=0$ and noting
that the derivatives of the pressure only appear as the 
speed of sound,  $ 
 \left(\frac{dp}{de}\right)_{n_{B}/s} \equiv \,c_{s}^{2}=
 \left(\frac{\partial p}{\partial e} \right)_{n_B} 
  + \frac{n_B}{e + p} 
 \left(\frac{\partial p}{\partial n_B} \right)_{e} 
$.

Strictly speaking,  
transverse shock waves develop near the phase transition
and invalidate the assumption of entropy conservation. 
However,
numerical and analytical evidence has shown that entropy production in hydrodynamic
simulations of nucleus-nucleus collisions is at most a few percent \cite{Ollitrault-RiemmanMixed}.
Below, entropy production is ignored  and the
pressure is specified along the trajectory $n_B/s = C_{n_B}/C_{s}$.

\begin{figure}
\begin{center}
   \includegraphics[height=3.0in,width=3.0in]{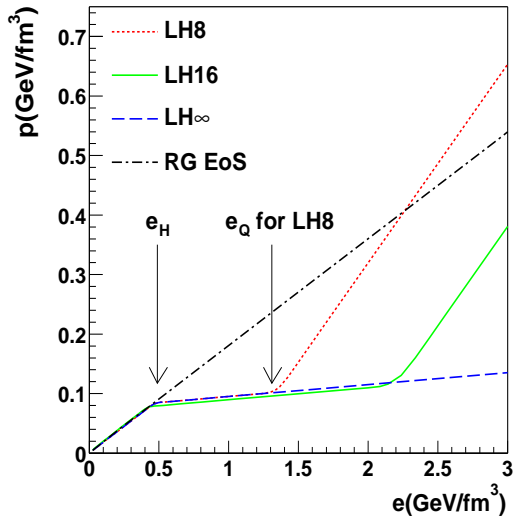}
\end{center}
\caption[Pressure as a function of energy density for the
different EOS used in this work]
{
The pressure ($p$) versus energy density ($e$) for different EOS. 
EOS LH8, LH16 and LH$\infty$ become increasingly soft and 
have  latent heats $0.8\,\mbox{GeV/fm}^{3}$, $1.6\,\mbox{GeV/fm}^{3}$, and $\infty$.
 The EOS are shown along
the adiabatic path for SPS initial conditions, $s/n_{B}=42$. 
For RHIC initial conditions, $s/n_{B}=150$,  the changes are
small.
}
\label{psEOS}
\end{figure}

The EOS consists of three pieces: a hadronic
phase, a mixed phase, and a QGP phase. 
In strong interactions, Baryon number (B), Strangeness
(S), and Isospin (I) are conserved  and
therefore the EOS depends on $T$ and $\mu_{B},\mu_{S}$,and $\mu_{I}$.
In the hadronic phase,   
the thermodynamic quantities --the pressure (p), the energy
density ($e$), the entropy density (s), 
and number densities ($n_{Q}$ where Q=B,S,I)-- are 
taken as ideal gas mixtures of the lowest  SU(3) multiplets
of mesons and baryons.
The mix includes  
the pseudo-scalar meson octet ($\pi,\eta,K$) and singlet ($\eta'$),
the vector meson octet ($\rho,K^{*},\omega$) and singlet ($\phi$),
the $\frac{1}{2}^{-}$ baryon and anti-baryon octets
and the $\frac{3}{2}^{-}$ baryon and the anti-baryon decuplets.
Specifically, $p,\, e,\, s$ and  $n_{Q}$ are given by
\begin{eqnarray}
   n_{Q}    &=& \sum_i{ Q_i \,n_{id}^{\sigma_i}\, ( T , \mu_B B_i + \mu_S S_i + \mu_I  I_i ) }\\
   p        &=& \sum_i{ p_{id}^{\sigma_i} \, ( T , \mu_B B_i + \mu_S S_i + \mu_I  I_i ) }\\
   e &=& \sum_i{ e_{id}^{\sigma_i} \, ( T , \mu_B B_i + \mu_S S_i + \mu_I I_i ) }\\
   s        &=& \sum_i{ s_{id}^{\sigma_i}\,  (T , \mu_B B_i + \mu_S S_i + \mu_I  I_i ) } \,.
\end{eqnarray} 
Here the sum is over the hadrons species, $B_i,S_i,I_i$ are the
quantum numbers of the i-th hadron,
 $\sigma_{i}$ is $+$  
for bosons  but $-$ for fermions, and  for example, 
$p_{id}^{+}(T,\mu)$ is  the pressure of a simple
ideal Bose gas.
A fast numerical method for evaluating the thermodynamic
quantities of simple Bose/Fermi gases has been  
constructed \cite{Thermo-Pons}.  
For a given T, $(\mu_{B}, \mu_{S}, \mu_{I})$ are determined
by the requirements that total strangeness ($n_{S}$) and 
isospin($n_{I}$) be zero and that $n_{B}/s = C_{n_B}/C_s$.
The thermodynamic quantities
are then  taken as functions of $e$ along the
adiabatic path where $n_{B}/S = C_{n_B}/C_s$. 
This hadronic EOS is taken up to a temperature of
$T_{c} = 165~\mbox{MeV}$ or an energy density 
$e_H \approx 0.45~\mbox{GeV/fm}^{3}$ (see Fig.\,\ref{psEOS}). 
The squared speed of sound is approximately  1/5 
in this hadronic gas. 

Above the hadronic phase, only the speed of sound squared, $c_{s}^{2}$, is
specified. 
For the mixed phase the speed of sound was made approximately
zero, $c_{s}^2 = 0.02\,c$.  
The width of the mixed phase (see Fig.\,\ref{psEOS}) is
the Latent Heat (LH), LH$=e_Q - e_H$.
LH is taken as a parameter  and is adjusted to 
form phase diagrams LH8, LH16,... 
with latent heats,  $0.80~\mbox{GeV/fm}^{3}$, $1.6~\mbox{GeV/fm}^{3}...$\,.
Above the mixed phase, $e > e_Q$, 
the degrees of freedom are taken as massless and the 
speed of sound is accordingly, $c_{s} = \sqrt{1/3}$. 
We also consider two limiting cases: a Resonance
Gas (RG) EOS and LH$\infty$. For a RG EOS, the speed of sound is constant above $e_H$. For LH$\infty$, the mixed phase
continues forever ($e_Q = \infty$) and 
there is no ideal plasma phase.

With the speed of sound specified in all phases, 
Eq.\,\ref{thermal-eoscs2} and \ref{thermal-eosds} 
are  integrated to find the pressure and entropy
along the adiabatic path specified by the initial conditions, 
$n_B/s = C_{n_B}/C_{s}$. The
full phase diagram for SPS initial conditions is shown in 
Fig.\,\ref{psEOS}.  In Sect.~\ref{FLRadialFlow} and Sect.~\ref{FLEllipticFlow},
 the subset of the EOS consistent  
with the available radial and elliptic flow data is found. 

\subsection{RQMD and the Cooper Frye Formula}
\label{CooperFryeSection}

Given the initial conditions and the EOS,  the
equations of motion are integrated in time.  
As the system expands and cools, the mean free path  
becomes much less than the nuclear radius, and the system 
breaks up into free-streaming particles. 
Typically, hydro practitioners \cite{Sollfrank-BigHydro,Schlei-BigHydro}
model the breakup or $freezeout$ stage by finding a 
hypersurface in space and time
where the temperature  equals some freezeout temperature, 
$T = T_{f}$.  

This simple picture in which all particles are emitted from a 
single space-time surface is however unrealistic. Different particles
have different hadronic cross sections and suffer 
their last interaction at different times. They are emitted
over a space-time region rather than on a sharp surface.
Further, particles in 
the center rescatter for a longer time than particles in the periphery.  To
model this physics, the spectrum of particles $exiting$ the space-time
surface is taken as the input  to the hadronic cascade,  
RQMD \cite{Sorge-RQMD}. Subsequently, the particles rescatter. 
Below, the space-time 
surface is referred to as the $switching$ surface rather than
the $freezeout$ surface. 
The  attending problems with this
approach are described below after the details of the input distribution
to RQMD are described.  

For the family of EOS 
discussed above, for $T<T_{c}$, the fluid is made up of a collection
of ideal gases of fermions and bosons. In
order to conserve energy and momentum across the surface, the
spectrum of $(\mbox{species})_{i}$ crossing the surface is given 
by \cite{CooperFrye,CooperFrye-HBT},
\begin{eqnarray}
\label{CooperFryeFormula}
   E\frac{d^{3}N_{i}}{d^{3}p} &=& 
   f^{\sigma_i}_{id}(p \cdot U, \mu_B B_i + \mu_S S_i + \mu_I  I_i) \nonumber \\
    & &\times \, p^{\alpha}\,d\Sigma_{\alpha} 
\end{eqnarray}
where   
 $d\Sigma_{\alpha}$
is a differential element of the freezeout hypersurface 
and $f^{\pm}_{id}(E,\mu)$ is the Bose/Fermi distribution function 
$\frac { g_J } 
{
   \exp( \frac{ E-\mu }{T} 
     ) 
\mp 1 }$. 
The particle 
index $i$  runs over all the species in the EOS -- no more, no less. 
If the quasi-particles 
are interacting and the
EOS is non-ideal, due to viscosity, mean fields, particle lifetimes, 
etc., then the $f^{\sigma_{i}}_{id}$ should  be modified accordingly.  

The differential elements  of the hypersurface $d\Sigma_{\alpha}$ 
can be separated into time-like 
($d\Sigma_{\alpha} d\Sigma^{\alpha} > 0$) and space-like 
($d\Sigma_{\alpha} d\Sigma^{\alpha} < 0$) surface elements.
For time-like surfaces, the integrand in Eq.\,\ref{CooperFryeFormula} 
is positive and there is a frame (the rest frame of the
surface), where  $d\Sigma_{\alpha}$  
= (dV,0,0,0). The spectrum of Eq.~\ref{CooperFryeFormula} is 
simply a thermal spectrum boosted by the flow velocity  
in the frame of the surface. (In practice, the 
surface velocity is small for most time-like surfaces). 
The yield of $(\mbox{species})_i$
leaving a  surface element,  
$d\Sigma_{\alpha}$, is simply $n_{id}^{\sigma_i}(T,\mu_{B}B_i 
+ \mu_{S}S_i+\mu_{I}I_{i}) U^{\alpha}d\Sigma_{\alpha}$, as may be found by 
integrating the left and right sides with 
$\frac{d^{3}p}{E}$ and going to the rest frame of the matter.

For space-like surfaces, the integrand in Eq.\,\ref{CooperFryeFormula}
is both positive and negative depending on the momentum of the
particle. When the integrand is positive, the particle is
leaving the surface and when it is negative the particle is entering
the surface. We reject particles entering the hydrodynamic
surface.
The distribution of particles exiting a space-like surface is 
\begin{eqnarray}
\label{CooperFryeFormulaPlus}
   E\frac{d^{3}N_{i}}{d^{3}p} &=& 
   f^{\sigma_i}_{id}( p \cdot U, \mu_B B_i + \mu_S S_i + \mu_I I_i)\, \nonumber
	\\& &
    \times\,  p^{\alpha}\,d\Sigma_{\alpha}\, \Theta(p^{\alpha}d\Sigma{\alpha}) \, .
\end{eqnarray}
It is this distribution that we take  as the input distribution
for RQMD. For a discussion of the problem of space-like surfaces  
see \cite{Freezeout-Theta,Freezeout-Bugaev}.
The number of particles leaving a differential surface 
element is given by a more complicated formula which
is again found by integrating both sides of the equation 
with $\frac{d^{3}p}{E}$ \cite{Teaney-Thesis}.
 For a stationary surface 
in the rest frame, the formula has the simple interpretation 
as the number of particles evaporating from the surface per area per 
unit time. A consequence of the theta-function is that energy, 
 momentum, and particle number are not exactly conserved across the
transition surface. However, the
error is only $\approx 2.0\%(\approx 5.0\%)$ for central (peripheral
b=8.0\,fm) AuAu collisions at RHIC.  

%

\section{The Space-Time Evolution}
\label{FLSpaceTime}

\subsection{The Hydrodynamic Solution}

\begin{figure}[!tbp]
   \begin{center}
      \includegraphics[height=1.7in,width=1.7in]{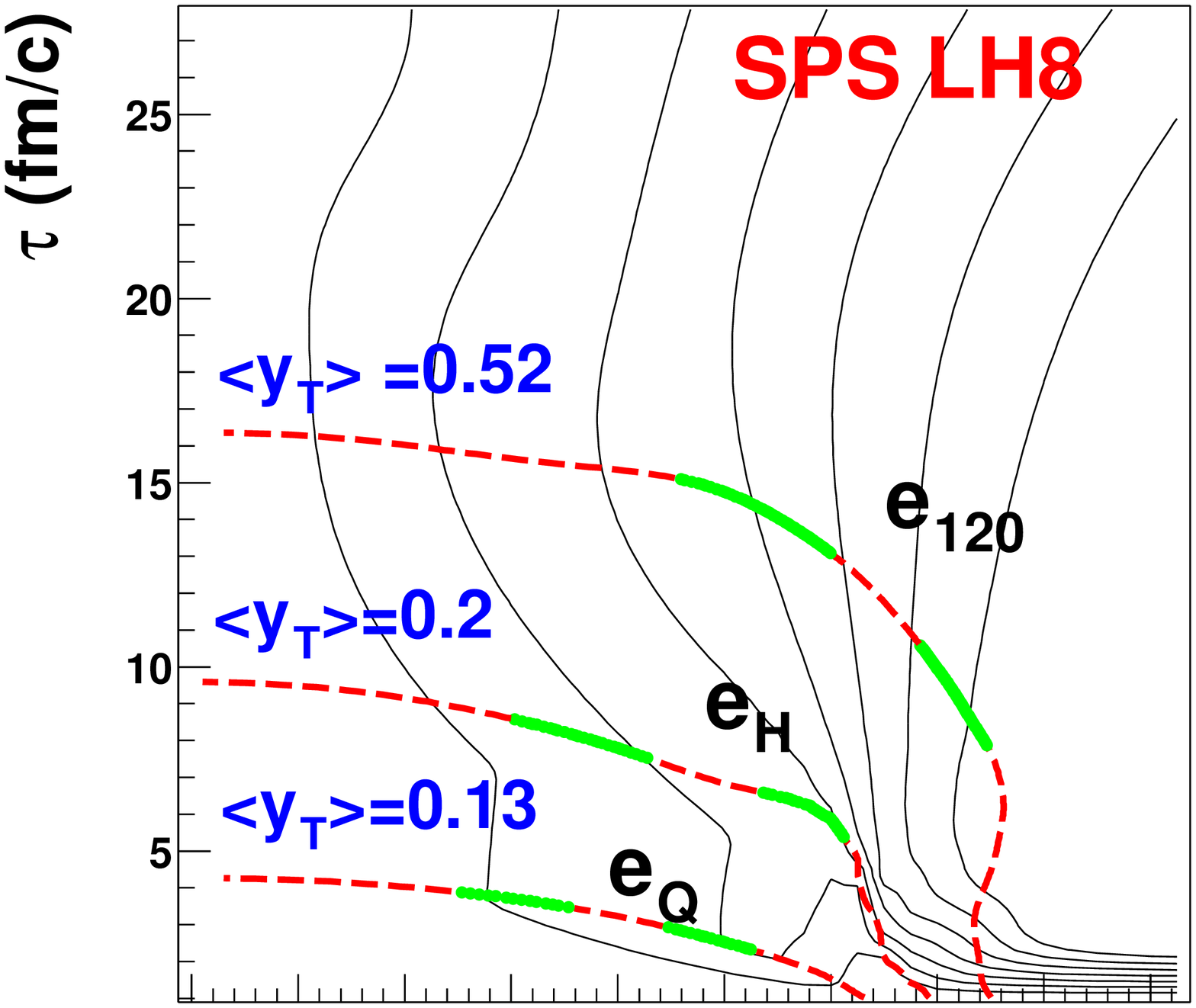}
      \includegraphics[height=1.7in,width=1.6in]{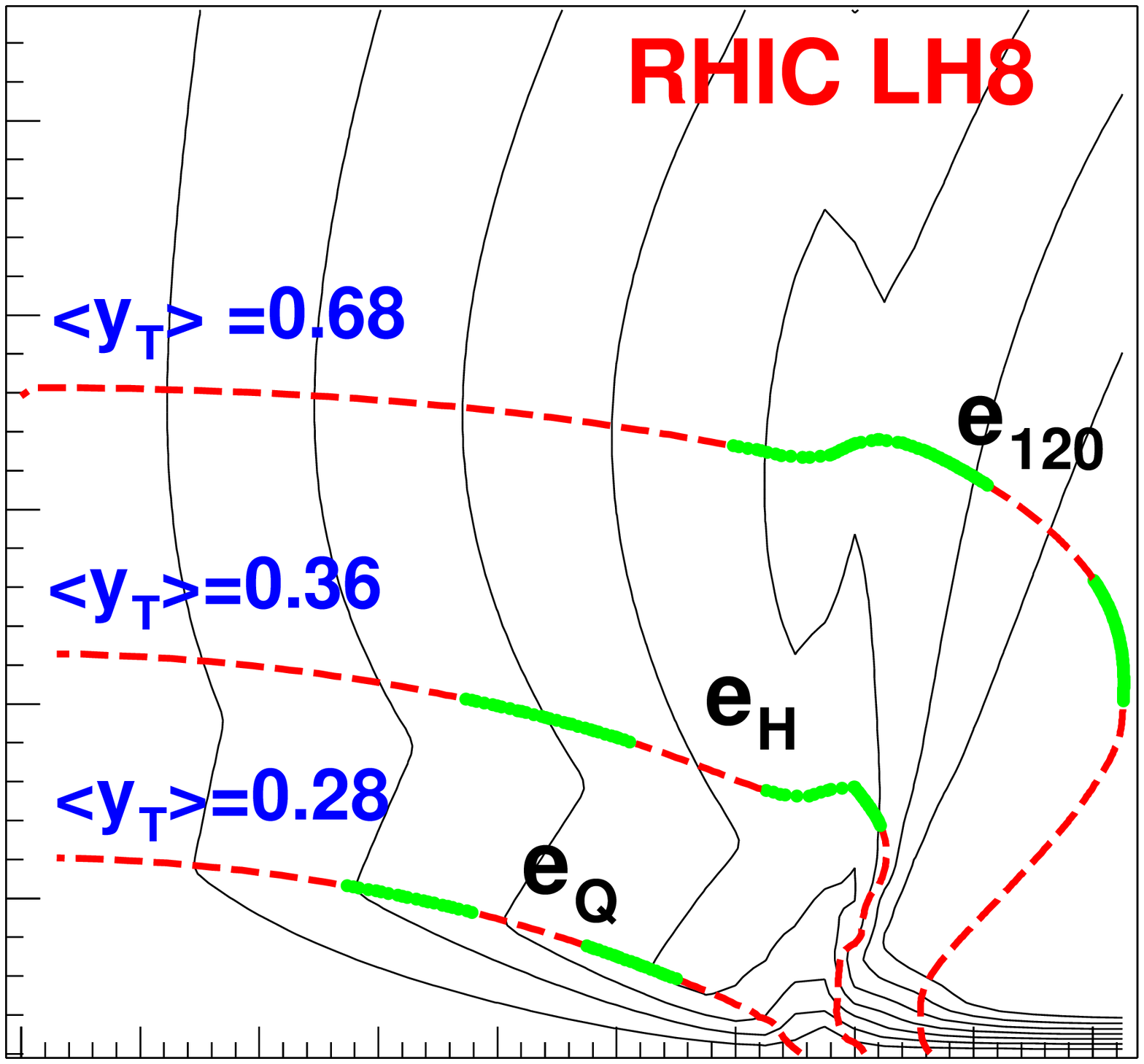}
      \includegraphics[height=1.7in,width=1.7in]{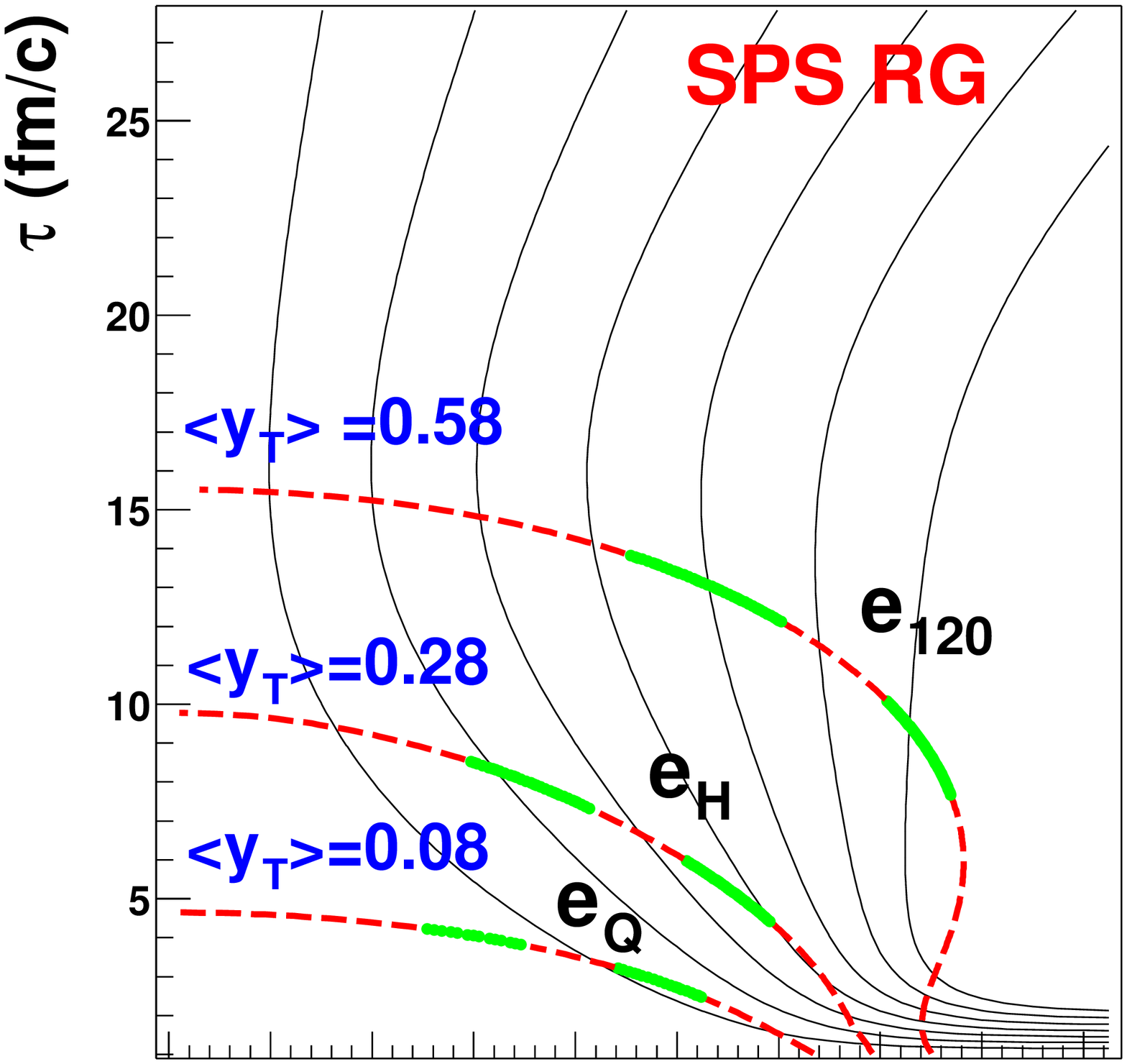}
      \includegraphics[height=1.7in,width=1.6in]{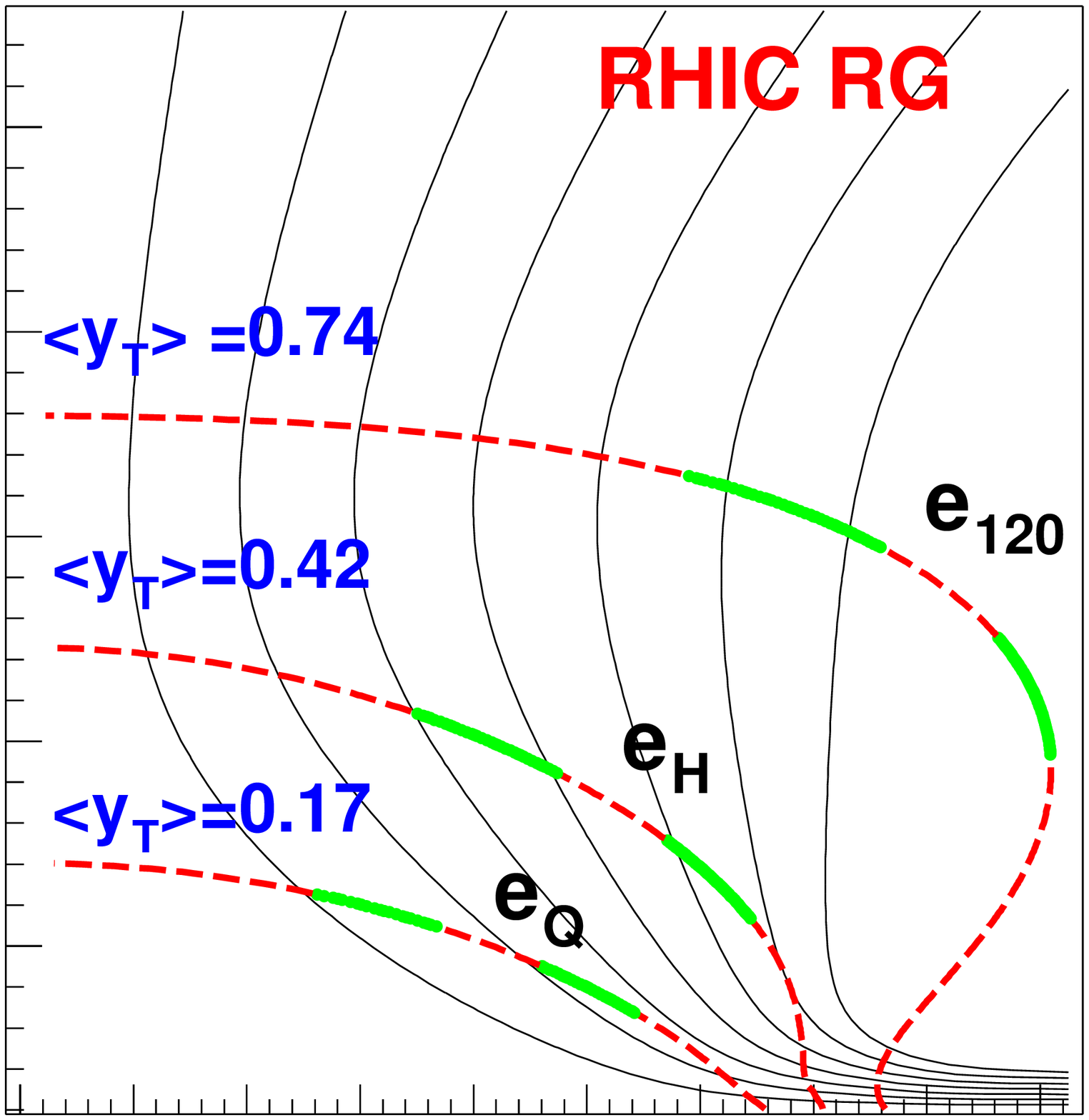}
      \includegraphics[height=1.7in,width=1.7in]{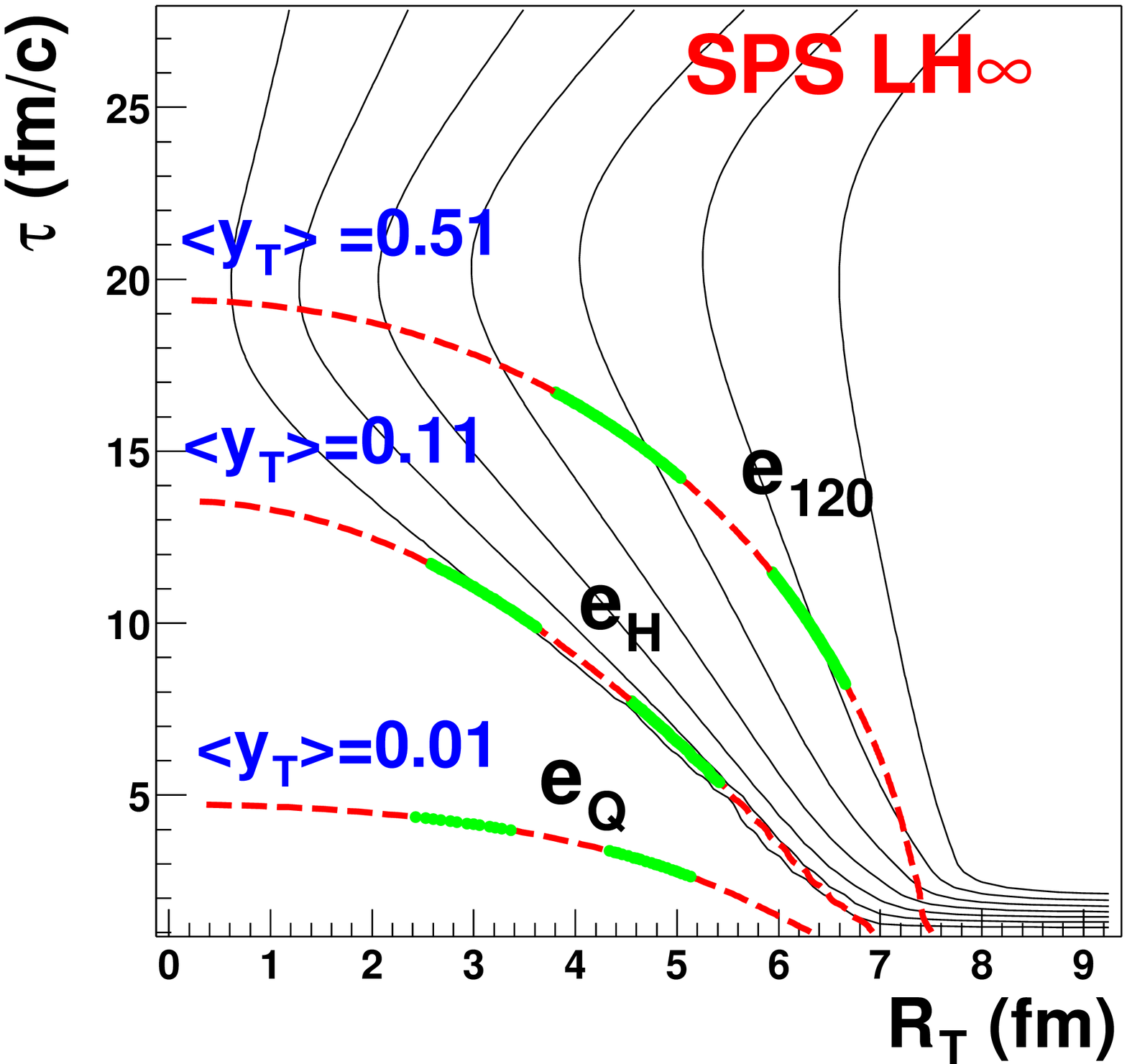}
      \includegraphics[height=1.7in,width=1.6in]{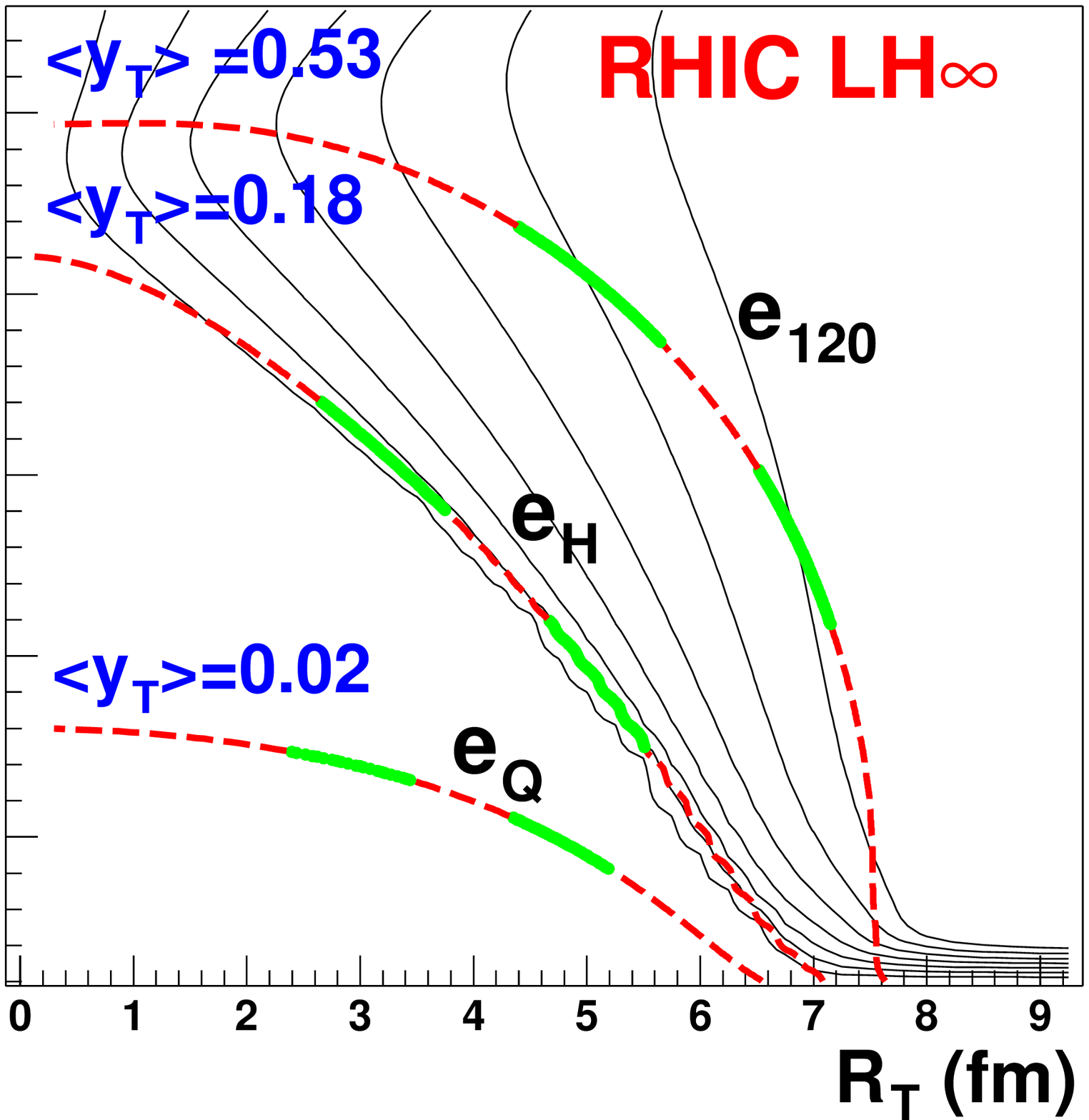}
   \end{center}
   \vspace{-0.1in}
   \caption[An illustration of the hydrodynamic solution
   at the SPS and RHIC for three different EOS.]{
   \label{FSurface}
      The left and right sides show the hydrodynamic solution at 
      the SPS and RHIC for different EOS.
      The thin lines show
      contours of constant transverse fluid rapidity ($v_{T} = \tanh(y_{T})$) 
      with values 0.1,0.2,...,0.7\,.  The thick 
      lines show contours of constant energy density. $e_{120}$
      denotes the energy density where $T=120\,\mbox{MeV}$. $e_{H}$ 
      and $e_{Q}$
      denote the energy density (for LH8) where the matter shifts from
      hadronic to mixed and mixed to a QGP, respectively. 
      The shift to RQMD is made at $e_{H}$. $\langle y_{T} \rangle$ 
		denotes the mean transverse rapidity weighted with 
		the total entropy flowing through the energy density contours.
      Walking along these contours, the line
      is broken into segments by dashed and then solid lines.
      20\% of the total entropy 
      passing through the entire arc passes through each segment.
   }
\end{figure}

This section reviews the hydrodynamic evolution for 
different EOS used in this work.
The evolution at the SPS and RHIC is summarized in  
Fig.~\ref{FSurface}.
The switching isotherm,
$e_{H}$ (shown in the middle), is particularly important
since in the hydro+cascade approach, the particles
are injected into RQMD with the velocity distribution of this
isotherm.  

For EOS with a phase transition (LH8) there are three
phases and three corresponding stages in the acceleration
history.
(i) an explosive QGP phase ($e>e_Q$), in which  the 
matter accelerates rapidly, 
(ii) a soft mixed phase ($e_{H}<e<e_{Q}$), 
in which the matter free streams with constant velocity 
and (iii) a hadronic phase ($e<e_H$), in which the 
hadronic pressure produces additional acceleration.  

The QGP phase 
dictates the duration and transverse size of the
collision. At RHIC, the QGP pressure drives the matter
outward, rapidly increasing the radius, which in turn shortens the
overall lifetime. Therefore, approximately doubling the
total multiplicity from the SPS to RHIC increases the total lifetime
only slightly, from $10\,\mbox{fm/c}$ to $11\,\mbox{fm/c}$. 
All the additional multiplicity is
absorbed by a slightly larger transverse radius.
Similarly, for a RG EOS the acceleration is robust and continuous 
and increasing the total multiplicity only slightly increases 
the radius and lifetime. In bulk, the radii and lifetimes of RG are
similar to LH8.

By contrast, for LH$\infty$,  the stiff QGP phase is 
absent and the mixed phase is dominant at high energy densities.
The strong transverse acceleration associated with LH8,
is replaced with a slow evaporative process.
The radius of the system slowly shrinks as a function
of time.  Unlike LH8, increasing the total multiplicity increases
the lifetime rather than the radius. Between the 
SPS and RHIC the lifetime increases from from $14\,\mbox{fm/c}$ 
to $21\,~\mbox{fm/c}$.  Summarizing, the QGP drives a
transverse expansion; the transverse expansion increases the
radius and shortens the overall lifetime compared to 
an EOS without the QGP push.

\begin{figure}
   \begin{center}
      \includegraphics[height=2.5in,width=2.5in]{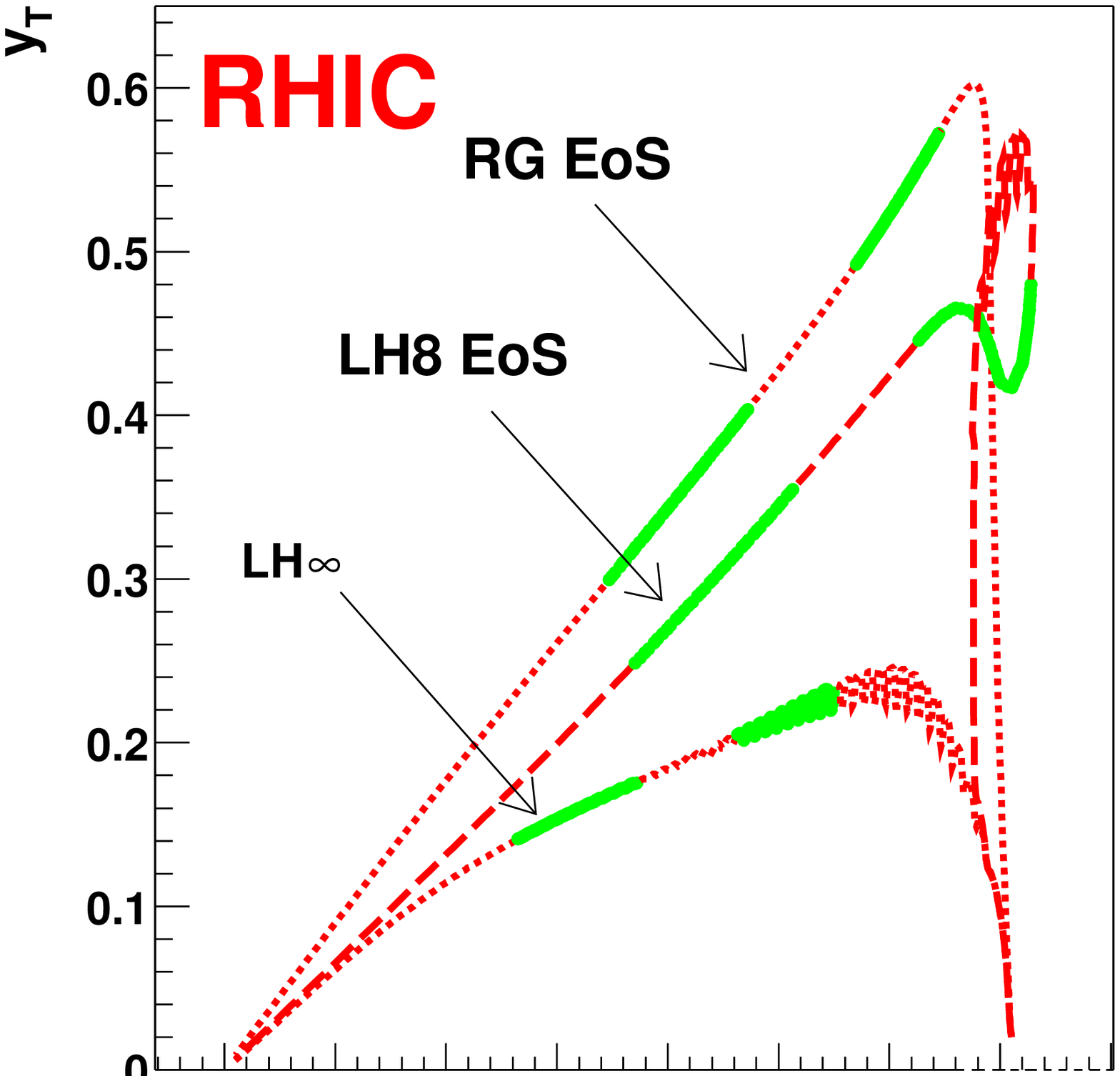}
      \includegraphics[height=2.5in,width=2.5in]{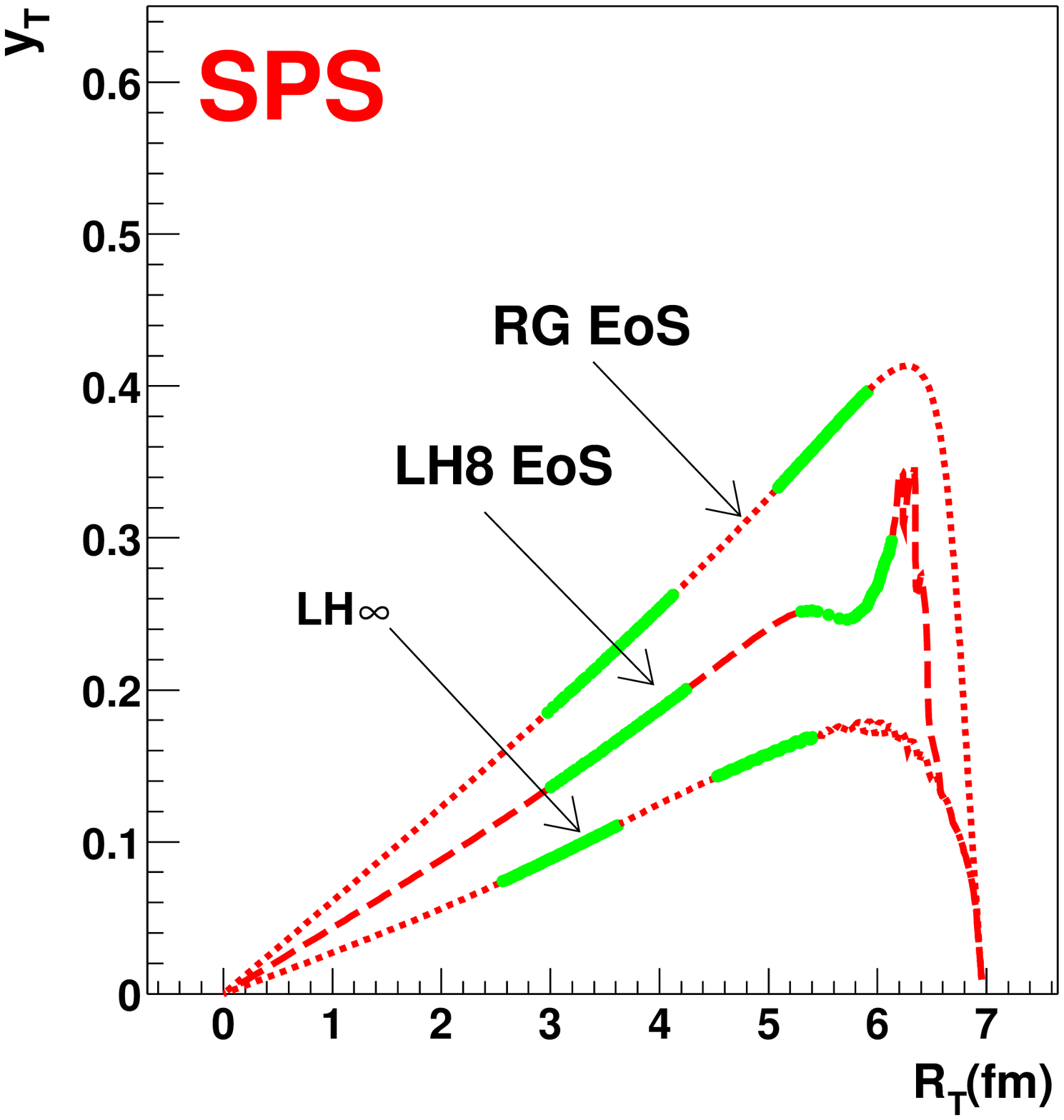}
   \end{center}
   \caption[The flow profile upon input into RQMD for different
   EOS]{
   \label{Switching}
      Walking along the $e_H$ contours in Fig.~\ref{FSurface} (where
      the switch to RQMD is made), 
      the transverse rapidity is traced as a function of 
      radius at (a) the SPS  and (b) RHIC for the three EOS.   
      See Fig.~\ref{FSurface} for an explanation of the 
      dashed and solid segments.
   }
\end{figure}

To quantify the input velocity distributions 
into RQMD, Fig.~\ref{Switching} plots the transverse fluid 
rapidity versus the 
radius along the switching isotherm $e_H$ at the SPS and RHIC.
For LH8 and RG, the transverse
rapidity shows a linear rise with radius.
A linear flow  profile is 
often used in phenomenological fits to the particle 
spectra \cite{SSH-FlowProfile}; this
calculation validates this approach. 

For an EOS with a phase transition to the
QGP (LH8), the acceleration is initially large
but subsequently stalls in the mixed phase. By contrast,
for an EOS without the phase transition (RG),
the acceleration is robust and continuous.
Therefore, although the initial transverse acceleration is
smaller for a RG than for LH8, the  RG velocity at 
the end of the SPS mixed phase is $50\%$ larger than for
LH8. At RHIC, where for LH8 the QGP phase lives longer, 
the RG velocity is only $\approx 15\%$ larger.
Although mean lifetimes and radii of the RG EOS
are similar to LH8, the change in the velocity
distributions from the SPS to RHIC are markedly 
different.  Comparing LH8 to LH$\infty$ at RHIC,the LH8 flow
velocity is approximately twice as large as the LH$\infty$
flow velocity.

Nevertheless, it should be noted that if freezeout
is taken as $T_{f}\approx120\,\mbox{MeV}$,
then the differences between the flow velocities of
the EOS is smeared out by the hadron phase, as 
can be seen by examining the mean flow velocities on
the $e_{120}$ curves in Fig.\,\ref{FSurface}. 
Indeed, the hadronic phase of LH$\infty$ 
(which lives longer, since it is
born with no transverse velocity) can partially compensate for the
weak initial acceleration. Along the $T_{f}\approx120\,\mbox{MeV}$
isotherm,
the flow velocities of LH$\infty$, RG and LH8 are 
roughly comparable.

To characterize the flow in non-central collisions for 
the EOS used in this work,
we follow Kolb $et$ $al.$ \cite{Kolb-UU} and calculate a
quantity derived from the stress tensor for 
\begin{eqnarray}
      \epsilon_{p}  &\equiv&
                \frac{
                  \left\langle T^{xx} - T^{yy} \right\rangle_{S}
                }{
                  \left\langle T^{xx} + T^{yy}\right\rangle_{S}
                } \,,
\end{eqnarray}
where the  $\langle \rangle_{S}$ denotes an average over
the transverse plane weighted with the entropy per area 
per unit spatial rapidity,  $s\gamma\,\tau dx\,dy$. 
$\epsilon_{p}$ is related to 
$\frac{
   \left\langle p_{x}^2 - p_{y}^2 \right\rangle
}{
   \left\langle p_{x}^2 + p_{y}^2 \right\rangle
},
$
the final momentum anisotropy of the particle 
distribution \cite{Ollitrault-Elliptic} which ultimately
is related to $v_{2}(p_T)$ \cite{Kolb-Flow}.
\begin{figure}[!tbp]
   \begin{center}
     \includegraphics[height=3.0in,width=3.0in]{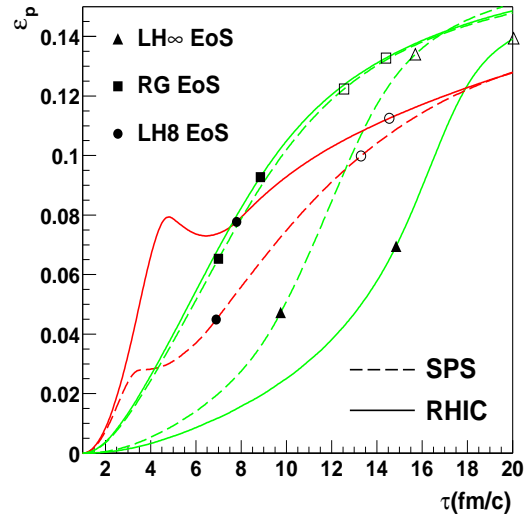}
   \end{center}
   \caption[The transverse momentum and anisotropy of the
   stress tensor as a function of time in the hydrodynamic 
   simulation for different EOS]{
   \label{TimeVt}
 Following Kolb $et$ $al.$ \cite{Kolb-UU}, we show the anisotropy of the stress tensor, $\epsilon_{p}$ (a 
measure of elliptic flow) for the EOS used 
in this work. The solid (dashed) curves 
are for RHIC (SPS). The solid (open) symbols indicate when
the center of the fluid passes through a temperature of 
160\,MeV (120\,MeV). The solid symbols are therefore
representative of the switching temperature to RQMD or $e_H$
in Fig.~\ref{FSurface}. 
}
\end{figure}

The general trends 
seen in $\epsilon_{p}$ follow from
the discussion above on the hydrodynamic
response of each EOS. LH8 shows a strong early response
followed by a stall and subsequent flattening as the matter
distribution becomes almost spherical. At RHIC the
strong early response lives substantially longer as the
matter spends a larger fraction of its total lifetime 
in the QGP phase.
For LH$\infty$ the matter is initially stalled
but rapidly accelerates as it slowly
enters the  hadronic phase (see Fig.~\ref{FSurface} (e) and (f)).

\subsection{Qualitative Predictions of the Hydrodynamic Response}
We can now make some qualitative predictions from the  hydrodynamic solution.
Assume momentarily that  the 
final hadron momentum distributions reflect 
the boundary between the mixed and hadronic
phases or $e_H$ in Fig.~\ref{FSurface}.  
Then with the curves presented in the last section, 
LH8 predicts two qualitative 
changes.
First with Fig.~\ref{TimeVt}, between the
SPS and RHIC, $p_{T}^{2}$ weighted elliptic flow should increase by almost
a factor of two as the QGP replaces the mixed phase 
and dominates the early evolution.
Second with Fig.~\ref{Switching},  the total transverse momentum should increase by 
 30\% as the QGP drives additional transverse motion.
The flow differentiates LH8 from a RG EOS and from LH$\infty$.
Since RG EOS accelerates continuously and does not 
stall in the mixed phase, the transverse momentum
is larger at both the SPS and RHIC. In addition, for SPS 
collision energies, the
elliptic flow ($\epsilon_{p}$) for a RG EOS is almost a factor of two 
larger than for LH8. For 
LH$\infty$, the transverse momentum is very low 
until the very end. 

To make these qualitative predictions quantitative  
and to compare the hydrodynamic solution to experimental data,
it is essential to model hadronic freezeout. Between the
time when $T_{f}=160\,\mbox{MeV}$ (the solid symbols)
and $T_{f}=120\,\mbox{MeV}$ (the open symbols) elliptic
flow changes dramatically for each EOS.  This
is especially true for LH8 at the SPS, where the mixed phase
abruptly stalls the development of elliptic flow but then
the hadronic phase rapidly completes the development.
The differences in the
early acceleration tend to get washed out by the hadronic stage. Indeed,
even LH$\infty$ has a reasonable radial and elliptic flow
by $T_{f}=120\,\mbox{MeV}$.  The extent to which signatures of the
early QGP acceleration remain in the final spectra depends
on whether the freezeout temperature should be taken
as $T_{f}=120\,\mbox{MeV}$ or $T_{f}=160\,\mbox{MeV}$. The 
breakup of a heavy ion collision can only be addressed
with hadronic cross sections and expansion rates.

\subsection{RQMD -- Input and Response}

Relativistic Quantum Molecular Dynamics (RQMD) \cite{Sorge-RQMD} 
is a hadronic transport computer code which incorporates many known
hadronic cross sections. RQMD has been used extensively
to model the heavy ion dynamics
\cite{Sorge-Strange,Sorge-elliptic,Sorge-PreEquilibrium,Sorge-kink,Sorge-Time}. 
Briefly, when two particles come within
$d < \sqrt{\sigma/\pi}$, they elastically scatter or form 
a resonance. Resonance formation and decay dominate the
evolution. The principal reactions are
$\pi\pi \rightarrow \rho$, 
$\pi\,N \rightarrow \Delta$, $\pi\,K \rightarrow K^{*}$, 
and $\pi\,\Lambda \rightarrow \Sigma^{*}$.  
Only binary collisions are considered in this hadron cascade.
Before discussing the response of RQMD, we first consider 
the input.  

\begin{figure}
\begin{center}
   \includegraphics[height=3.0in,width=3.0in]{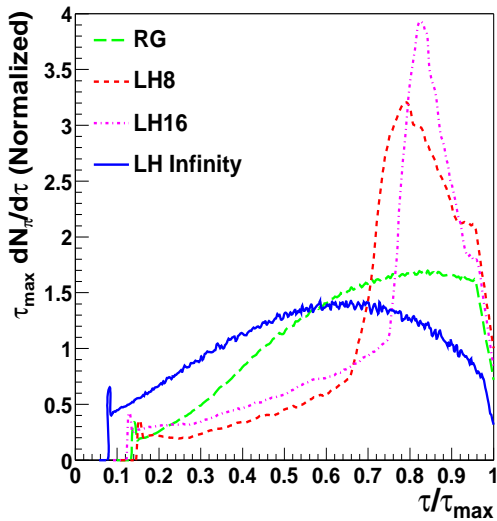}
   \end{center}
   \caption[The fraction of pions emitted per unit time into
   the RQMD stage of the model]{
   \label{TAUX_DSFREEZEDTAUX}
   The fraction of pions emitted into RQMD per unit of time
   relative to the last instant($\tau_{max}$) that the matter
   is evolved by hydrodynamics.
   }
\end{figure}

The time distributions are found
by projecting the entropy
distribution on the switching isotherm $e_H$ in Fig.~\ref{FSurface} 
(or $T=160\,\mbox{MeV}$) onto the $\tau$ axis. 
In Fig.~\ref{TAUX_DSFREEZEDTAUX}
the fraction of pions (or entropy) injected into RQMD per unit $\tau$ 
is plotted as a function $\tau/\tau_{max}$ for each EOS.  
For LH8, very few particles are evaporated
from space-like surfaces at early times, and  at 
$\tau/\tau_{max}\approx 0.8$ ($\tau\approx 9\,\mbox{fm/c}$)
particles are emitted in bulk from  the time component of the switching
surface. For LH$\infty$, particles are continuously 
evaporated from the transition surface and
the radius slowly decreases. Therefore,  
the time distribution is relatively  uniform.
Finally for a RG EOS, the freezeout surface is not box-like and 
particles are also emitted into RQMD slowly and continuously.  

\begin{figure}[!tbp]
   \begin{center}
      \includegraphics[height=1.8in,width=2.8in]{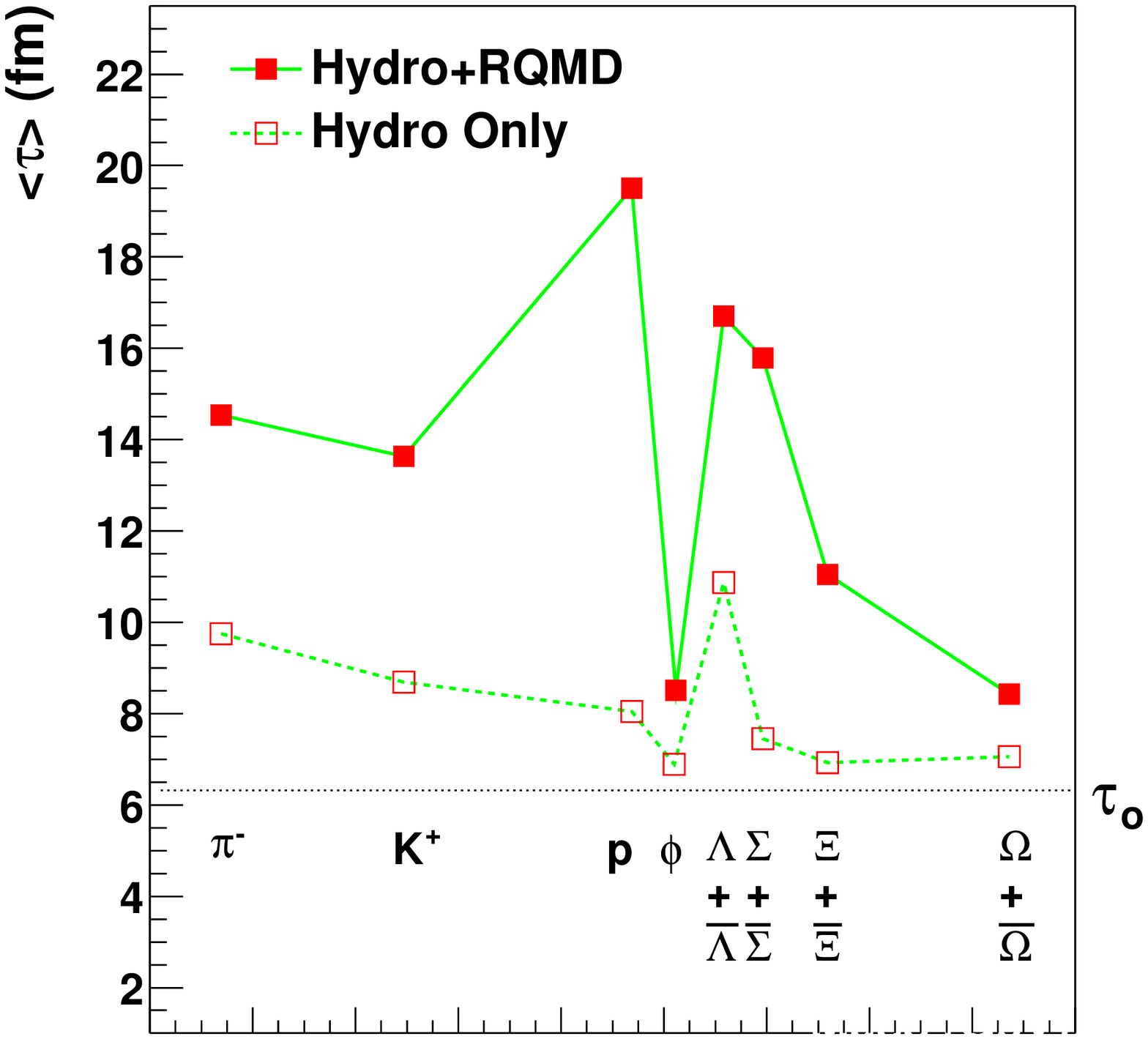}
      \includegraphics[height=1.8in,width=2.8in]{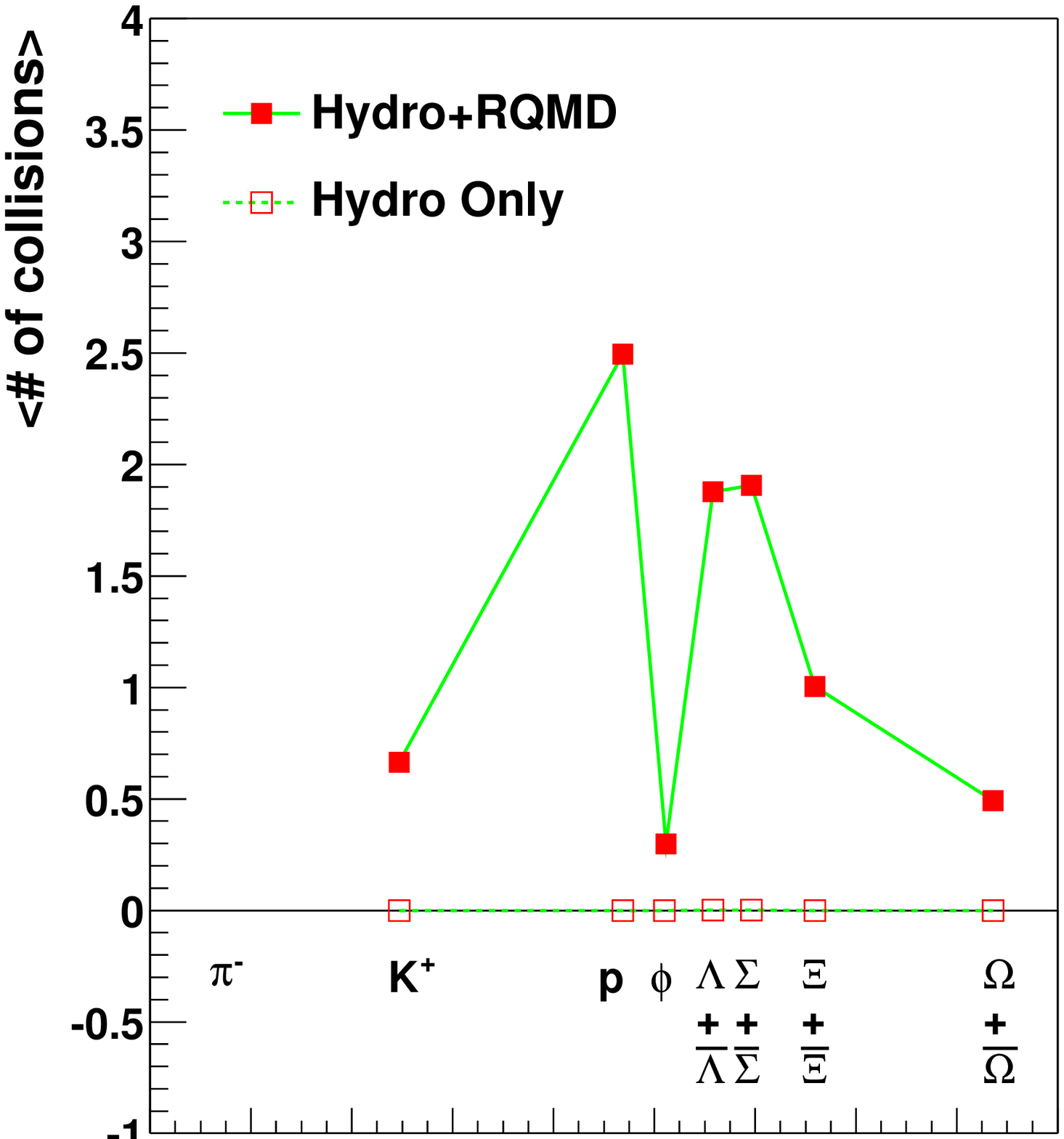}
      \includegraphics[height=1.8in,width=2.8in]{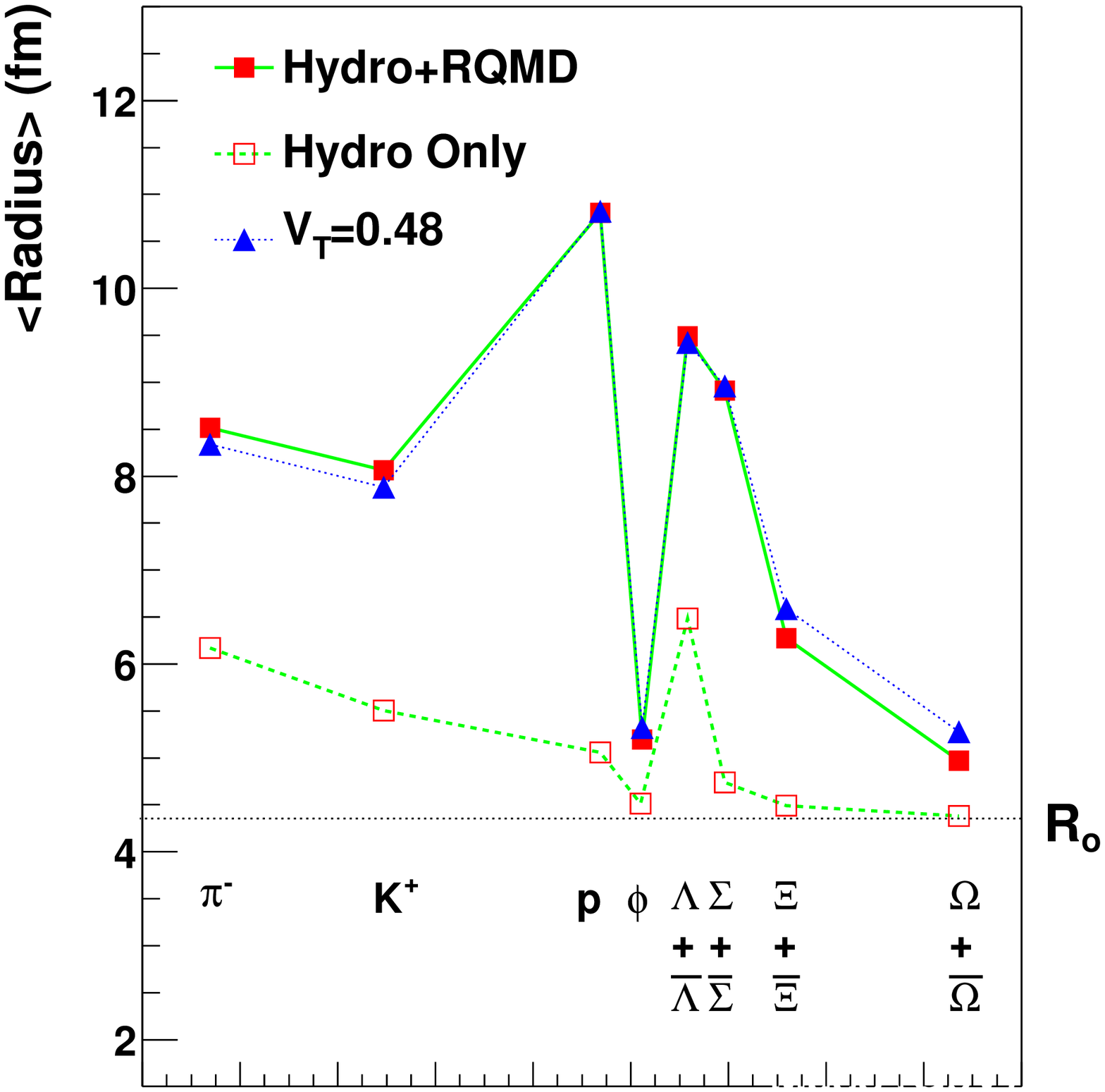}
      \includegraphics[height=1.8in,width=2.8in]{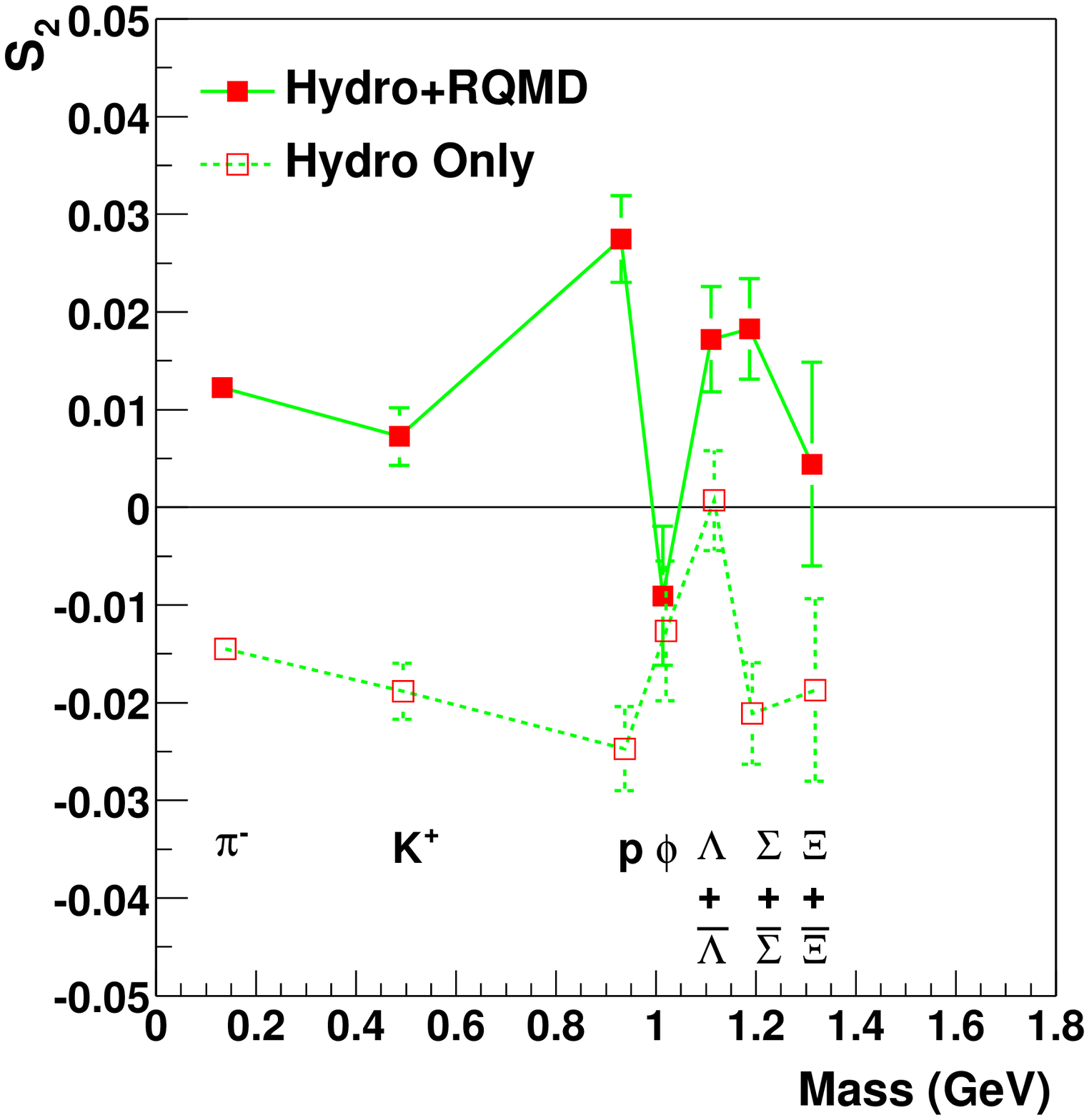}
      \vspace{-0.1in}
   \end{center}
   \caption[The mean emission time, number of collisions, emission radius,
   and spatial anisotropy,  
   as a function of particle mass with and without hadronic rescattering]{
   \label{timeFreezeout}
   The mean (a) emission time, (b) number of collisions, (c) 
   emission radius, and (d)
   spatial anisotropy 
   $s_{2}$ (see Eq.~\ref{s2}) as a function of particle mass with
   and without the RQMD hadronic after-burner. 
   The averages are taken over the points of last interaction for 
	AuAu collisions at b=6\,fm at RHIC. 
   In (a)  and (c),
 $\tau_{o}$,  and $R_{o}$ 
label the
   mean emission time and radius of the $\phi$ meson.  $v_{T}$ denotes 
   the freezeout drift velocity of Eq.~\ref{vdrift}.
   }
\end{figure}

Now consider the dynamic response of the hadronic cascade. 
For LH8, the hydrodynamic input into RQMD can
be characterized as a simple thermal model with a linearly
rising flow profile with a uniform radial distribution except
at the edge of the distribution where there is a small maximum.  
Once this input distribution is taken, the hadrons re-scatter 
within RQMD and different particles decouple from the cascade at different 
times.
Fig.~\ref{timeFreezeout} (a) plots the
mean emission time $<\tau>$ (the time of last interaction)
versus the mass  of the particle species.
Also shown is $<\tau>$, when all
collisions in RQMD are switched
off and only resonance decays are allowed.
The mean number of collisions experienced by a particle is shown
in Fig.~\ref{timeFreezeout} (b).
The  mesons  scatter approximately once after their principal
resonances
($\rho,\,K^{*},$\,etc.) decay and decouple around $\tau \approx 14\,\mbox{fm/c}$.
In contrast, due to strong meson-baryon resonances
$\Delta,~\Sigma^{*},...$ ,
nucleons and  hyperons ($\Lambda$ and $\Sigma$)
scatter approximately  twice 
and decouple around $\tau \approx 18\,\mbox{fm/c}$.
The $\phi$ and $\Omega^{-}$
are emitted directly from the phase boundary since they have small
hadronic cross sections.

The duration of the hadronic stage dictates the spatial extent
of the final source.
In Fig \ref{timeFreezeout} (c), the
mean radius is shown as a function of particle mass with
and without re-scattering in the hadronic cascade.
For comparison, we  apply the simple formulas:
We assume all particles are emitted from the switching surface at
a mean radius $R_{o}$ and a mean time $\tau_{o}$, with a constant
radial velocity $v_{T}$
(see Fig. \ref{timeFreezeout} (a) and (c)). Since $\phi$ is emitted
directly from the switching surface, $R_{o}$ =$\langle R_{\phi}\rangle$  and
$\tau_{o}=\langle\tau_{\phi}\rangle$. With the formula distance = velocity $\times$
time, we have 
\begin{eqnarray}
\label{vdrift}
\langle R\rangle_{x} = R_{o} + v_{T} (\langle \tau_{x}\rangle -\tau_{o}),
\end{eqnarray}
where $\langle R\rangle_{x}$ ($\tau_{x}$) is the freezeout radius (time)  of particle x
and $v_{T}$ is the freezeout drift velocity.
 This velocity
incorporates a thermal drift velocity and the flow velocity of the 
source.    
Given a constant velocity as a function of mass $v_{T}\approx.42\,c$, a 
very simple fit to the freezeout radii is obtained,
 as shown in 
Fig.~\ref{timeFreezeout}(c).  Thus, hadronic cross sections
dictate the final freezeout radii of the source. 

Hadronic cross-sections also dictate the spatial geometry in non-central 
collisions. In non-central collisions the ellipticity of the source
at freezeout is quantified by the spatial anisotropy,
\begin{eqnarray}
\label{s2}
s_{2} = \left\langle \frac {x^2 -y^2} {x^2 + y^2} \right\rangle \,.
\end{eqnarray}
Here, the averages are taken over points of last interaction in the
cascade.
$s_{2}$ is negative for the initial almond-shaped distribution but
positive for a cucumber-shaped distribution. 
Fig.~\ref{timeFreezeout}(d) shows $s_{2}$ without re-scattering but
with resonance decays (Hydro Only) and with hadronic re-scattering
(Hydro+RQMD).  The initial elliptic flow ($v_{2}$) changes the overall geometry ($s_2$)
by the end of the RQMD stage. At the end of the hydrodynamic stage 
 $s_{2}$ is negative, indicating that the 
source retains at least some of its initial almond distribution. 
$s_{2}$ becomes positive as the system evolves and the momentum
asymmetry changes the source geometry. For nucleons, $s_{2}$ is
almost +3\% for modest impact parameters; this may have observable
consequences \cite{STAR-EllipticParticle}.  

\subsection{Impact Parameter Dependence of the Space Time Evolution}

\begin{figure}[!tbp]
   \begin{center}

\includegraphics[height=3.0in,width=3.0in]{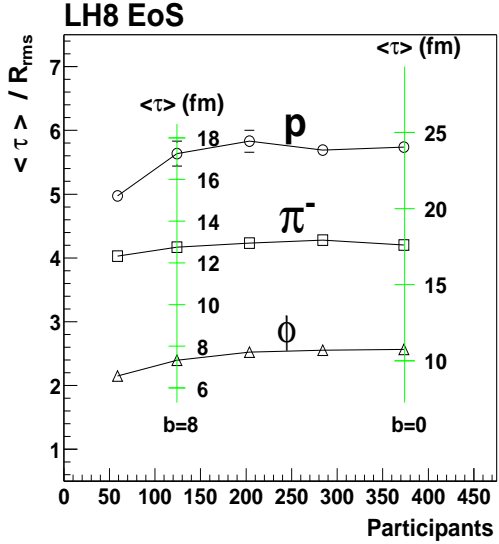}
\includegraphics[height=3.0in,width=3.0in]{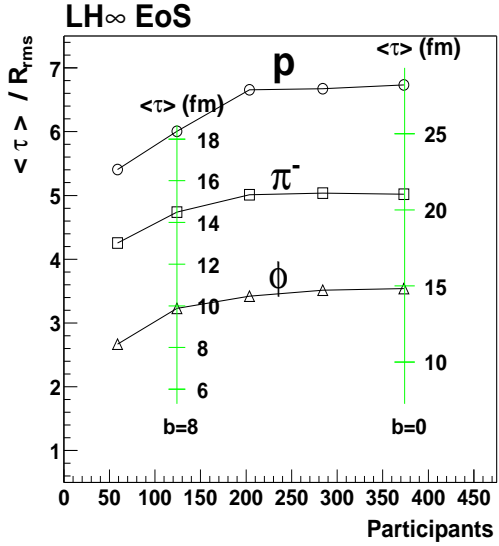}
   \end{center}
   \caption[The mean emission time as a function of participants
   for different particle types and EOS]{
   \label{FreezeoutScaling}
   The mean emission time $\langle \tau \rangle$  relative to the 
   Glauber r.m.s. radius (see Eq.~\ref{GlauberEquRms}) 
   as a function of the number of participating nucleons for
   (a) LH8  and (b) LH$\infty$.  The free axes show the
   absolute lifetime at impact parameters b=0\,fm and 
   b=8\,fm. 
   }
\end{figure}

In the previous section, we discussed how hadronic cross sections
control the lifetime and geometry of the final hadronic
distributions. Now the impact parameter is varied and
the freezeout distributions are modified. 
In non-central collisions, the number of
charged particles scales as the number of participants;  
therefore the lifetime of the hadronic stage should 
also scale as the number of participants. However, the
hadronic lifetime is also a function of the cross section, the
radius, and the expansion rate ($\partial_{\mu}U^{\mu}$). These depend
respectively on the particle species, the r.m.s. radius
of the initial geometry, and the EOS.  
In Fig.~\ref{FreezeoutScaling}, the different contributions 
to the total lifetime are studied.
We plot the mean emission time $<\tau>$,
divided by size $R_{rms}^{Glauber}$, as a function of the number
of participants for different particles and EOS.
To set the absolute scale, the ``free'' axes show $<\tau>$ directly
at two impact parameters.

Consider first the LH8 curves (a):
The total lifetime for all particle species falls by approximately 30\%
from central (b=0\,fm) to peripheral (b=8\,fm) collisions.
The order of particle emission remains as the impact parameter
is varied: First rare species ($\phi,\Omega$) are emitted, 
then mesons ($\pi,K$) and finally baryons ($N,\Lambda$). 
For the $\phi$, which is representative of the
hydrodynamic stage, the curves in Fig.~\ref{FreezeoutScaling}(a) 
are flat at the 15\% level, indicating
that the total lifetime scales roughly with the size of the overlap region.
For pions, the total lifetime also scales with $R_{rms}$. 
For protons, indicative of baryon emission,
the total lifetime does not quite scale as $R_{rms}$ but rather
depends on the absolute number of charged particles in addition to
the geometry. This is natural since the freezeout of 
protons is controlled by the formation of $\Delta$ resonances.

For LH$\infty$, 
$\langle\tau\rangle/R_{rms}$ does change more rapidly than for LH8
This is especially true for nucleons. 
However, for $\phi$ and $\pi$ the difference in the 
$N_p$ dependence of  $\langle\tau\rangle/R_{rms}$ is small and 
to a reasonable approximation,
the lifetimes of  $\phi$ and $\pi$ scale 
with $R_{rms}^{Glauber}$ for all EOS.
Changing the EOS simply moves the various curves up and down on Fig.
\ref{FreezeoutScaling} (a) and (b). 
A RG EOS was also studied (not shown) and the 
lifetime and $N_{p}$ dependence were quite similar to LH8. 

Eq.~\ref{vdrift} relates the freezeout radii and geometry of the different
particles to a freezeout time and a
single freezeout drift velocity ($v_{T}$)  and a single freezeout 
radius ($R_{\phi}$). 
This simple formula was
found to be applicable to all impact parameters with 
approximately the same drift velocity as in central collisions. 
The lifetimes and radii ($R_{\phi}$) all scale with the
root mean square radius.

\begin{figure}[!tbp]
   \begin{center}
      \includegraphics[height=3.0in,width=3.0in]{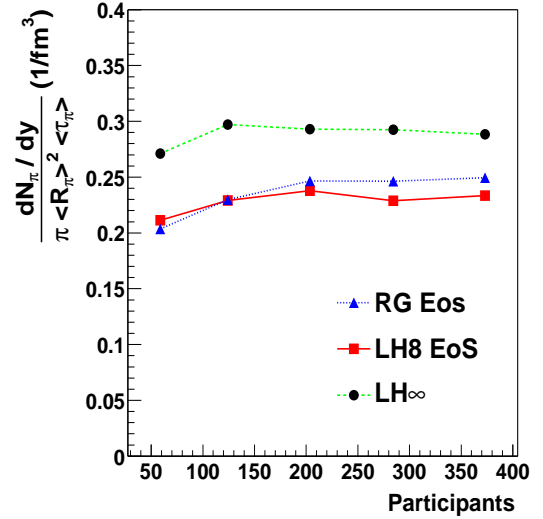}
   \end{center}
   \caption[The density of pions at freezeout  as a function
   of participants for different EOS]{
   \label{TimeEnergy}
   The density of pions at freezeout (see text) versus
   the number of participants for different EOS.
   Here $<\tau_{\pi}>$ and  $<R_{\pi}>$ 
   denote the mean pion emission time and radius.
   }
\end{figure}

Given the rather simple scaling of lifetimes and radii as 
a function of impact parameter, it is natural to 
consider the density of pions at freezeout  as first done in \cite{HydroUrqmd}.
Since pion number is approximately conserved during the cascading process, 
we have $s \propto n_{\pi}$ and  the freezeout  entropy density is  
\begin{eqnarray}
   \frac{dN_{\pi}/dy}{\pi <\tau> <R>^{2}} \sim s_{f} \, .
\end{eqnarray}
This quantity is shown as a function of the number of participants
for 
RHIC collisions in Fig.\,\ref{TimeEnergy}. 
The freezeout entropy density is roughly
constant as a function of impact parameter.
In addition,
the freezeout density is independent of EOS in spite of
differences in transverse velocity gradients.
It has been argued that central PbPb collisions
cool to a lower temperature than peripheral collisions since
transverse and longitudinal velocity
gradients are larger in peripheral 
collisions \cite{Hung-Freezeout}.
However at least in the model,  freezeout is
not driven by the expansion rate; rather, the freezeout 
condition reflects a density where the 
mean free path 
becomes comparable to the radius of the 
nucleus.

\begin{figure}[!tbp]
   \begin{center}
      \includegraphics[height=2.5in,width=2.7in]{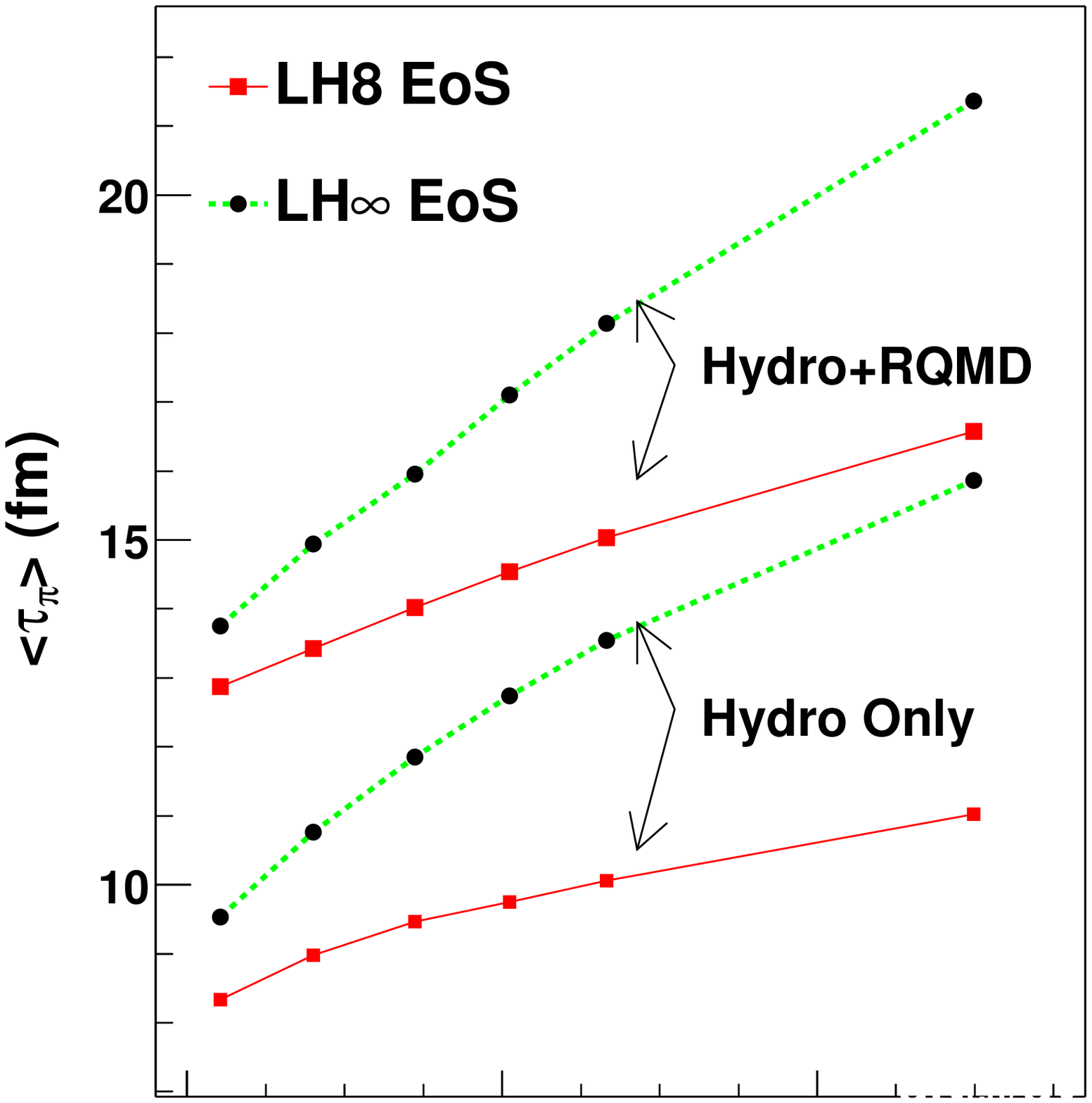}
      \hspace{-0.4in}
      \includegraphics[height=2.0in,width=2.7in]{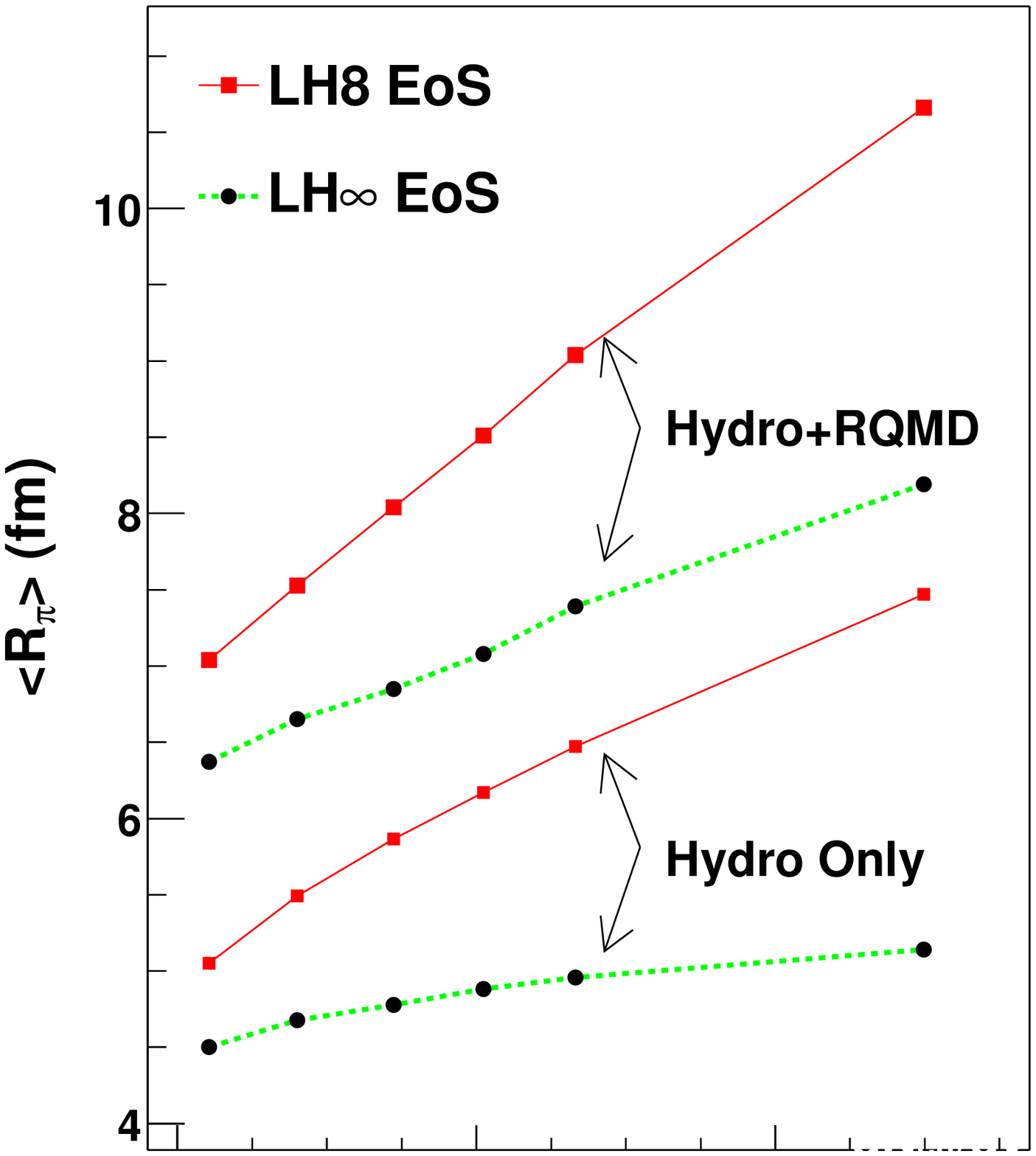}
      \includegraphics[height=2.5in,width=2.7in]{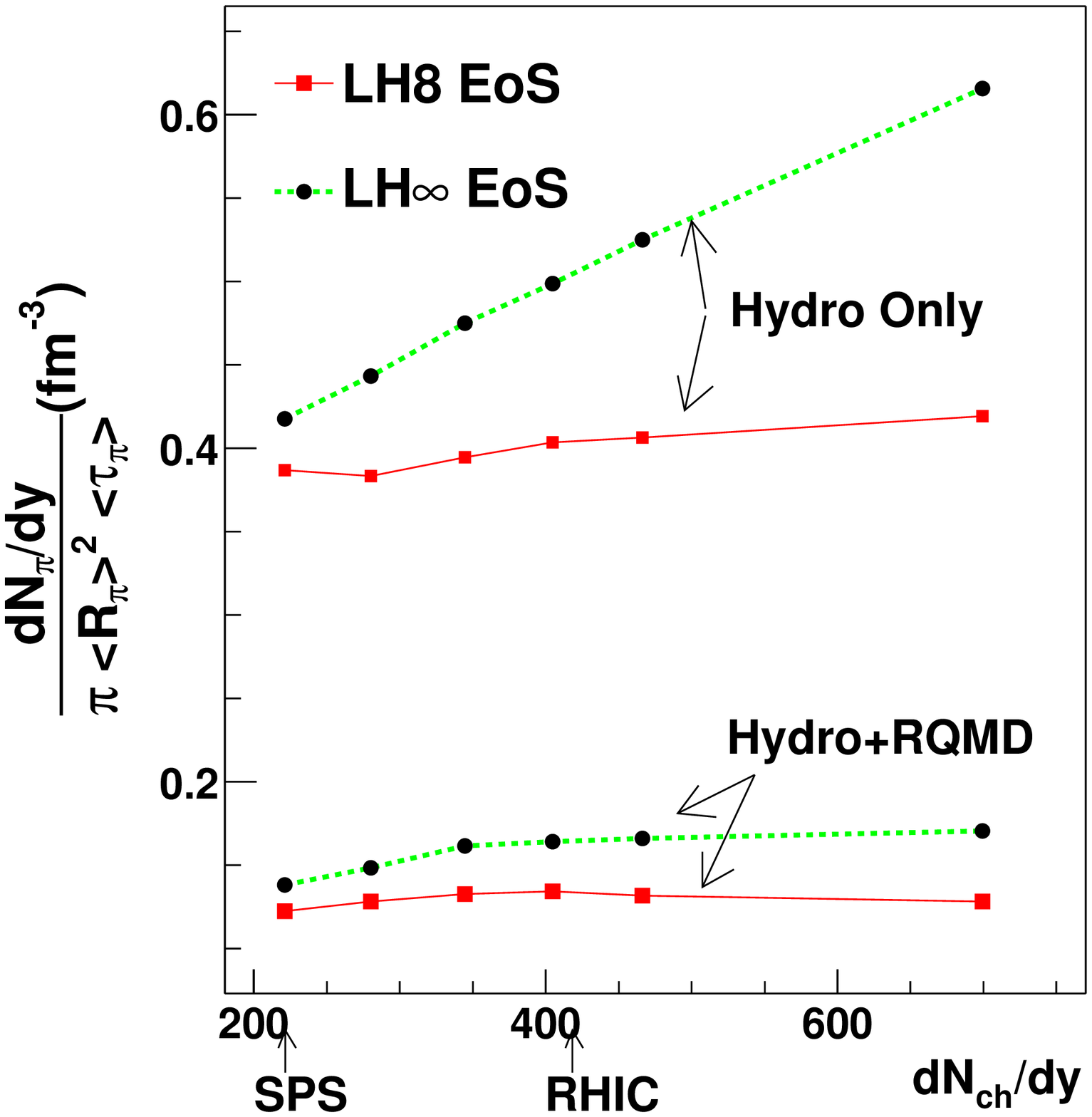}
   \end{center}
   \caption[The mean pion emission time, radius, and density
   as a function of the total multiplicity in the collision]{
   \label{rtaublk}
   The mean pion (a) emission time, (b) emission radius, and
   (c) freezeout density as a function of the
   total multiplicity in a PbPb collision at b=6\,fm. The
   SPS and RHIC arrows denote where the charged particle per participant per rapidity 
   matches SPS and RHIC initial conditions respectively. 
   }
\end{figure}

Next we hold the collision geometry fixed and examine the
changes in lifetime, radius and freezeout density 
as the initial entropy density ($i.e.$,
the collision multiplicity) is increased.
Fig.~\ref{rtaublk} (a) and (b) shows the lifetime and emission
radius for a PbPb collision at b=6\,fm as the multiplicity per
participant is increased from the SPS to the RHIC domain.
For LH8, the radius increases but the emission time does not, while
for LH$\infty$ the situation is exactly reversed.

This behavior is readily understood as 
entropy conservation in the transverse plane.
Entropy conservation (see Eq.~\ref{total_entropy}) 
relates the
entropy density to the Bjorken time ($\tau$), the total conserved
entropy per unit rapidity ($dS_{tot}/dy$),  
and the effective area ($A_{eff}$) of the source with
the schematic relation,
\begin{eqnarray}
\label{sschematic}
    <s> \sim  \frac{dS_{tot}/dy}{\tau A_{eff}(\tau)}.
\end{eqnarray}
As seen with Fig.\,\ref{rtaublk}(c), the
entropy density at freezeout ($s_{f}$) is roughly constant 
as a function of beam energy. 
The freezeout time $\tau_{f}$ 
may therefore be related to freezeout entropy
density $s_{f}$ with 
\begin{eqnarray}
   \tau_{f} \sim  \frac{dS_{tot}/dy}{s_{f} A_{eff}(\tau_{f})} \,.
\end{eqnarray}
For LH8, the strong transverse acceleration rapidly increases the 
area and lowers the entropy density $s$ to $s_{f}$.
Consequently, as the multiplicity is doubled, the total
lifetime increases by only 20\%.
For LH$\infty$, the radius does not increase but the lifetime
increases significantly.  Thus the transverse expansion,
together with entropy conservation, ultimately determine
the total lifetime.

\section{Radial Flow From the SPS to RHIC}
\label{FLRadialFlow}

\subsection{The SPS}
\label{FLRadialFlow-SPS}

In a traditional hydrodynamic calculation 
 \cite{Sollfrank-BigHydro,Schlei-BigHydro,Kolb-Flow}, the pion and 
nucleon yields  fix the total entropy and baryon number 
in the initial conditions.
The freezeout temperature is adjusted to fit the 
pion and proton $p_{T}$ spectra.  In the hydro+cascade approach
advocated here,
the freezeout temperature is not a parameter 
since particles decouple from the cascade when their collision
rates become small. Therefore, the pion and
nucleon yields set the total entropy and baryon number as before,
but the slope parameters provide  significant information about the EOS. 
In particular, the latent heat of
the phase transition which best matches the pion and nucleon  
$p_{T}$ spectrum is  $\mbox{LH}\approx 0.8\,\mbox{GeV/fm}^{3}$.

\begin{figure}[!tbp]
   \begin{center}
      \includegraphics[height=3.0in,width=3.0in]{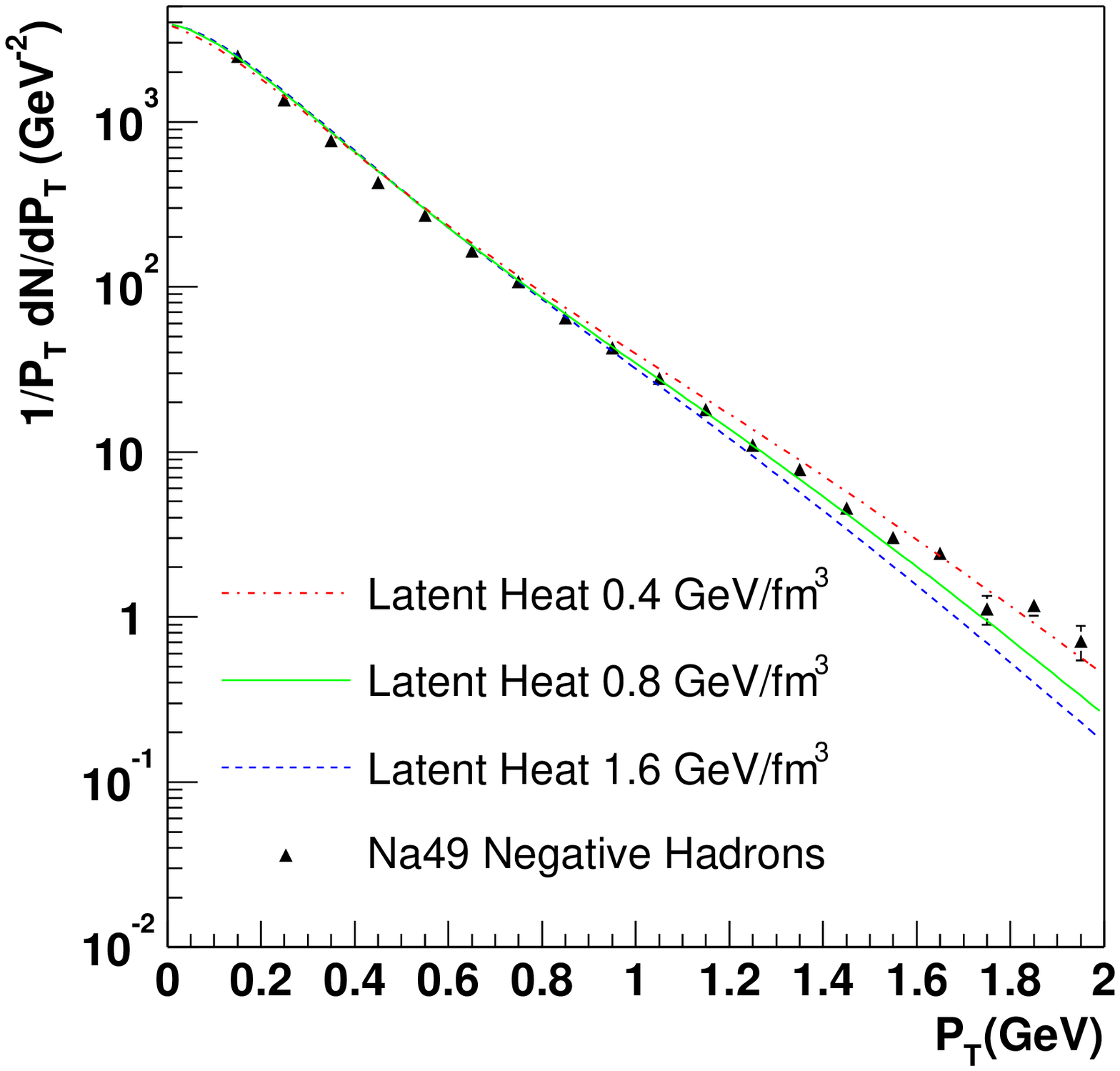}
      \includegraphics[height=3.0in,width=3.0in]{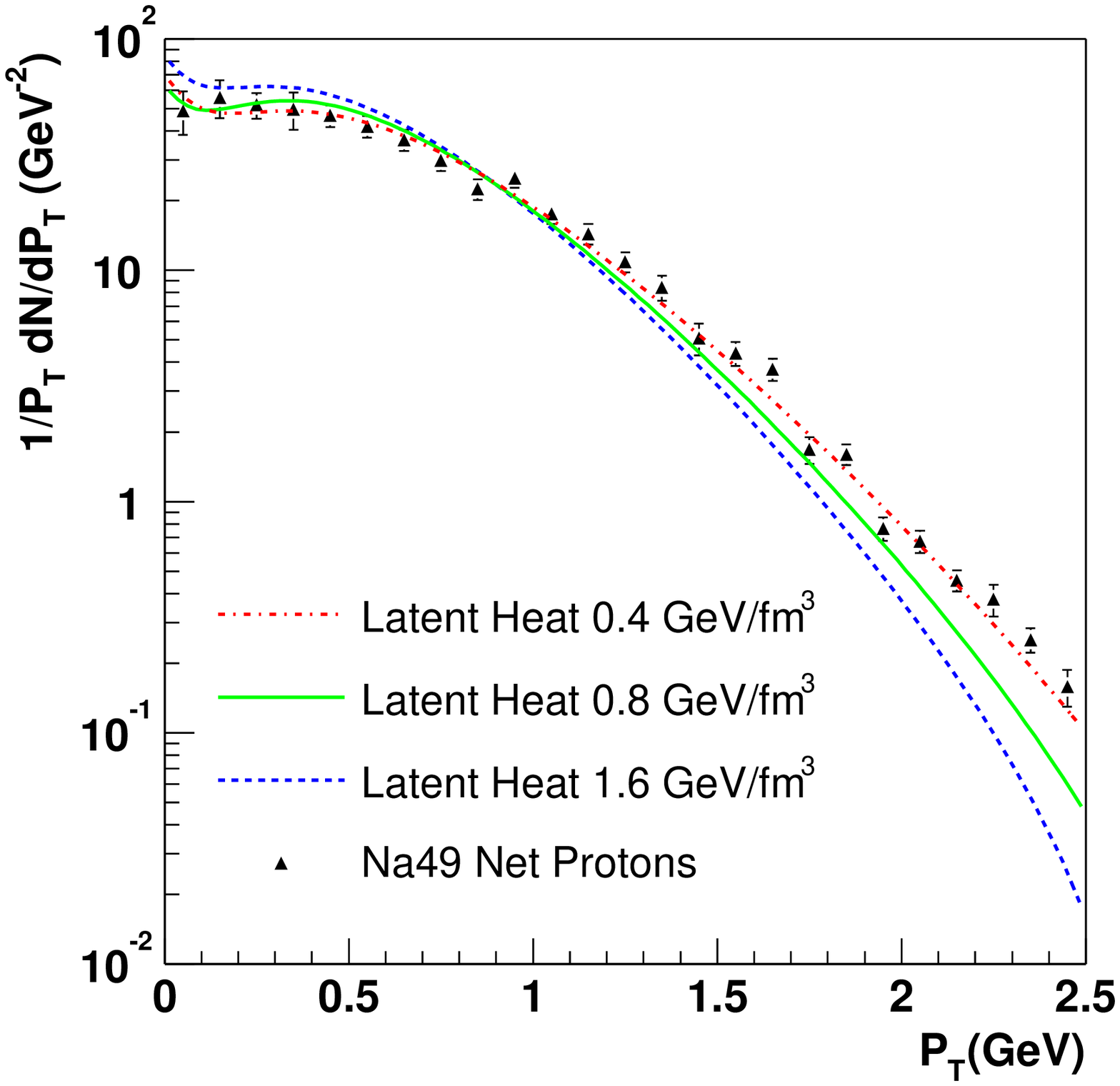}
   \end{center}
   \caption[A comparison  of NA49 negative hadron and net-proton spectra 
   to model results ]{
      A comparison to NA49 \cite{NA49-Spectra} (a) negative hadron and (b) net-proton spectra for different EOS. 
   \label{lhfig}
   }
\end{figure}

In the previous section, we discussed a family of EOS, each 
with a different latent heat. Now we show in Fig.~\ref{lhfig},
the calculated net proton spectrum for the resonance gas EOS and for
EOS LH4-LH16. The experimental  and theoretical spectra 
are absolutely normalized.
The two numbers parameterizing 
the initial conditions $C_{S}$ and $C_{n_B}$ (see Table~\ref{snbtable}) 
are adjusted to
match the height of these spectra. The model curves have been multiplied by
a factor of 0.93  to account for the fact that the data is 5\%
central. Once the height of the spectrum is tuned, the
shape of the spectrum is determined by the course of the
hydrodynamic evolution, or more generally, by 
pressure gradients and the duration
of the collision.  Therefore, it is significant that
hydrodynamics generates a flow $v_T\approx 0.5$, which
is needed to explain the spectra. This flow velocity
was extracted from a variety of thermal  
analyses \cite{Uli-Expand}.

For EOS with large latent heats (e.g. LH16), 
the $p_T$ spectrum is too soft. This is because
the hydrodynamic system spends 
a long time in the mixed phase in which  
pressure gradients do not generate collective
motion.  Bag-model equations of state, employed in many hydrodynamic
calculations \cite{Sollfrank-BigHydro,HydroUrqmd}, typically have a latent heat from $1-1.5\,\mbox{GeV/fm}^{3}$ which makes the EOS rather soft. 
This large latent heat is usually compensated by adjusting
the freezeout temperature \cite{Sollfrank-BigHydro}.
An EOS with a modest first order phase transition (e.g. LH4 and LH8) 
generally reproduces the shape of the $p_{T}$ spectra in Fig\,\ref{lhfig}.
Unfortunately, a RG EOS can also reproduce the shape of the $p_{T}$ spectrum
and additional experimental information is needed to separate EOS.

\begin{figure}[!tbp]
   \begin{center}
      \includegraphics[height=2.7in,width=3.0in]{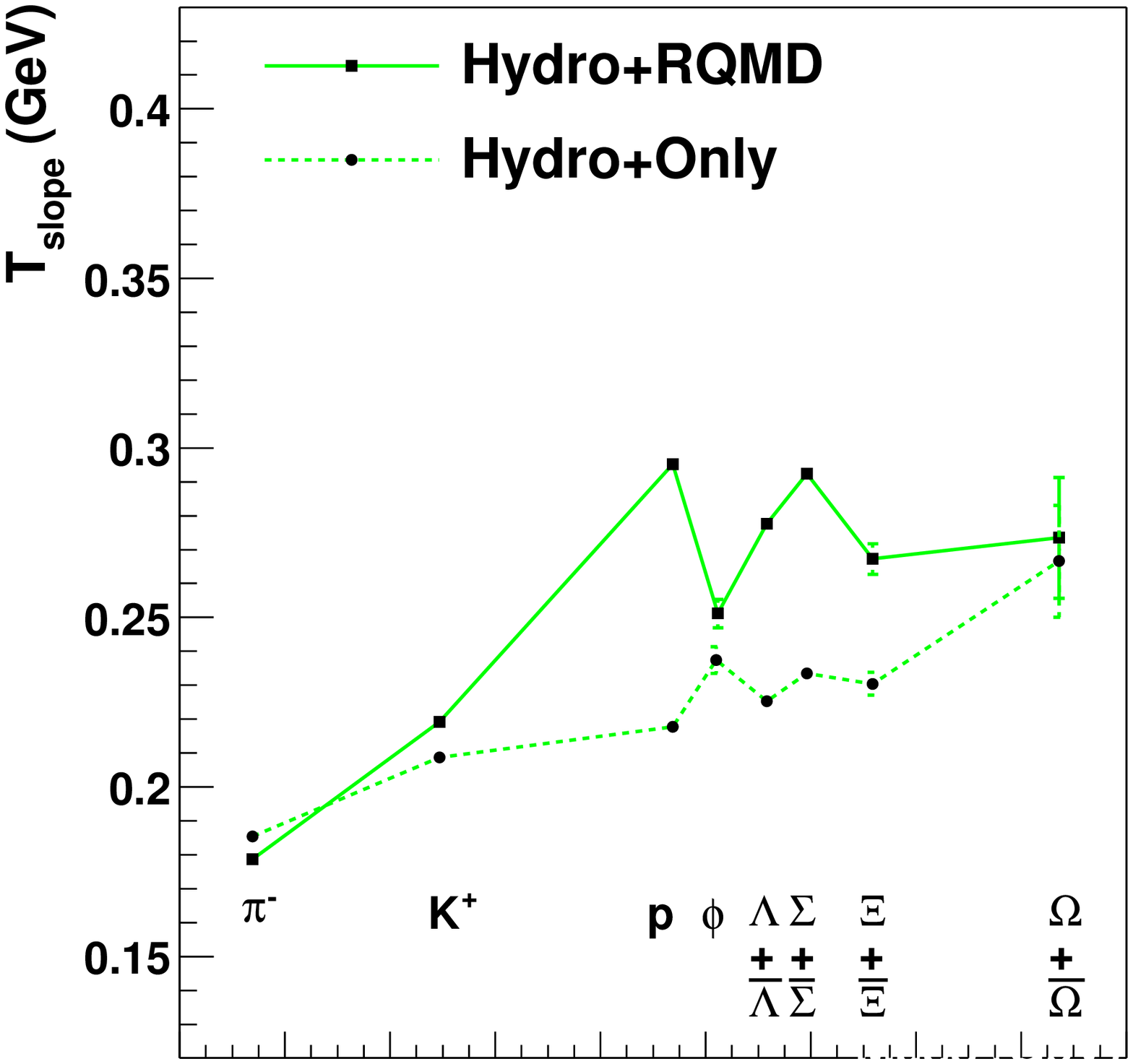}
      \includegraphics[height=2.4in,width=3.0in]{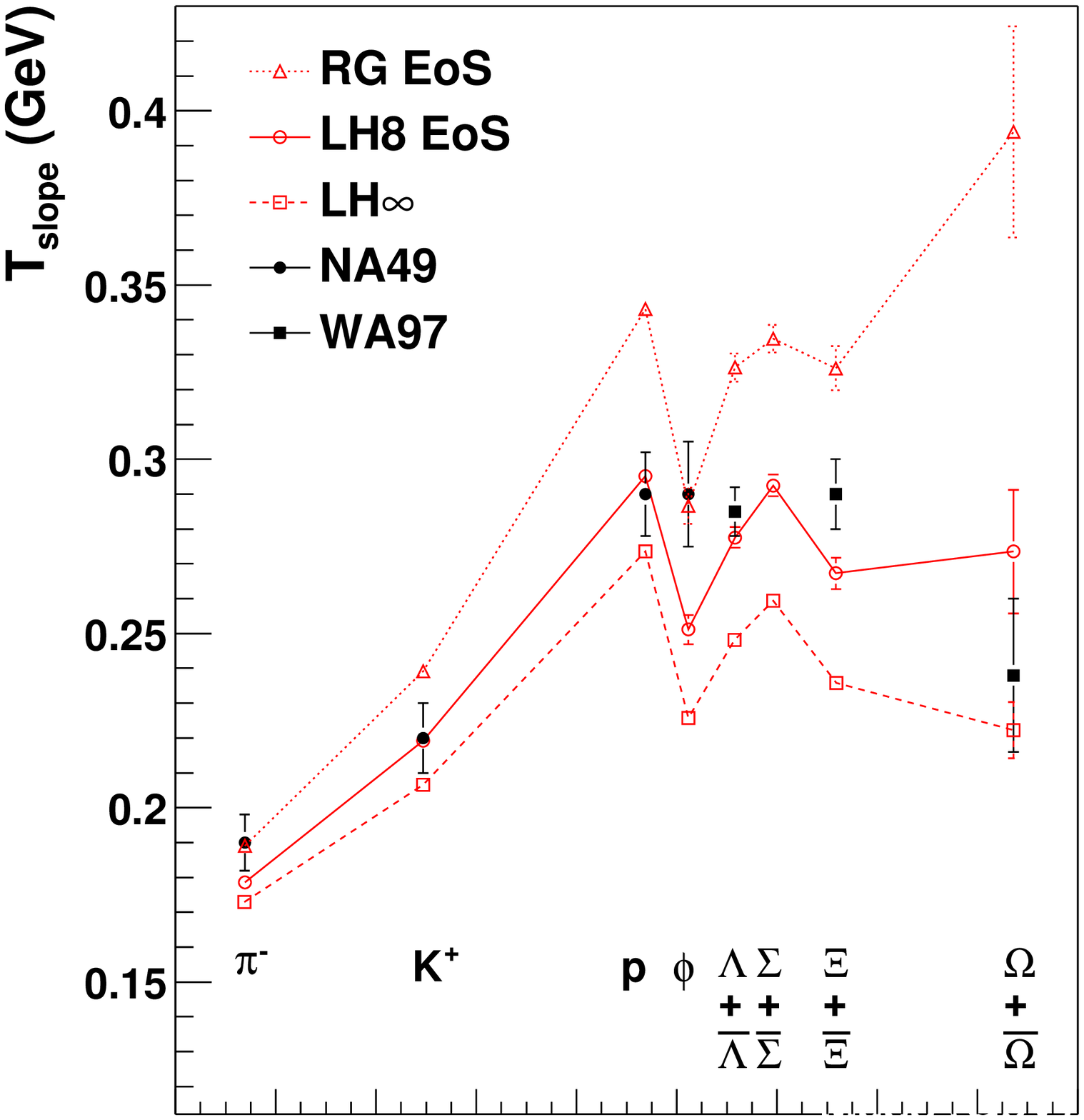}
      \includegraphics[height=2.6in,width=3.0in]{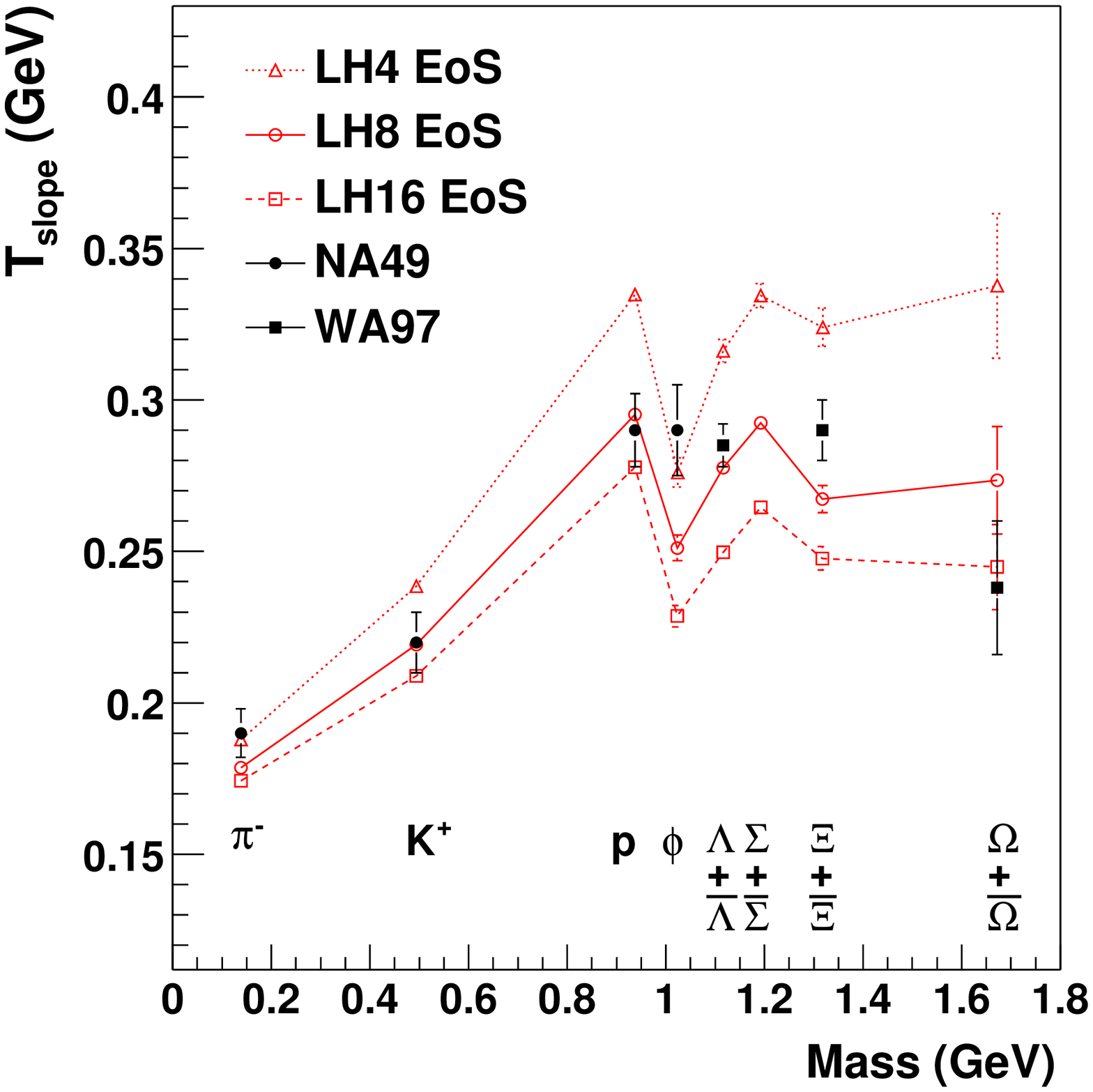}
   \end{center}
   \caption[A comparison of model and experimental slope parameters at the
   SPS. ]{
      \label{slopessps}
      A compilation of slope parameters from the SPS experiments and
      a comparison to the model for different EOS. (a) 
      shows the model slope parameters with and without hadronic
      rescattering
      (b) 
      shows the slope parameters for three qualitatively different
      EOS: RG, LH8, and LH$\infty$. (c) shows the sensitivity
      to the latent heat in the LH family of EOS. NA49
      data points are from \cite{NA49-Slopes} while WA97 data points are 
      from \cite{WA97-Slopes}. 
		The model spectra are  fit with Eq.~\ref{slopeparam} over the
		range corresponding to the WA97 \cite{WA97-Slopes} experimental acceptance 
		 ($M_{T} - m = 0.0-0.9\,\mbox{GeV}$).
   }
\end{figure}

The slope systematics of $\Lambda$, $\Xi$ and $\Omega$ provide
the necessary information.
The presence of a phase transition stalls the 
acceleration \cite{Rischke-Lifetime,HS-soft}; 
therefore information about the velocity at the 
end of the mixed phase can separate  a RG EOS from LH8.
At the SPS, the spectra of the different particle species are all
reasonably exponential and a slope parameter, $T_{slope}$ is extracted.
Specifically the data are fit to the following form,
\begin{eqnarray}
   \label{slopeparam}
   \frac{1}{M_{T}} \frac{dN}{dM_T}&=& C \exp(-\frac{M_{T}}{T_{slope}}),
\end{eqnarray}
where $M_{T}=\sqrt{m^2 + p_T^2}$. In this parameterization, the 
slope parameter $T_{slope}$,  
is directly related to the $<M_T>=2 T_{slope} + \frac{m^{2}}{T_{slope} + m}$.
The model spectra are  fit with Eq.~\ref{slopeparam} over the
range corresponding to the WA97 \cite{WA97-Slopes} experimental acceptance 
 ($M_{T} - m = 0.0-0.9\,\mbox{GeV}$);
the slope 
parameters are shown versus particle mass with and without RQMD
in Fig.~\ref{slopessps}(a). 
First look at the ``Hydro Only'' curves. The slopes increase
approximately linearly  
with mass as is expected in a thermal expanding source
model \cite{SZ-FlowProfile,SSH-FlowProfile,SR-FlowProfile}. 
The non-monotonous increasing mass in the ``Hydro Only'' curves is
due to resonance decays and the baryon content of the particles.
Once RQMD is included, the slopes are modified  by hadronic rescattering
leading to mass dependence characteristic of differential
freezeout \cite{Sorge-Strange}.
Note the following features. First, in the model the $\Omega$
gives a 
good measure of the flow velocity at the end of the mixed phase.
Second, note the $\approx40\%$ 
increase in the nucleon and $\Lambda$ 
slope parameters and  the small 
$decrease$ in the pion slope parameter, due to cooling. As the hadron
gas expands, the pions excite $\Delta$ and $\Sigma^{*}$ 
resonances and drive  
additional transverse motion in the nucleon and hyperon sectors. However,
the pions increase the nucleon $\langle M_{T} \rangle$ only 
at the expense of their own kinetic energy.
In a traditional hydrodynamic approach, the   hydrodynamic evolution
would be continued to match the slope of the nucleon spectrum. Judging
from Fig.~\ref{slopessps}(a), this is
misguided, as a nucleon receives much of its momentum 
after pions have decoupled. As shown below, 
the nucleon receives about 20\% of its transverse 
kinetic energy from the pion ``wind'', irrespective of the 
 colliding energy. 
To incorporate the rich freezeout dynamics of a cascade,
different freezeout 
temperatures  and velocities should be taken for different particles
\cite{Hung-Freezeout}.

A comparison to the available data on slope systematics is given in 
Fig.\,\ref{slopessps}: (b) shows the slopes for different
types of EOS while (c) shows the sensitivity to the latent heat.
Although RG and LH4  are capable
of reproducing the pion  and nucleon spectra,  they significantly 
over-predict the slope parameters of  $\Lambda$, $\Xi$ and $\Omega^{-}$.
This is because  
LH4 and RG already have developed a substantial 
flow velocity at the end of the mixed phase. 
The slope parameter of the $\Omega$ is a 
sensitive measure of the flow velocity
at the end of the mixed phase since the flow velocity is
amplified by the mass in the approximate formula 
$T_{slope}=T_{th} + m\,<v>^{2}$.  
LH$\infty$,  by contrast, under-predicts the slope
parameters of $\Lambda$ and $\Xi $, indicating that 
$\langle \mbox{v}_{T} \rangle$  is too small at the end of the mixed phase.
With LH8, the acceleration is modest -- but significant -- and the
slope systematics are generally reproduced. 
In the model, only an EOS which has both 
a stiff and soft piece is capable of reproducing general trends
seen in the particle spectra.

\subsection{Qualitative Changes at RHIC}
\label{FLChanges}

\begin{figure}[!tbp]
   \begin{center}
      \includegraphics[height=3.4in,width=3.4in]{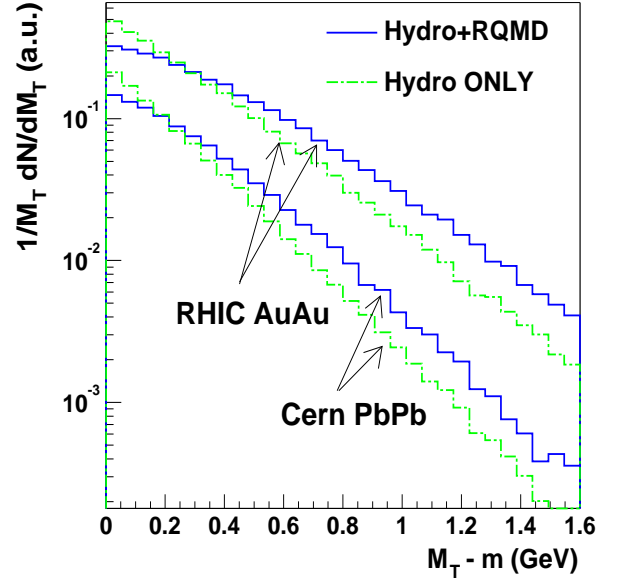}
   \end{center}
   \caption[The nucleon $M_{T}$ spectrum with and without hadronic
   rescattering]{
   \label{mt_spectra}
     The nucleon $M_{T}$ spectrum at the SPS and RHIC with and without 
     the RQMD after-burner.
   }
\end{figure}

 It was argued above that
LH8 provides the best description of the  radial flow
at SPS:
Now the same EOS is used to make predictions for RHIC. 
At RHIC, the initial energy density is well above the
phase transition, 
and the large early pressure is expected to
drive collective  motion. 
In Fig.~\ref{mt_spectra}, the nucleon
$M_{T}$ spectrum for the SPS and RHIC are shown with and without
the hadronic rescattering in RQMD. Two features are immediately
observed:
1.  The $\langle M_T \rangle$ increases as the collision energy 
is increased from the SPS to
RHIC \cite{Kataja-MixedPhase,HydroUrqmd,Ollitrault-MixedPhase}.  
2. The spectra without hadronic rescattering are reasonably well
described by a single exponential ($i.e.,$ they look linear
on the log plot shown). Once rescattering is included the spectra
are curved; this  curvature grows from the SPS to RHIC. Describing
the spectra with a single slope parameter, although useful in 
summarizing a large variety of data, is only approximate.

\subsection{The $\langle M_T \rangle$ from the SPS to RHIC and Beyond}
\label{sectMtmm}

To summarize the bulk energy transport in the model we show in
Fig.\,\ref{mtmm}:
\begin{figure}[!tbp]
   \begin{center}
      \includegraphics[height=2.9in,width=2.9in]{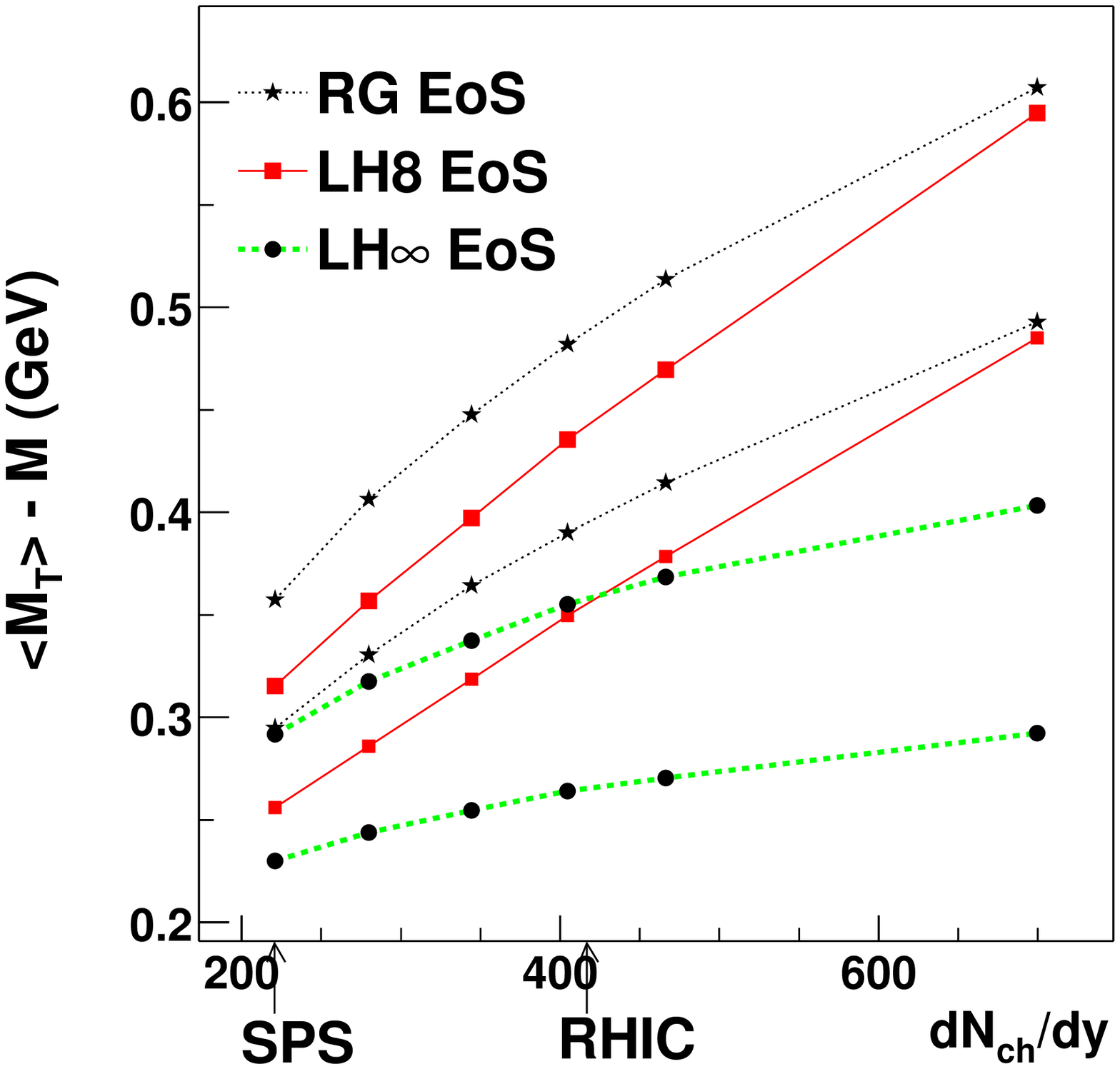}
      \hspace{-0.4in}
      \includegraphics[height=2.9in,width=2.9in]{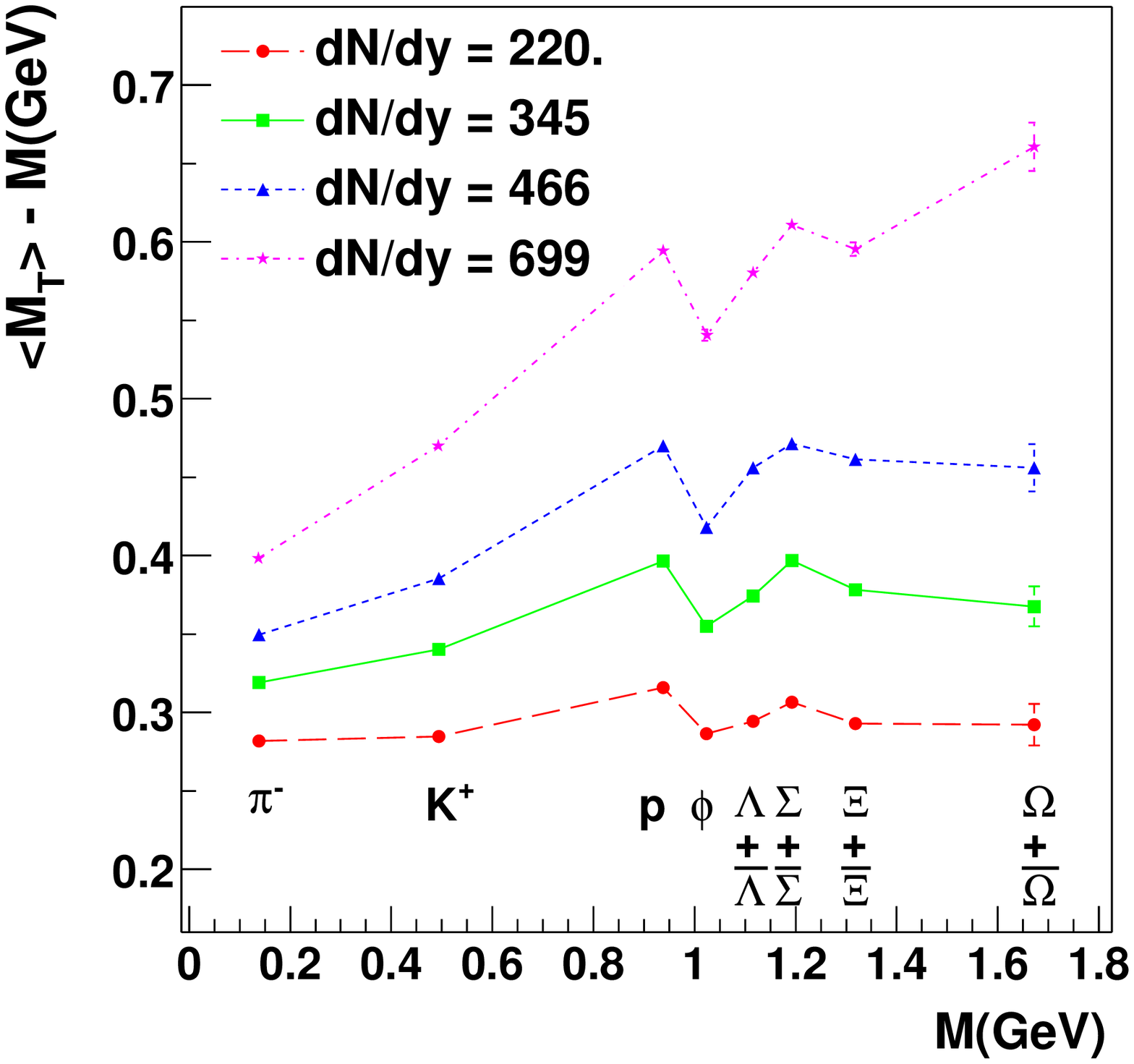}
   \end{center}
   \caption[The dependence of $\langle M_{T}\rangle$ on 
     EOS,  hadronic rescattering, particle type and total multiplicity.
   ]
   {
   \label{mtmm}
     For PbPb collisions at b=6\,fm,
     (a) shows $\langle M_T \rangle - M$ for nucleons versus the charged
     particle multiplicity ($dN_{ch}/dy$). For each EOS the 
     upper curve is with hadronic rescattering (Hydro+RQMD) and
     the lower curve is without hadronic rescattering (Hydro Only).
     (b) shows $\langle M_T \rangle - M$ for different particle species as the    
total multiplicity is increased from the SPS to RHIC and beyond. 

   }
\end{figure}
(a) the $\langle M_T \rangle-M$ for nucleons as a
function of collision energy ($dN_{ch}/dy$) for different EOS, and
(b) the $\langle M_T \rangle-M$ for different particle species as a
function of mass for different collision energies (particle multiplicities).

Fig.~\ref{mtmm}(a) demonstrates that the additional entropy gets converted
into additional transverse motion for each EOS. For each EOS,
the hadronic contribution to the mean $\langle M_T \rangle$
remains constant and is approximately 20\% for LH8 and  
RG, but is approximately 30\% for LH$\infty$. LH$\infty$ is a special
case, and we may say that the RQMD contribution
is approximately 20\% and is independent of the underlying EOS.

Fig.~\ref{mtmm}(b) 
demonstrates how  the increase in the 
mean $M_T$ influences  the spectra of different particles
by plotting $\langle M_{T} \rangle$ versus mass \cite{Sorge-Strange}. 
At the SPS the flow velocity at the end
of the mixed phase is relatively small -- $v_{T} \approx 0.4$\,.
The slopes before the RQMD phase show a  linear rise characteristic
of hydro, $T_{slope}= T_{th} + m\,\langle v^{2} \rangle$. 
When the flow velocity is small, hadronic rescattering changes
the linear mass dependence significantly, giving the
characteristic shape observed at the SPS. As the flow velocity
increases from the SPS to RHIC and beyond, 
the linear rise with mass becomes  increasingly steep and 
hadronic rescattering, while still contributing to  
20\% of the $\langle M_T \rangle$ for nucleons,  does not change
the overall mass dependence. The qualitative shape of the 
mass dependence of $\langle M_T \rangle$ therefore gives
a good measure of the flow velocity at the end of the mixed
phase. Since this flow velocity is 
different for different EOS,
the mass dependence of the $\langle M_T \rangle$
can therefore separate the different EOS studied.
 

\subsection{The Flow Profile from the SPS to RHIC }
\label{Profile}

 The curvature
in the $M_T$ spectrum is a signature of a radially flowing thermal
source. The general features can be understood from a simple 
thermal model. For a cylindrically symmetric shell, which 
expands longitudinally
in a boost invariant fashion and  which freezes out in an instant  
with constant temperature T, and a radial velocity 
$v_{T}=\tanh{\rho}$, 
the $M_T$ spectrum is given by \cite{SZ-FlowProfile,SSH-FlowProfile,SR-FlowProfile} 
\begin{eqnarray}
\label{Bessel}
   \frac{1}{M_{T}}\frac{dN}{dM_{T}} &\propto& 
   M_{T}I_{0}\left(\frac{M_{T} \sinh \rho}{T_{th}}\right) 
   K_{1}\left(\frac{M_{T} \cosh \rho}{T_{th}}\right)  \, . \nonumber \\
\end{eqnarray}
For $p_T \gg m$ we have 
\begin{eqnarray}
\label{blueT}
   T_{slope} &=& T_{th} \sqrt{\frac{1+v}{1-v}}
\end{eqnarray}

Generally, increasing the velocity increases the curvature of the
final spectrum for heavy particles. Increasing the mass also 
increases the curvature of the flow profile.
The shape of the spectrum, together with the mass dependence 
of the observed particle, may provide a good experimental 
measure of the velocity of the source at hadronization.

\begin{figure}[!tbp]
   \begin{center}
      \includegraphics[height=3.0in,width=3.0in]{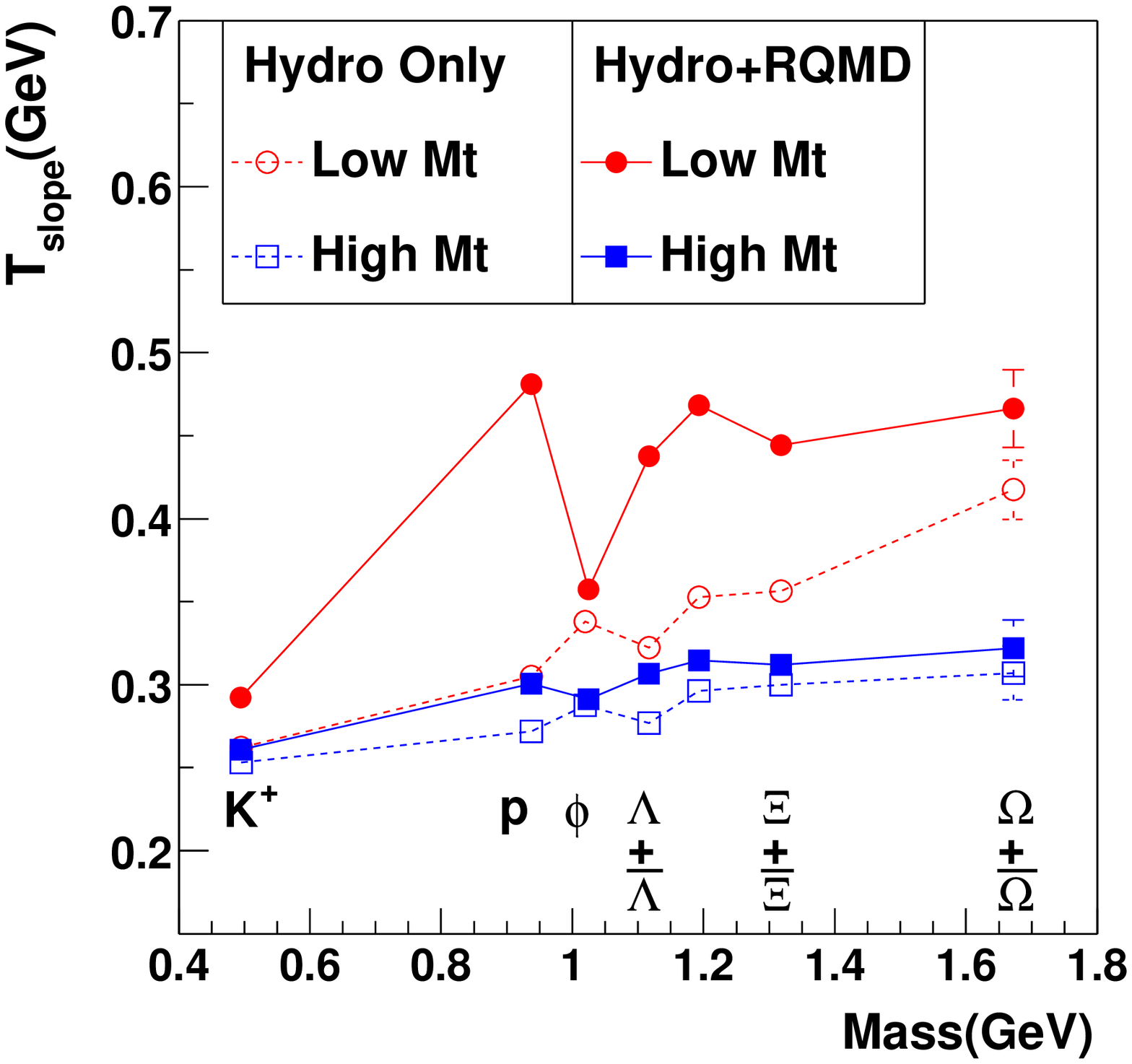}
      \includegraphics[height=3.0in,width=3.0in]{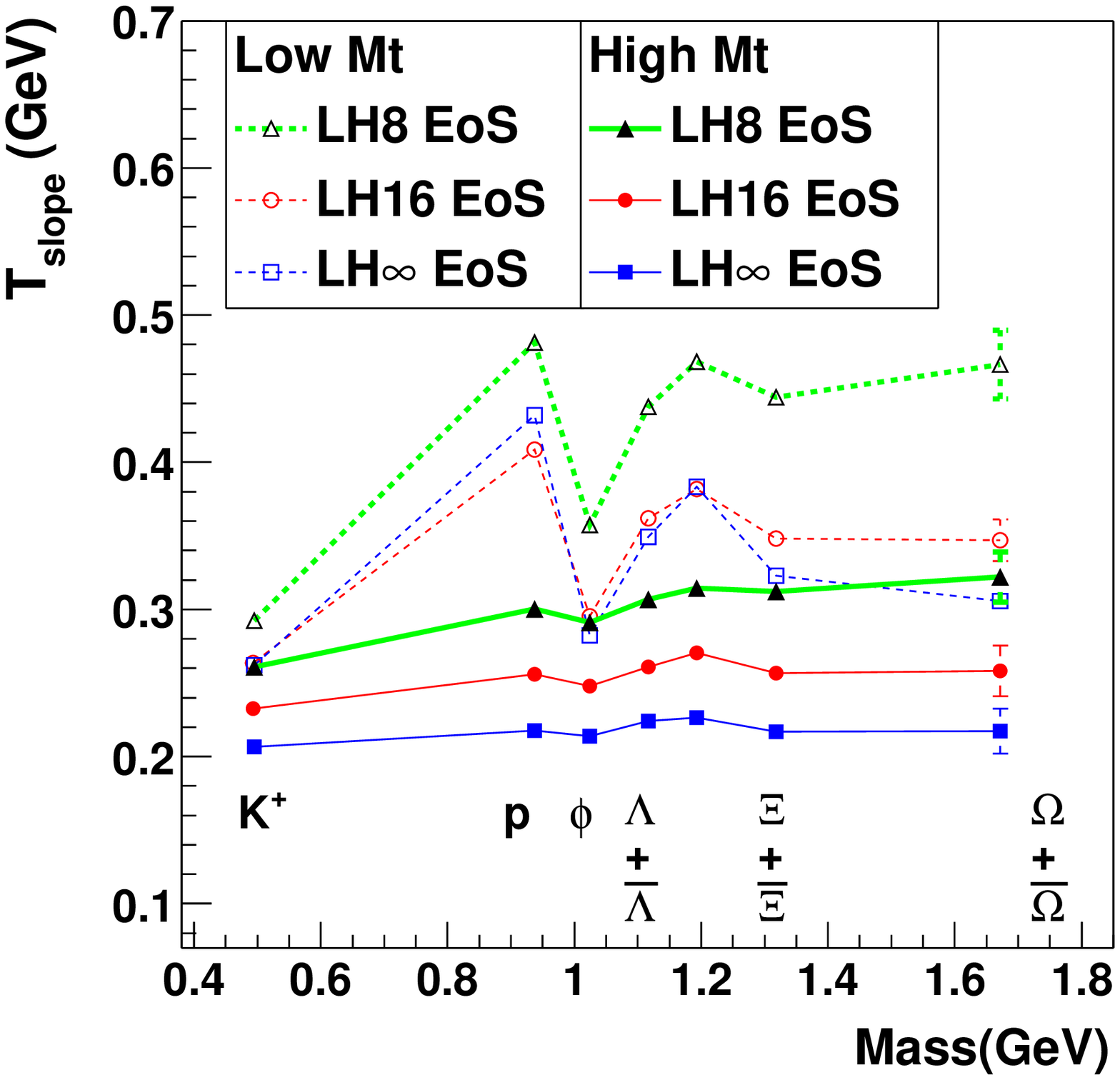}
   \end{center}
   \caption[ The dependence of model slope parameters on particle mass, 
   hadronic rescattering, $M_{T}$ range, and EOS]{
   \label{tmlh8}
Each particle spectrum in central AuAu collisions,
$\frac{dN}{M_{T}\,dM_{T}}$, is
fit with an exponential slope parameter  from 
$0<M_T<0.6\,\mbox{GeV}$ and from $0.6<M_{T}<1.6\,\mbox{GeV}$.
Thus there is a low and a 
high $M_T$ slope representative of the slopes measured by the 
STAR \cite{STAR-Spectra} and PHENIX collaborations 
 \cite{PHENIX-Spectra}. (a) shows the mass dependence
of the low and high $M_{T}$ slopes with and without the
RQMD after-burner for LH8. (b) shows the mass dependence of the 
low and high $M_{T}$ slopes for the different EOS with 
the RQMD after burner.
   }
\end{figure}

Hadronic rescattering changes the curvature seen in the
thermal spectra discussed above.
First, different particle types have different hadronic
cross sections and  therefore freezeout at different times
and with  different velocities.
The curvatures in the final spectra measure these different
freezeout velocities. Second, the cascade generates additional
transverse flow predominantly in the low $M_T$ region of the
spectra. To quantify these effects,
we divide the $M_{T}$ spectra into a low $M_{T}$ region,
$0 < M_{T} < 0.6 \,\mbox{GeV}$ and a high $M_{T}$ region, 
$0.6\,\mbox{GeV}\,< M_{T} < 1.6\,\mbox{GeV}$. We  then  fit an 
exponential in both domains.  Thus, there is a low $M_{T}$ slope 
and high $M_{T}$ slope.  We have checked that this parameterization gives
a good description of the shape for all the spectra considered. 
Fig. \ref{tmlh8} shows the low
and high $M_T$ slopes as a function of the particle mass, with 
and without the RQMD after-burner. The curves illustrate the collective
acceleration which occurs during hadronic rescattering, and illustrate
an interplay between freezeout and hydrodynamic behavior. First
look at the ``Hydro Only'' curves: the curvature (i.e. the 
difference between the low $M_T$ and the high $M_T$ slopes) 
increases with mass as expected from Eq.~\ref{Bessel}.
When the cascade is included, rescattering changes this mass dependence.
The curvature no longer increases but remains approximately constant
after the nucleon mass. It is useful to compare the 
flow of the nucleon and the $\Omega^{-}$. 
The nucleon has a 
smaller mass which, according to Eq.~\ref{Bessel}, decreases
the  curvature relative to the $\Omega^{-}$.  
However, the nucleon decouples 
later than the $\Omega^{-}$ 
and through hadronic rescattering develops larger transverse 
velocity,\, which increases the curvature. In the end, the $\Omega^{-}$ and the
nucleon have approximately the same spectral shape. 
In contrast, the $\phi$, which has 
approximately the same mass as the nucleon but which decouples
early, has  very little spectral curvature.  
To summarize, an interplay between the differential freezeout dynamics 
and the curved thermal spectra of Eq.~\ref{Bessel} 
results in rich features in the final
spectra of different particles. 

Before discussing their  impact parameter dependence 
of these rich features (see Sect.~\ref{Bdepend-RFlow}), 
we study the sensitivity of the spectra to the EOS. 
The mass dependence of the slopes is a feature of an expanding
thermal source and differential freezeout. It is not a feature
of the underlying EOS.  
In Fig.\,\ref{tmlh8}, the low $M_{T}$ and
high $M_{T}$ 
slopes are shown for three different EOS. 
For this discussion, the direct comparison of model and data nucleon spectra 
in Fig.\,\ref{rhic-sp2} may be helpful. 
For the high $M_{T}$ slopes for all particles,
there is a simple ordering, $T_{slope}^{LH8}> T_{slope}^{LH16}>
T_{slope}^{LH\infty}$, which reflects (through Eq.~\ref{blueT})  
the ordering of the transverse flow,
$v_{T}^{LH8}> v_{T}^{LH16} > v_{T}^{LH\infty}$.  
In the low $M_{T}$ region, the ordering is more complex and reflects
the space-time structure of the freezeout surface for different EOS.   
LH$\infty$ evaporates particles shrinking radially 
inward. This causes an enhancement of the particle yield
at low $M_{T}$ and gives LH$\infty$ a significant slope in
the low $M_{T}$ region.
Still, with LH8 
the $\Omega{-}$  shows much more flow at low $M_{T}$ than 
it does with 
LH16 and LH$\infty$ indicating a large velocity at the end
of the mixed phase.

\begin{figure}[!tbp]
   \begin{center}
      \includegraphics[height=2.7in,width=3.0in]{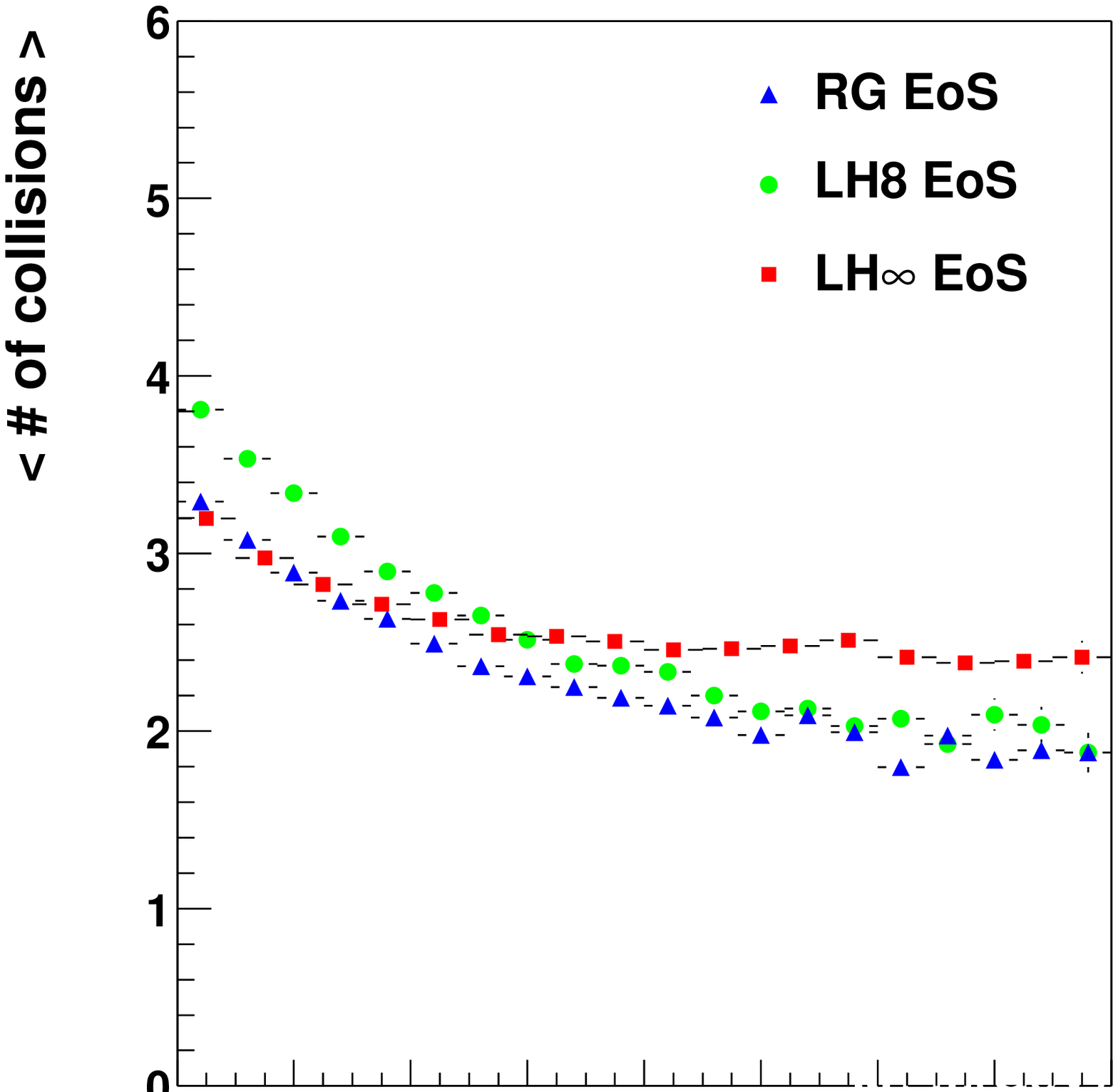}
      \includegraphics[height=2.7in,width=3.0in]{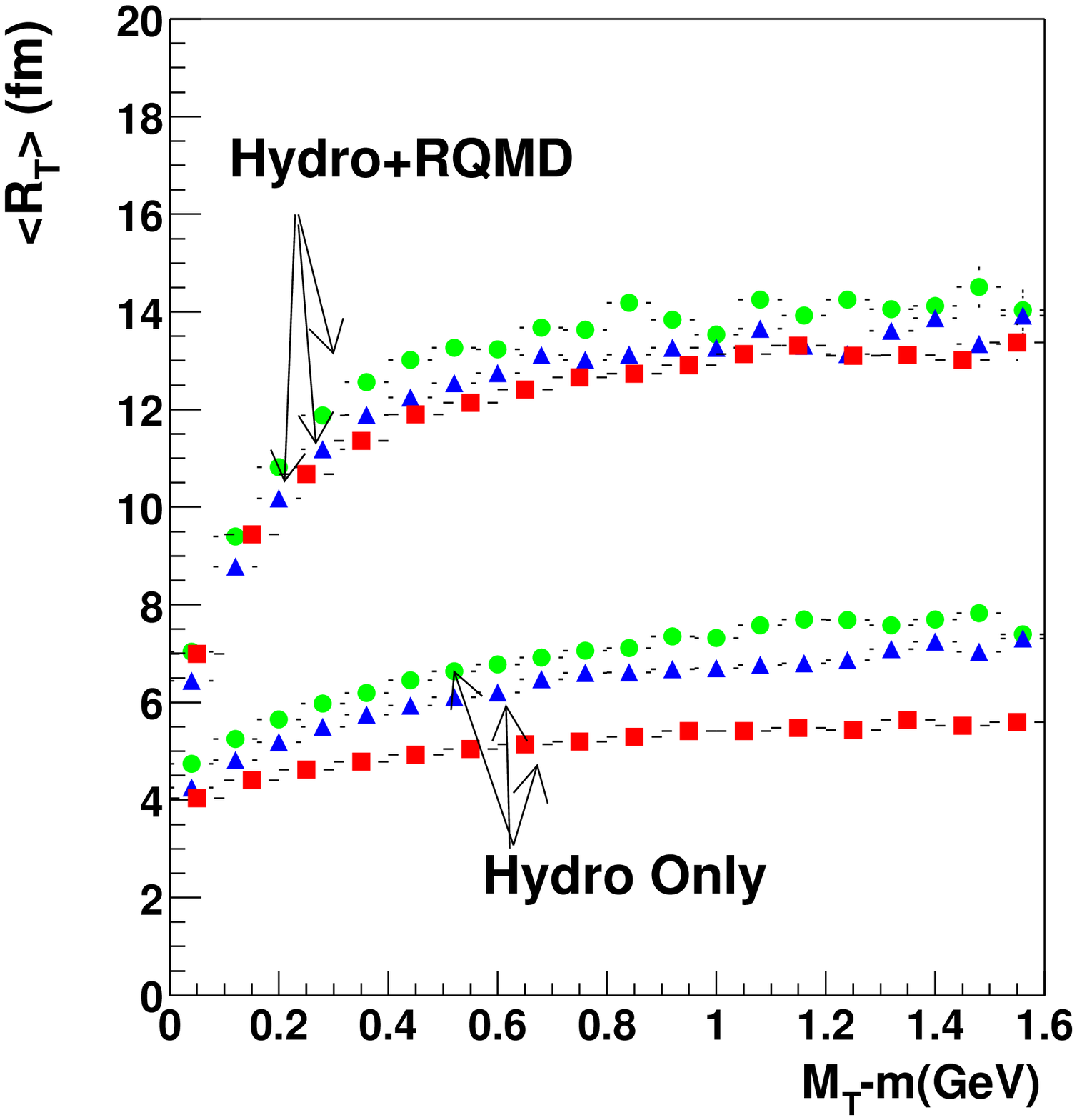}
   \end{center}
   \caption[The mean  number of collisions and emission radius 
   as function of $M_{T}$ for nucleons]{
   \label{mtanalysis}
For nucleons in central AuAu collisions, (a) analyzes the
number of collisions experienced by a nucleon  as function
of $M_T$ while (b) analyzes the mean 
freezeout radius of a nucleon as a function of $M_T$.
   }
\end{figure}

The curvature in the $M_{T}$ spectra at small $M_{T}$ is
a consequence of the mass dependence of Eq.~\ref{Bessel} and hadronic
rescattering, as discussed above.
Now the role of hadronic rescattering, or the ``pion wind'', 
is studied in detail with Fig.\,\ref{mtanalysis}. 
At low $M_T$, where the cascade is most effective,   
all the nucleons come from the center of the nucleus, as can be
seen in Fig.\,\ref{mtanalysis}(b).  Accordingly,
the number of collisions is larger and the nucleons are accelerated more. 
At high $M_{T}$, the nucleon spectrum (recall Fig.\,\ref{mt_spectra}) is
simply shifted with 2-3 collisions 
by a constant amount, approximately $300\,\mbox{MeV}$, which
increases the slope.  These collisions happen over a time scale of
$\approx10\,\mbox{fm}$ and the collision rate is therefore $\approx \frac{1}{5\,\mbox{fm}}$.
Collecting these observations, nucleons coming from the center
of the collision freezeout last,
populate the low $M_{T}$ region, and are kicked the most by the pion wind.  

\subsection{Comparison to Central RHIC Data}

Now we compare model predictions to the first RHIC spectra.
Keep in mind the two major predictions of hydrodynamics introduced
in Sect.~\ref{FLChanges}.
First,  $<M_T>$ should increase significantly.
Since LH8 was found to give the best agreement to SPS flow data,
LH8 should give the best agreement
at RHIC. Out of  all the EOS studied, the flow velocity 
increases the most for LH8.  
Second, the spectra should show the flow profile of Eq.~\ref{Bessel}.
This profile is sensitive to the particle mass and flow velocity. 
At RHIC therefore, LH8 predicts a significant change in slope from low $M_{T}$ 
to high $M_{T}$.

\begin{figure}[!tbp]
   \begin{center}
      \includegraphics[height=2.9in, width=2.9in]{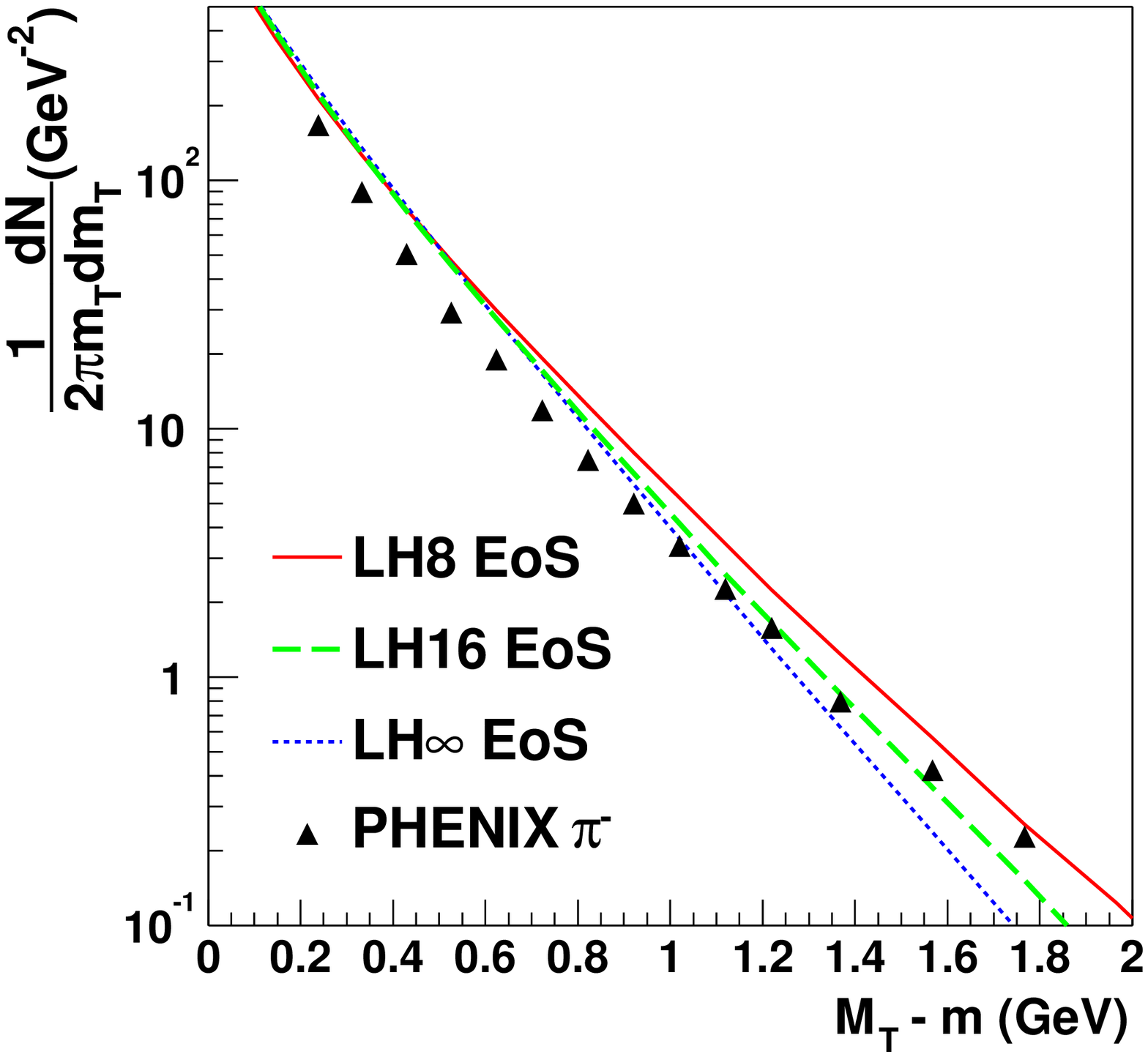}
      \hspace{-0.4in}
      \includegraphics[height=2.9in, width=2.9in]{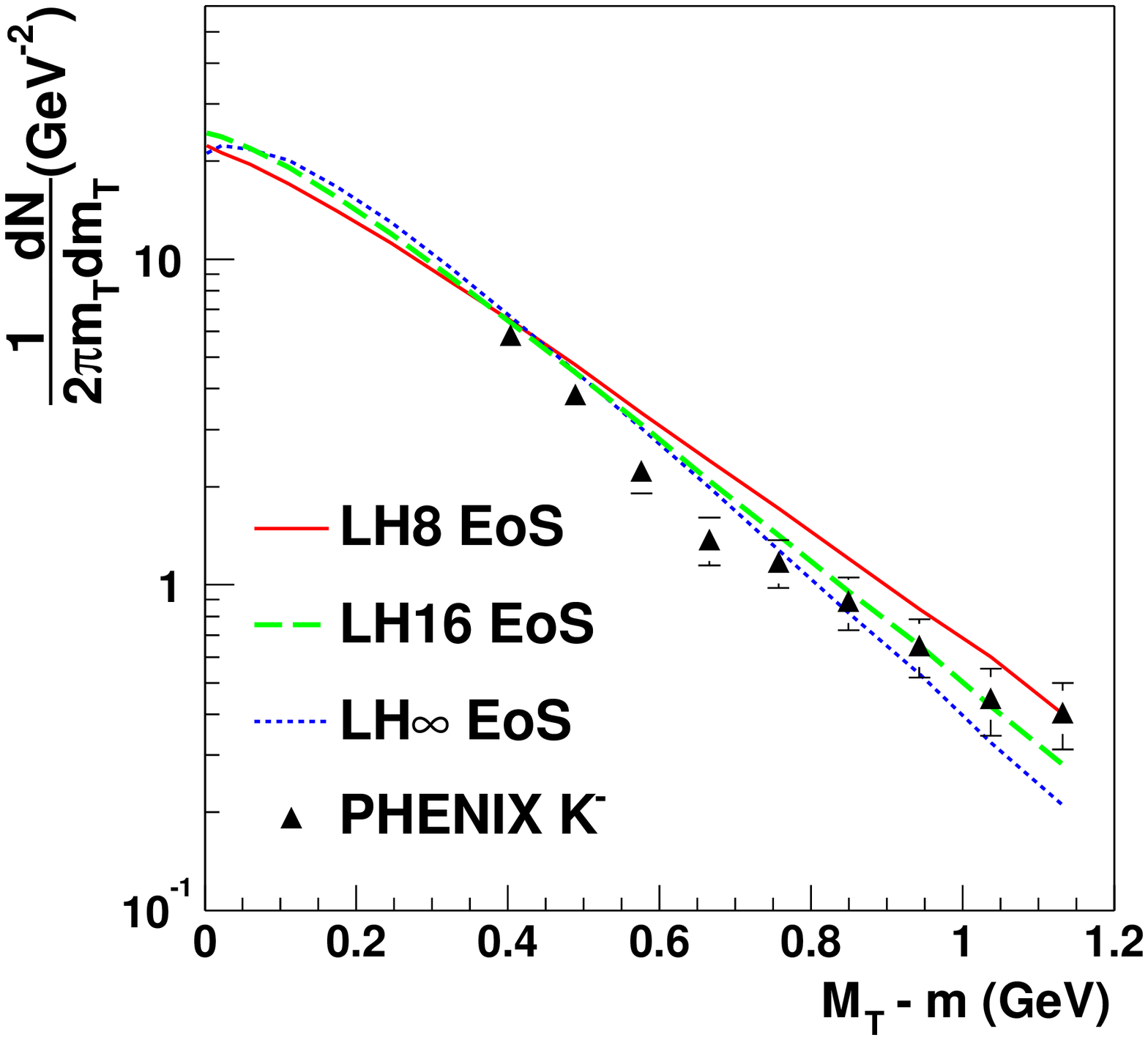}
      \end{center}
      \caption[A comparison of model and experimental $M_{T}$ spectra
      at RHIC]{
         \label{rhic-sp}
 A comparison to PHENIX spectra \cite{PHENIX-Spectra} :
(a) compares $\pi^{-}$ spectra. (b) compares $K^{-}$ spectra.
Both the model and the experimental
spectra are absolutely normalized.
      }
\end{figure}

\begin{figure}[!tbp]
   \begin{center}
      \includegraphics[height=2.9in, width=2.9in]{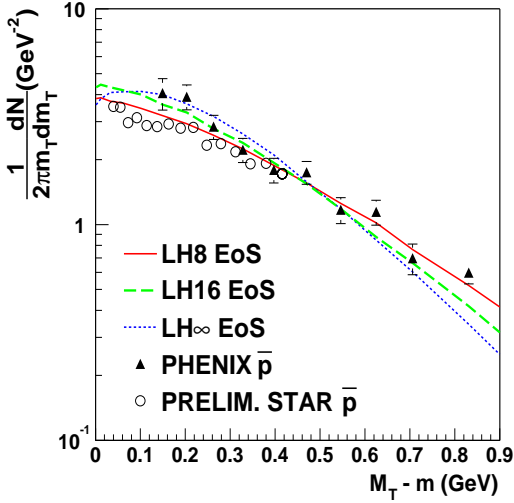}
      \hspace{-0.4in}
      \includegraphics[height=2.9in, width=2.9in]{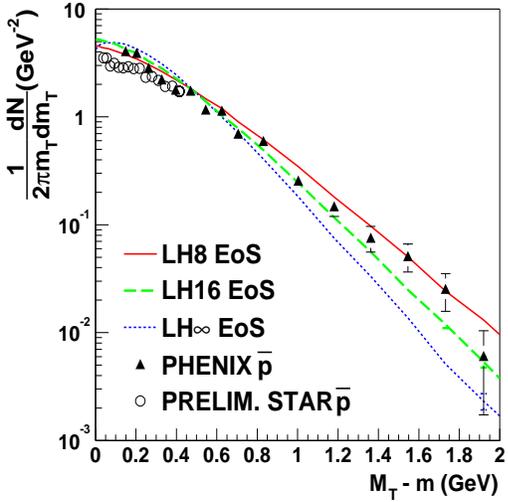}
      \end{center}
      \caption[A comparison of model and experimental $M_{T}$ spectra
      at RHIC]{
         \label{rhic-sp2}
 A comparison to PHENIX \cite{PHENIX-Spectra} and 
STAR \cite{STAR-Spectra} $\bar{p}$ spectra for 
(a) low and (b) high  $M_T$, respectively.
Both the model and the experimental
spectra are absolutely normalized.
      }
\end{figure}

Fig.\,\ref{rhic-sp} and Fig.~\ref{rhic-sp2} show the absolutely normalized 
model spectra for three different
EOS compared to data for $\pi^{-}$,$K^{-}$ and $\bar{p}$.
For each particle type, the transverse mass spectrum is strong.
Generally, LH$\infty$
under-predicts the flow profile 
while  LH8 reproduces the spectrum. The
data indicate a strong macroscopic transverse response, as
expected of an EOS with speed of sound $c_{s}^{2}\approx 1/3$. 
Some caveats must be mentioned. It is known that the
transverse mass spectrum is sensitive to the details of the 
initial profile and longitudinal expansion \cite{Sollfrank-BigHydro}. 
In particular, 
the transverse mass spectrum is modified if the initial
entropy density is distributed according to binary 
collisions \cite{Kolb-Centrality}. However, even when
the entropy is distributed entirely according to 
binary collisions, the change in the spectrum is small.
The strong increase in radial flow from the
SPS to RHIC is reproduced by the hydrodynamic response of LH8.
Thus, prediction (1) is borne out by the first spectra at RHIC.
Next, look at the shape of the $\bar{p}$ spectrum 
in Fig\,\ref{rhic-sp2}. This flattening
at low $M_T$ is characteristic of a flow profile. We 
expect a smaller   
flattening in the kaon spectrum. Thus, the rich
flow profile of Eq.\,\ref{Bessel} is also borne out 
in the first RHIC data and prediction (2) is correct. 

\begin{figure}[!tbp]
   \begin{center}
      \includegraphics[height=3.0in, width=3.0in]{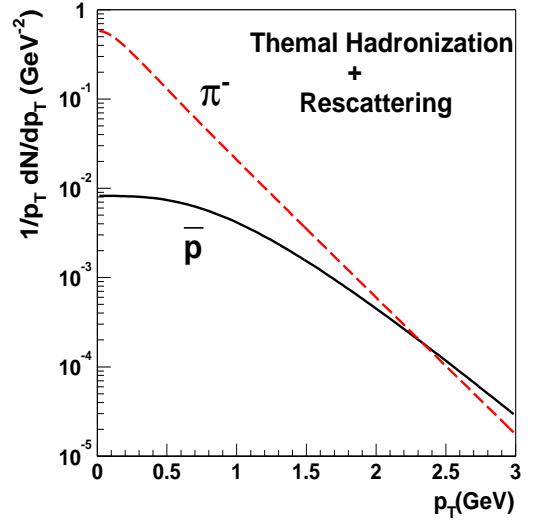}
      \includegraphics[height=3.0in, width=3.0in]{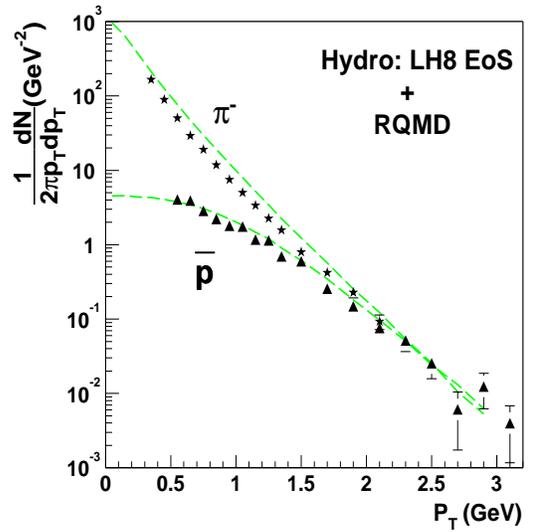}
    \end{center}
      \caption[A simple model of the $\pi^{-}$ and $\bar{p}$ 
       spectra]{
      \label{thermal-model}
      A comparison of $\pi^{-}$ and $\bar{p}$ spectra.
  (a) shows a simple thermal model with parameters discussed in 
      the text. The spectra are relatively normalized. 
  (b) shows an absolutely normalized comparison of the complete model
      spectra and PHENIX spectra \cite{PHENIX-Spectra}. 
      }
\end{figure}

It is worthwhile to plot the $\pi^{-}$ and $\bar{p}$ spectra on the same 
plot.  The spectra almost cross for $p_{T}\approx 2.3\,\mbox{GeV}$
It was recently pointed out that the measured $\pi^{-}/\bar{p}$ 
ratio is several times above the expected ratio
from jet fragmentation 
and from a hydrodynamic calculation that does not incorporate
chemical freezeout \cite{Vitev-Baryon}. 
The ratio is readily explained in a simple hydro/thermal model with additional hadronic scattering. The thermal 
input into RQMD is  roughly summarized by  
Eq.~\ref{Bessel}. Above $M_{T}>2.0\,\mbox{GeV}$ without rescattering, 
the slope parameters  of pions and nucleons approach the 
universal value $T_{slope}\approx250\,\mbox{MeV}$. This slope is
given by Eq.~\ref{blueT} with the parameters 
for $T_{th}=160\,\mbox{MeV}$ and $v_{T}=0.45\,c$. 
Accounting for hadronic rescattering, 
the nucleon slope at large $p_T$ approaches
$T_{slope}\approx 300\,\mbox{MeV}$ and is better described by
$T_{th}=160\,\mbox{MeV}$ and $v_{T}=0.55\,c$. 
Hadronic rescattering therefore increases the nucleon flow velocity
slightly, from $v_{T}\approx0.45\,c$ to $v_{T}\approx0.55\,c$ . 
We then adjust $\mu_B/T$ to match the experimental $\bar{p}/p$ 
ratio \cite{STAR-ppbar}, $\bar{p}/p=\exp(-2\mu_{B}/T)=0.65$.  
Then with all the parameters specified, we draw 
Eq.~\ref{Bessel} for $\bar{p}$ and $\pi^{-}$ 
in Fig.~\ref{thermal-model}.
A source expanding with a collective velocity
of $v_{T}\approx 0.5\,c$ and hadronizing according
to a thermal prescription at a temperature of $T\approx 160\,\mbox{MeV}$, 
generates the observed $\pi^{-}/\bar{p}$ ratio once
pion nucleon scattering is taken into account. 

\subsection{The Impact Parameter Dependence of Radial Flow}
\label{Bdepend-RFlow}

In peripheral collisions, hydrodynamic features
should disappear since the mean free path 
becomes comparable to the root mean square radius
$R_{rms}$.
Because ideal hydrodynamics is scale invariant, the
hydrodynamic
stage of the model does not capture finite size
effects
which become increasingly important at larger 
impact parameters. However, this does not mean that the
radial velocity is independent of impact parameter.
The
hydrodynamic lifetime scales approximately as $R_{rms}$;
the flow velocity reflects this lifetime.
Finite size effects make the total lifetime decrease
more quickly than $R_{rms}$.
Finite size effects are
most important during the freezeout stage which is 
modeled with RQMD and therefore the hydro+cascade
approach can capture some non-trivial features of
the impact parameter dependence.  

\begin{figure}[!tbp]
   \begin{center}
   \includegraphics[height=3.0in, width=3.0in]{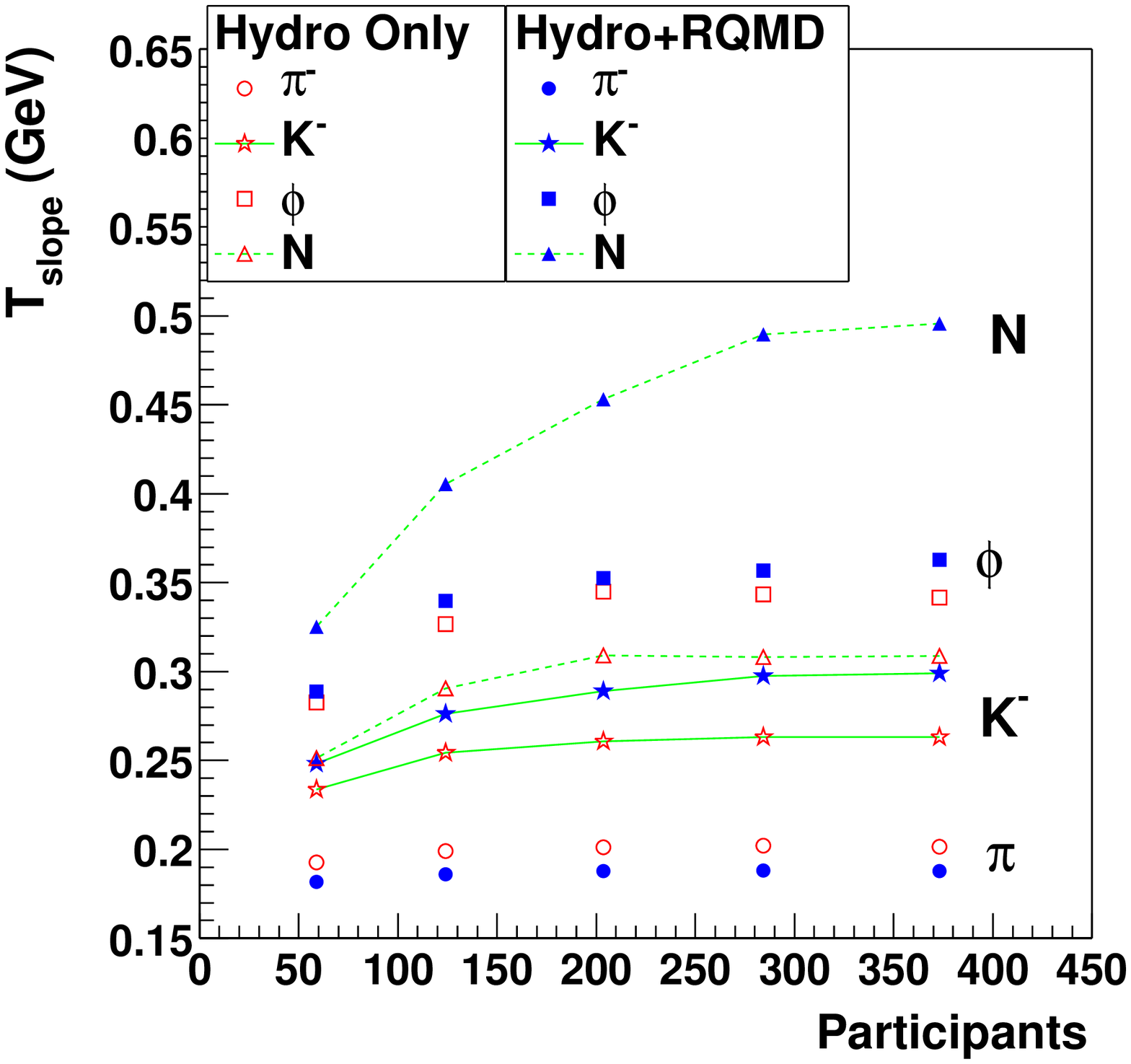}
   \includegraphics[height=3.0in, width=3.0in]{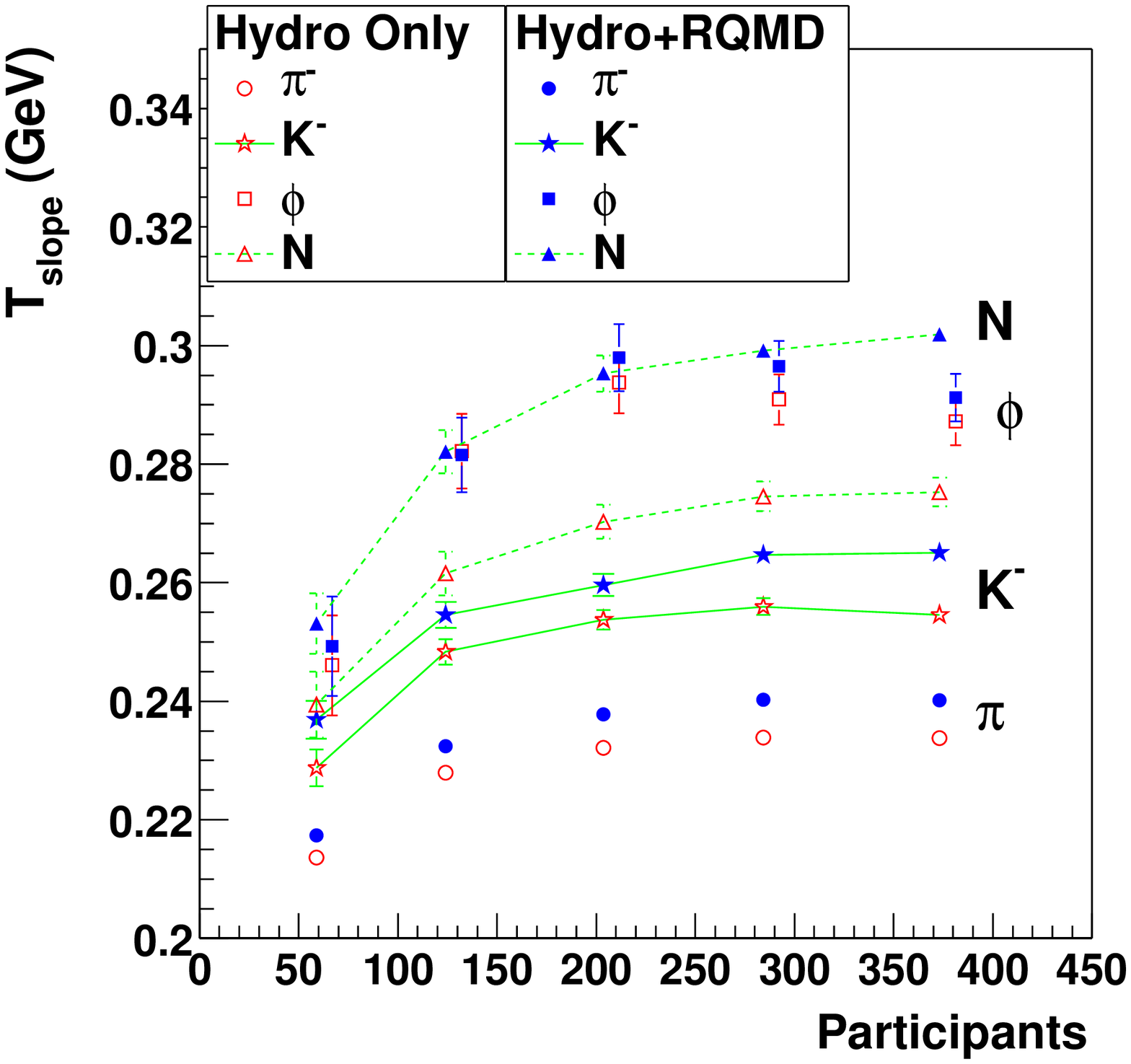}
   \end{center}
   \caption[Slope parameters for different particle species as
   a function of the $M_{T}$ range and the impact parameter]{
   \label{rhic-bdependence}
      The (a) low $M_T$   
      and (b) high $M_T$ slope parameters 
      (see Fig.~\ref{tmlh8})
      as a function of the number of participants in the collision.
   }
\end{figure}

In the low $M_{T}$ region,
pion-nucleon scattering in RQMD is largely
responsible for the $\approx 500\,\mbox{MeV}$ anti-proton 
slope measured by the
STAR collaboration \cite{STAR-Spectra}. In the high $M_{T}$ region
measured
by the PHENIX collaboration \cite{PHENIX-Spectra}, the hydrodynamic stage
of the model is  much more significant than hadronic rescattering. 
Fig.~\ref{rhic-bdependence} shows 
the low and high $M_T$ slope parameters as a function
of 
the number of participants in the collision.
For each particle, the open symbols show the slope
parameter
without RQMD and the closed symbols show the slope
parameter
with RQMD. For both pions and kaons in the low and
high
$M_{T}$ regions, only a small impact
parameter dependence is observed. The pion spectrum 
is cooled at all impact parameters.

Compare the nucleon and the $\phi$ slope parameters.
In the low $M_T$ region, shown in Fig.\,\ref{rhic-bdependence}(a),
the nucleon
slope exhibits a very rapid dependence on the number
of
participants. Approximately 40\% of this slope is
due to RQMD and  the ``Hydro Only'' curve for the
nucleon is relatively
flat as a function of impact parameter. The
nucleon and the $\phi$ have approximately the same
mass but the $\phi$ is devoid of the strong
$\Delta$
resonance which drives the flow in the nucleon system.
Therefore, the ``Hydro Only''
curve for nucleons is similar to the ``Hydro+RQMD'' 
curve for the $\phi$.  

Next, compare the nucleon and the $\phi$ slopes in the high $M_T$
region, shown in Fig.\,\ref{rhic-bdependence}(b).
Here $\pi N\rightarrow\Delta$ is less significant and
RQMD is responsible for only $\approx 15\%$ of the nucleon
slope.  Consequently, RQMD makes up the small difference
between the nucleon and the $\phi$ ``Hydro Only'' curves, and
the final slope parameters of the two particles are
similar in shape and magnitude. 
Experimentally, the difference in slope parameters
between the nucleons and the $\phi$ can be used to assess the
contribution of the hadronic phase in the low $M_T$ region.

\begin{figure}[!tbp]
   \begin{center}
      \includegraphics[height=3.0in,width=3.0in]{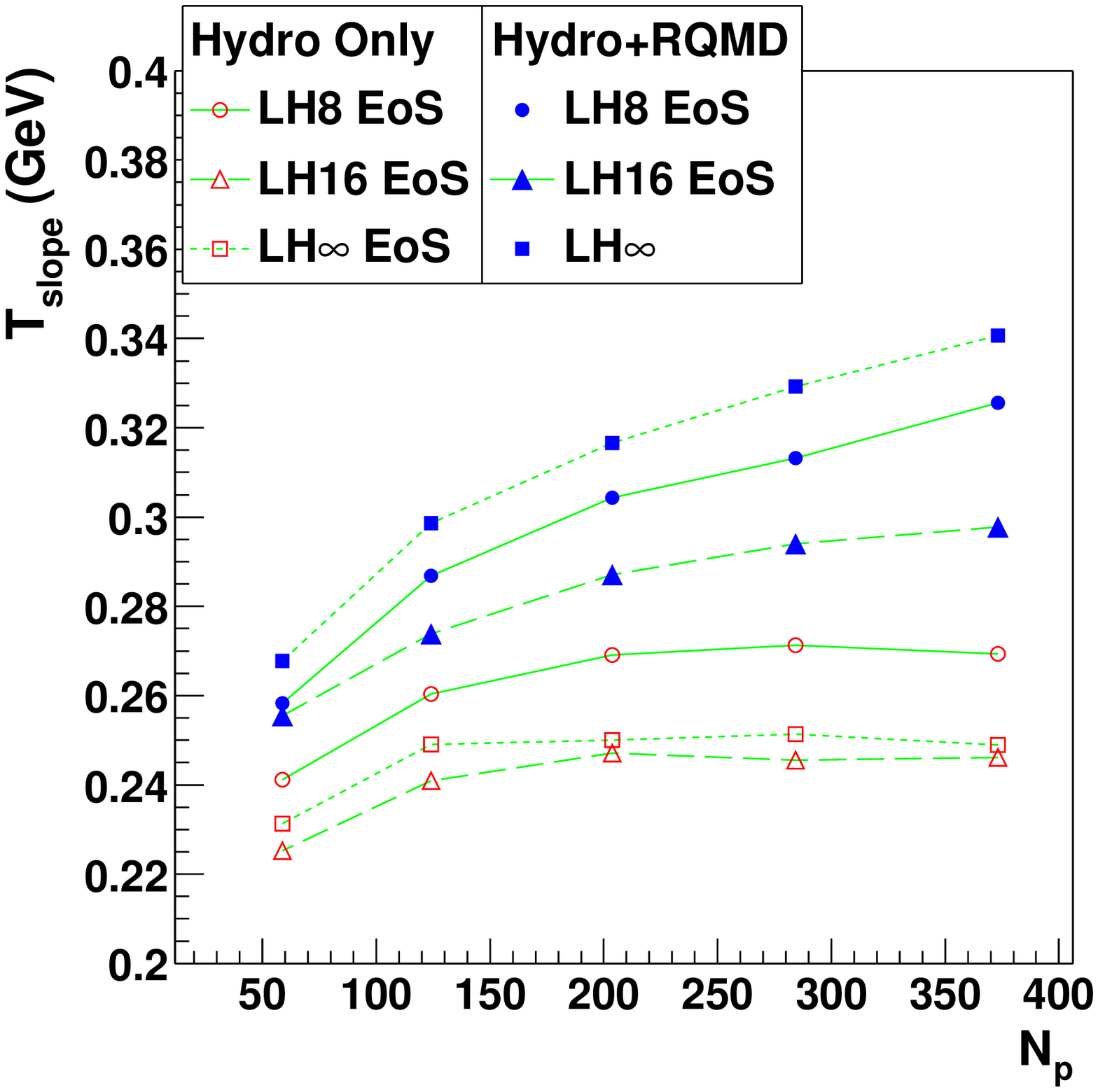}
      \includegraphics[height=3.0in,width=3.0in]{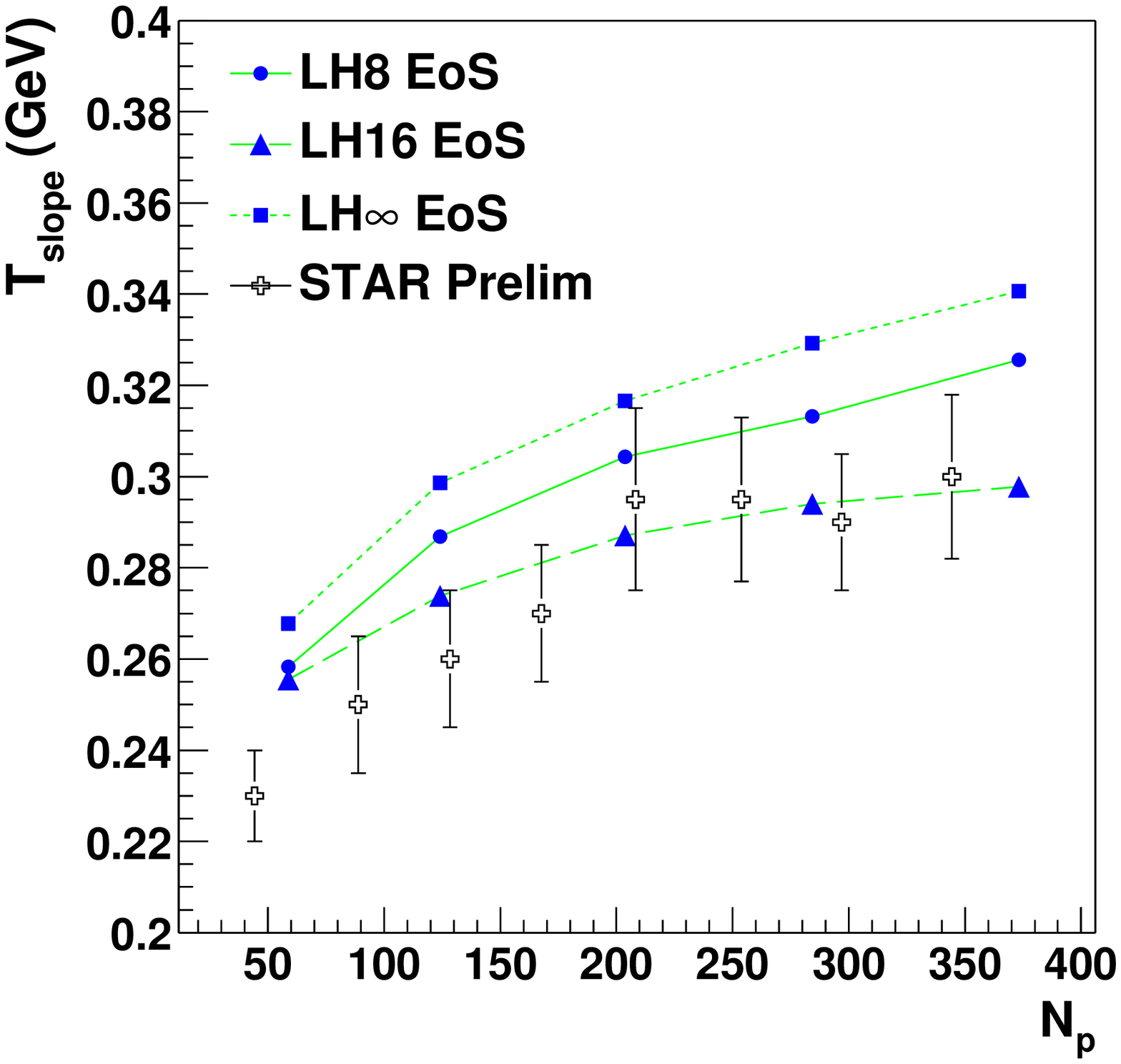}
   \end{center}
   \caption[A comparison of model $K^{-}$ slope parameters to
   preliminary data from the STAR Collaboration as a function of impact parameter]{ 
   \label{StKm}
     Model and preliminary STAR $K^{-}$ \cite{STAR-Spectra} 
     slope parameters  in the
     fit range $0<M_{T}-M<0.5\,\mbox{GeV}$ as a function of the number
     of participants for different EOS.
     (a) compares model predictions for each EOS with and without
     the RQMD after-burner; 
     (b) compares the model to data with the RQMD after-burner.
   }
\end{figure}

\begin{figure}[!tbp]
   \begin{center}
      \includegraphics[height=3.0in,width=3.0in]{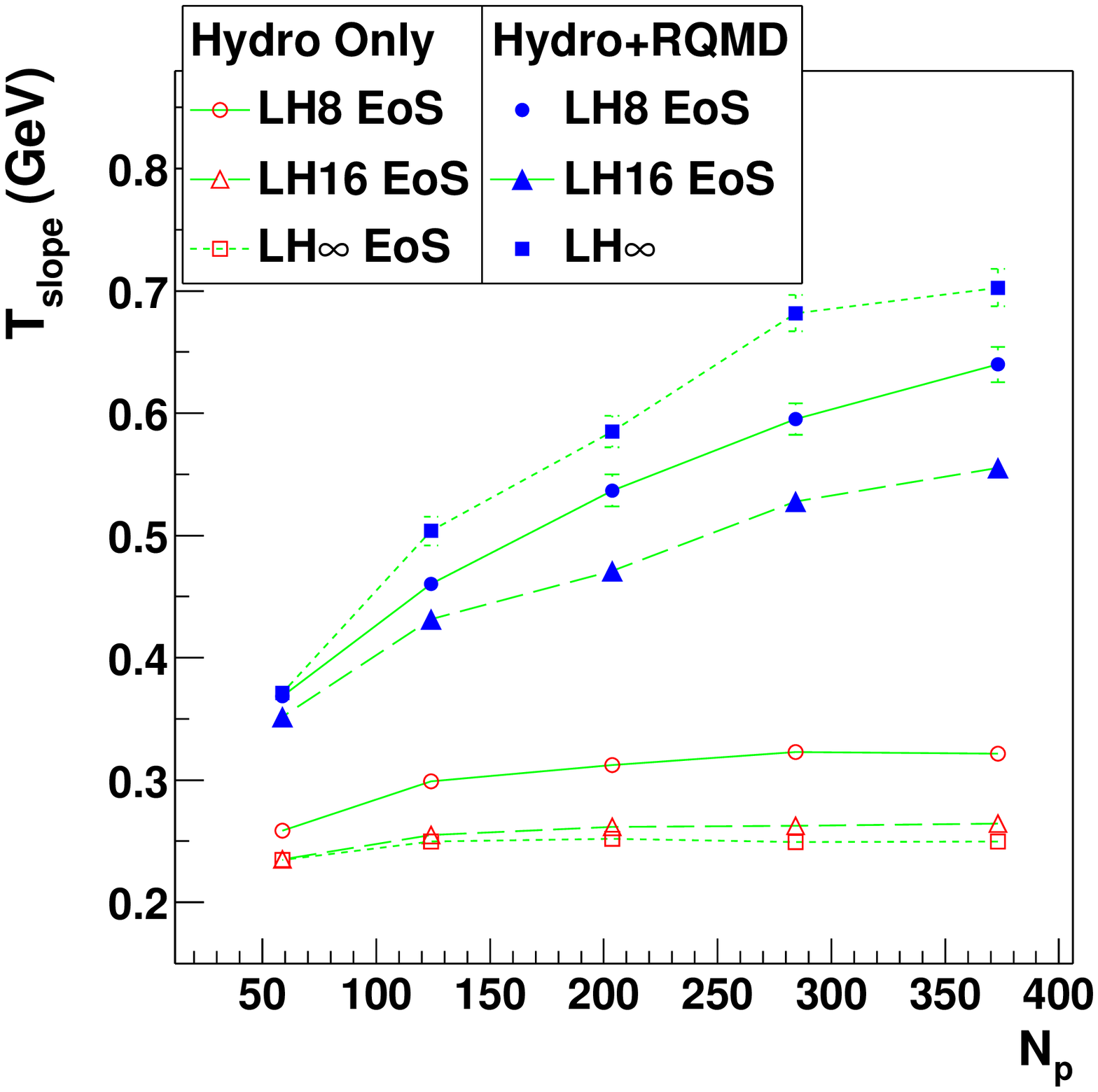}
      \includegraphics[height=3.0in,width=3.0in]{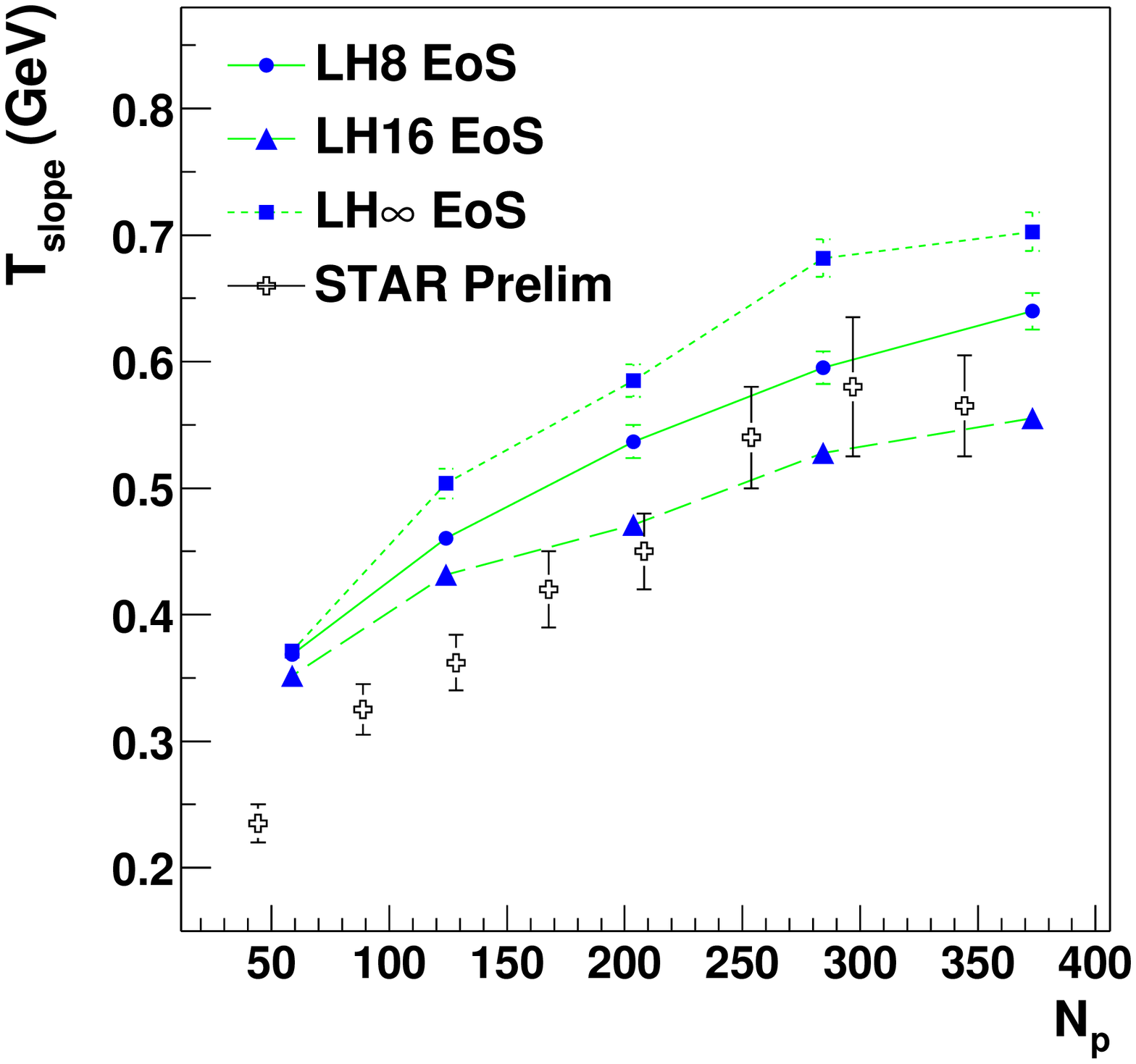}
   \end{center}
   \caption[A comparison of model $\bar{p}$ slope parameters to
   preliminary data from the STAR Collaboration as a function of impact parameter]{ 
   \label{StPbar}
     Model and preliminary STAR $\bar{p}$ \cite{STAR-Spectra} slope parameters  in the
     fit range $0<M_{T}-M<0.5\,\mbox{GeV}$ as a function of the number
     of participants for different EOS.
     (a) compares model predictions for each EOS with and without
     the RQMD after-burner; 
     (b) compares the model to data with the RQMD after-burner.
   }
\end{figure}

Turning to the experimental data, we first examine
the low $M_{T}$ region as measured by the STAR 
collaboration.
We compare
the model b-dependence of the 
anti-proton and $K^{-}$ slope
parameters to the experimental
data in Fig.~\ref{StKm} and ~\ref{StPbar}.  
Comparing the anti-proton slopes for different EOS,
we see that the rapid b-dependence is not a
consequence of a change in EOS.
Note that LH$\infty$  has a larger slope at small $M_T$ 
than LH8.  This is an artifact of  the exponential fit.
The spectrum for LH$\infty$ in central
collisions is shown in Fig\,\ref{rhic-sp2} and is not 
exponential. The measured spectra are
 well described by
an exponential in this region \cite{STAR-Spectra}. 
Coarsely,
the model reproduces the strong b-dependence of the
anti-protons
and the weaker b-dependence of the kaons.  However, for
more 
peripheral collisions, the
measured slope parameters fall somewhat faster than the 
model 
predicts. Naively, this indicates that in the most
peripheral collisions the hydrodynamic description is
only beginning to
work and finite size effects (e.g viscosity) should be
taken into account in the QGP and mixed phases which
are modeled with the hydrodynamics.
The ``Hydro Only'' curves presented in 
Fig.\,\ref{StKm} and \ref{StPbar} definitely do not reproduce the strong
b-dependence of the slopes. With RQMD on the 
other hand, finite size effects
during the late hadronic stages are modeled and most of the
rapid b-dependence seen in the data is reproduced. 

\begin{figure}[!tbp]
   \begin{center}
      \includegraphics[height=2.8in,width=2.8in]{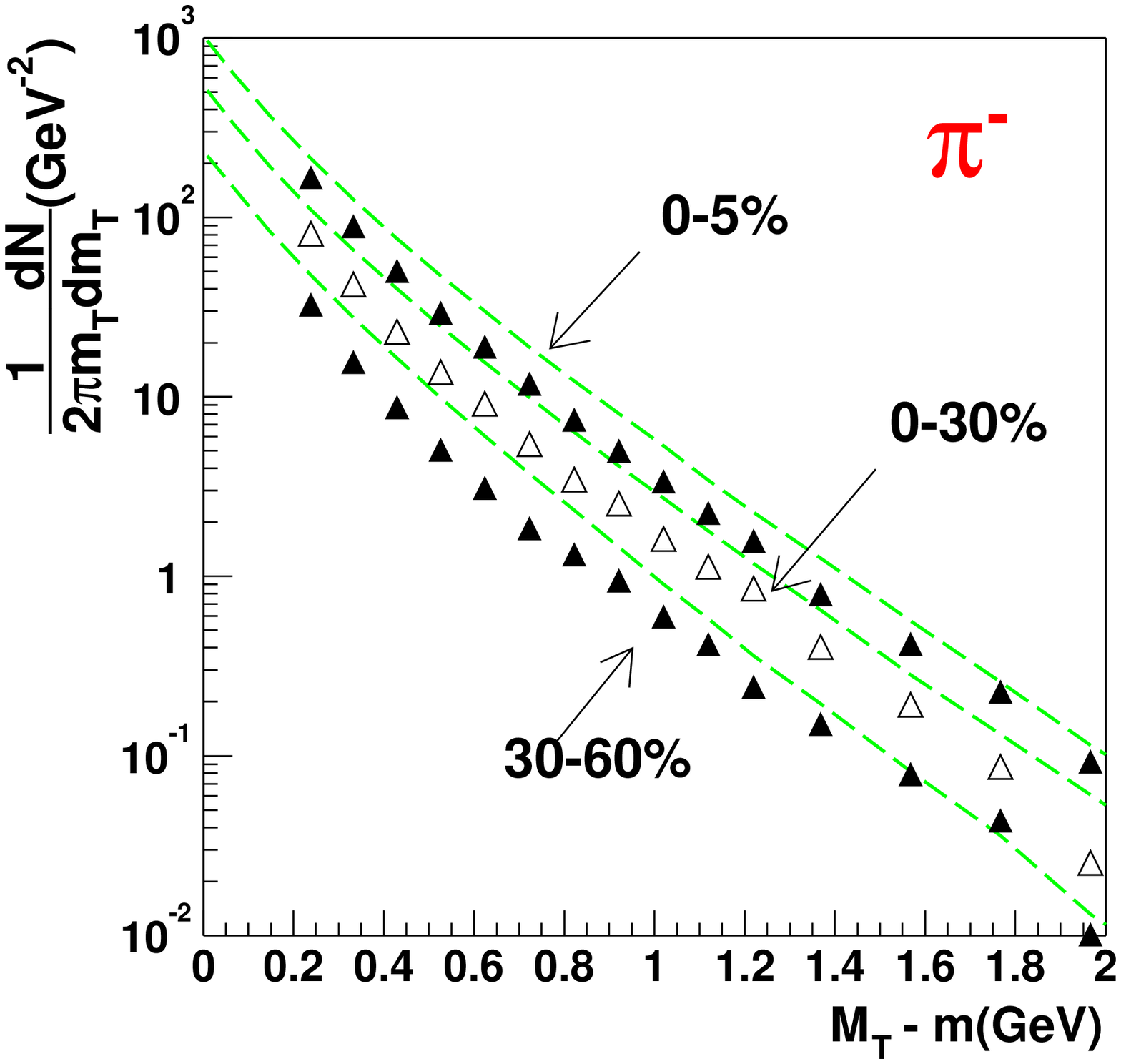}
      \hspace{-0.2in}
      \includegraphics[height=2.8in,width=2.8in]{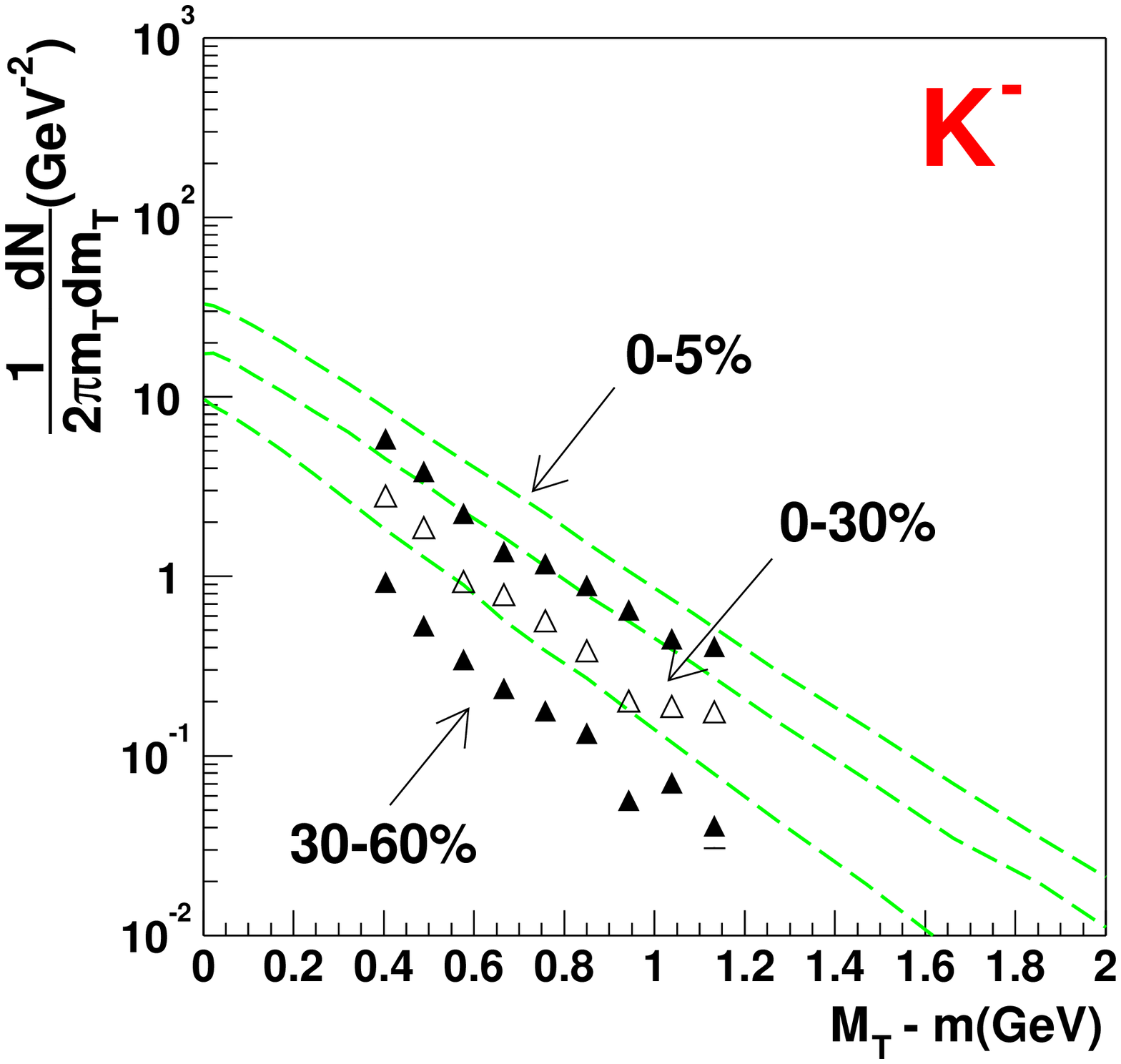}
      \includegraphics[height=2.8in,width=2.8in]{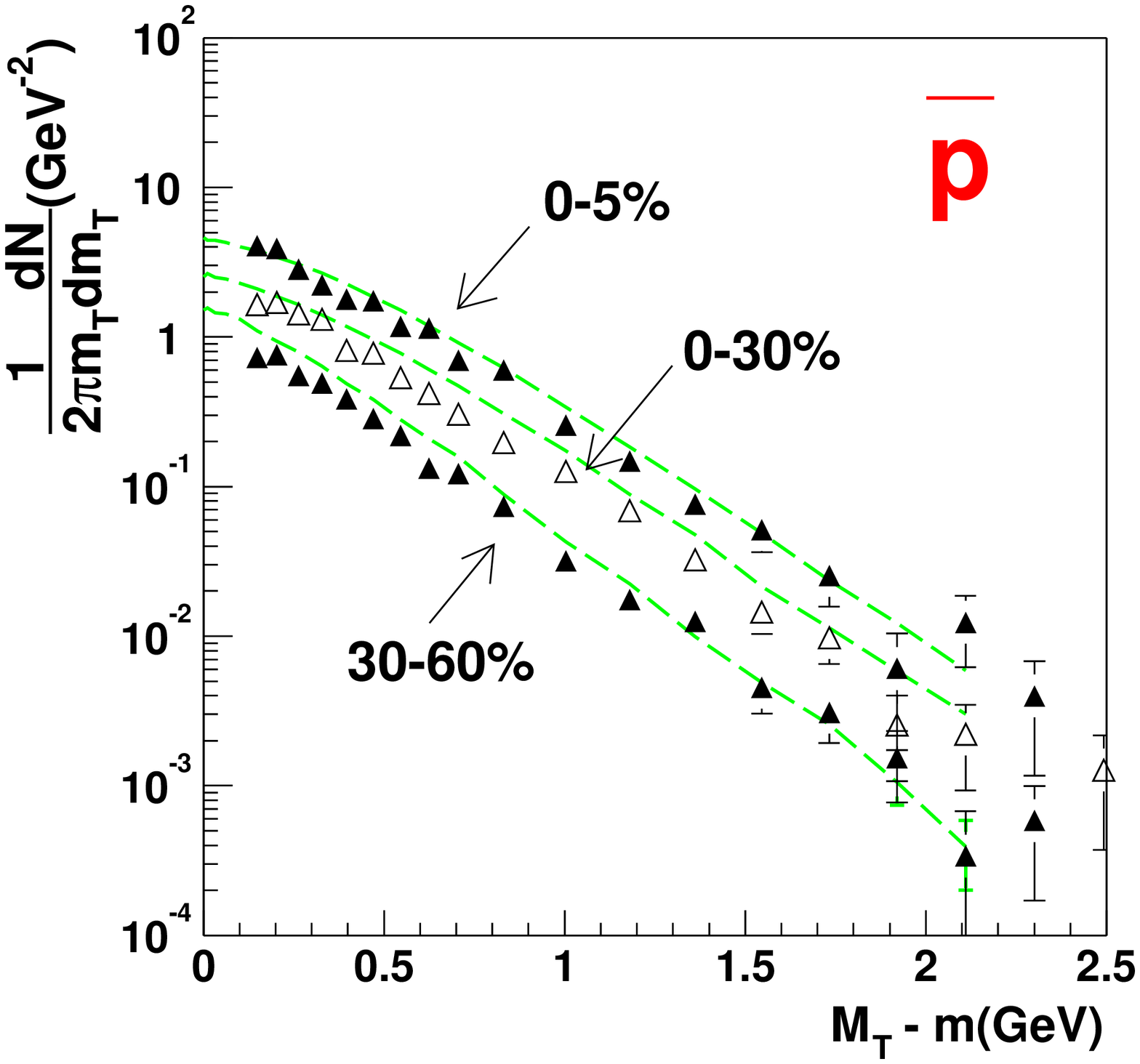}
   \end{center}
   \caption[Comparison of model $M_{T}$ spectra to
   PHENIX data at different centralities]{
   \label{phCent}
   The absolutely normalized model spectra for  LH8
   compared to PHENIX spectra  \cite{PHENIX-Spectra} at different
   centralities. (a) is for $\pi^{-}$, (b) is for $K^{-}$ and, 
   (c) is for $\bar{p}$.
   }
\end{figure}

Turning to the high $M_T$ region, the absolutely normalized 
model spectra are compared 
directly to the absolutely normalized PHENIX spectra
for different 
centralities and particles in Fig.\,\ref{phCent}. 
For $K^{-}$ and $\bar{p}$ the spectrum is reproduced
in the most central bin (0-5\% central with $<N_{p}>=350$) 
and in the semi-peripheral bin (30-60\% central with
$<N_{p}>=76$). For pions, the spectral shape in the most central bin is reproduced.  However, in the 
semi-peripheral bin, the pion spectrum resembles a
power-law rather
than a thermal spectrum. This change in shape from
peripheral to central has been attributed to 
jet-quenching \cite{Wang-JetQuenching,GDavid-JetQuenching}.



\section{Elliptic Flow from the SPS to RHIC}
\label{FLEllipticFlow}

In non-central collisions the particles emerge with
an elliptic flow. 
The spectator matter
flies down the beam pipe and the excited nuclear matter
is formed in the transverse plane with an almond shaped
distribution. Subsequently,
if pressure develops in the system, the pressure gradients
are larger in the impact parameter direction (the x-direction) 
than in the y-direction. Then, the
excited matter expands preferentially in the x-direction.
The magnitude of this elliptic response is quantified experimentally
by expanding the distributions in a Fourier series 
\begin{eqnarray}
\frac{dN}{p_{T}\,dp_{T}\,dy\,d\phi} &=& \frac{dN}{2\pi\, p_{T} dp_{T} dy}
(1 + 2\,v_{2}(p_T,y)\,\cos(2\phi) \nonumber \\
& & +...)
\end{eqnarray}
where $\phi$ is measured around the z-axis relative to the
impact parameter, which points in the x direction.
The $elliptic$ flow, $v_2(p_T,y)\equiv\langle \cos(2\phi) \rangle_{p_T,y}$,
gives
a measure of the dynamic response of the excited nuclear matter 
to the initial anisotropy.


The initial spatial anisotropy is quantified using the parameter
$\epsilon$ (see Eq.~\ref{GlauberEquEps}).
The hydrodynamic response is
linear in $\epsilon$ \cite{Ollitrault-Elliptic} and therefore 
$v_2$ is sometimes divided by $\epsilon$ to compare different
impact parameters and nuclei \cite{Sorge-kink,Voloshin-LowDensity}. 
As the system expands, the
eccentricity $\epsilon$ decreases.  Since $\epsilon$ 
is the driving force behind the elliptic flow, the  elliptic
development finishes before the radial development. 
Therefore, elliptic flow is generated by the early
pressure, although this statement must be qualified (see below). 
The
spatial anisotropy that remains after the collision is quantified
by $s_{2}$ (see Eq.~\ref{s2}).  
 $s_{2}$ measures how much of the initial spatial
anisotropy $\epsilon$ was not used during the collision for the
production of elliptic flow.


\subsection{Qualitative Changes from the SPS to RHIC}
\label{FLEllipticFlow-Qual}
In Sect.~\ref{FLRadialFlow-SPS} radial flow was used
to constrain the EOS. The best (though certainly not unique)
description of the data was given by LH8.
For  consistency, we require the same EOS to
describe the elliptic flow data.

\begin{figure}[!tbp]
   \begin{center}
   \includegraphics[height=2.8in, width=2.8in] {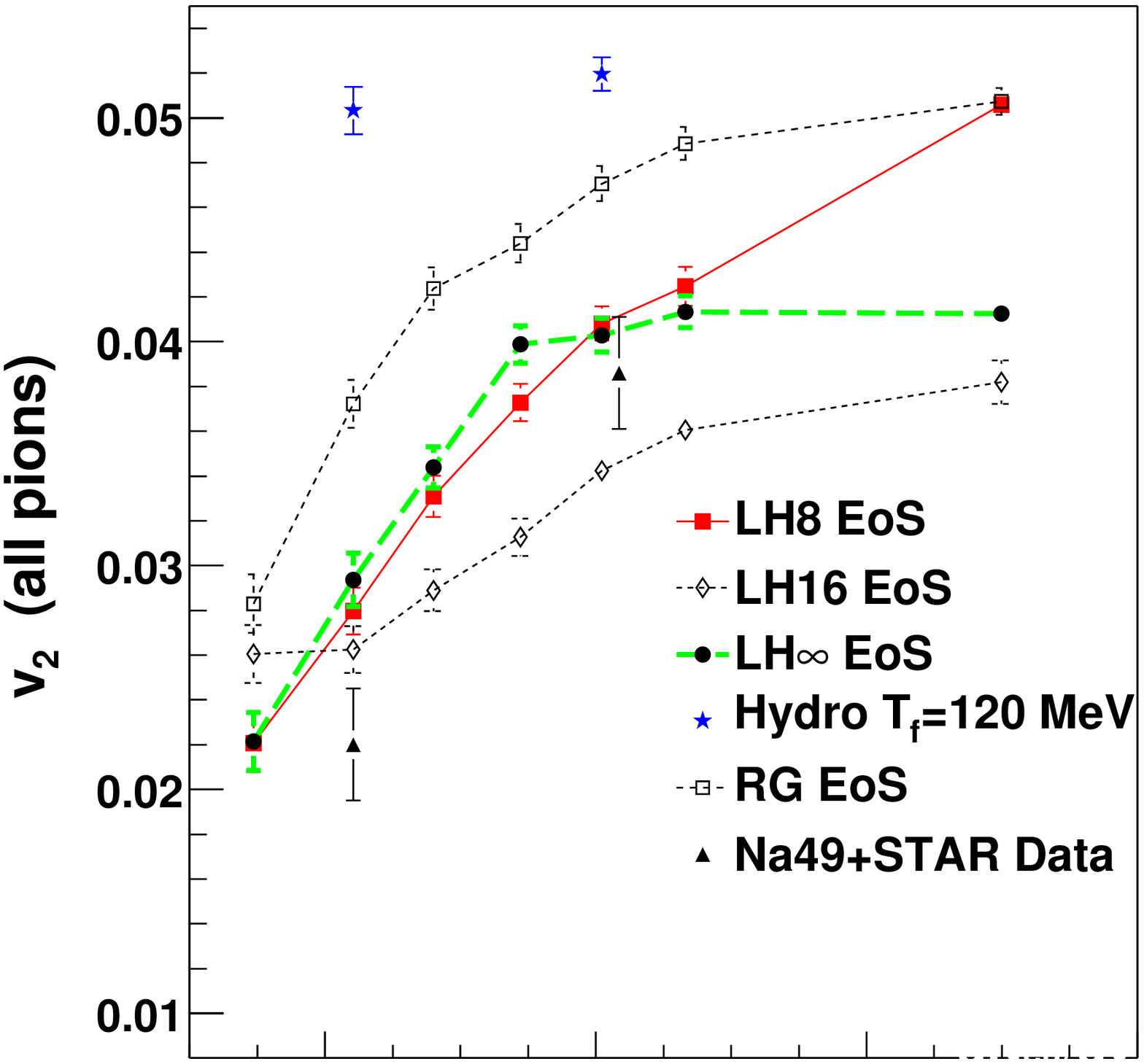}
   \hspace{-0.4in}
   \includegraphics[height=2.4in, width=2.8in] {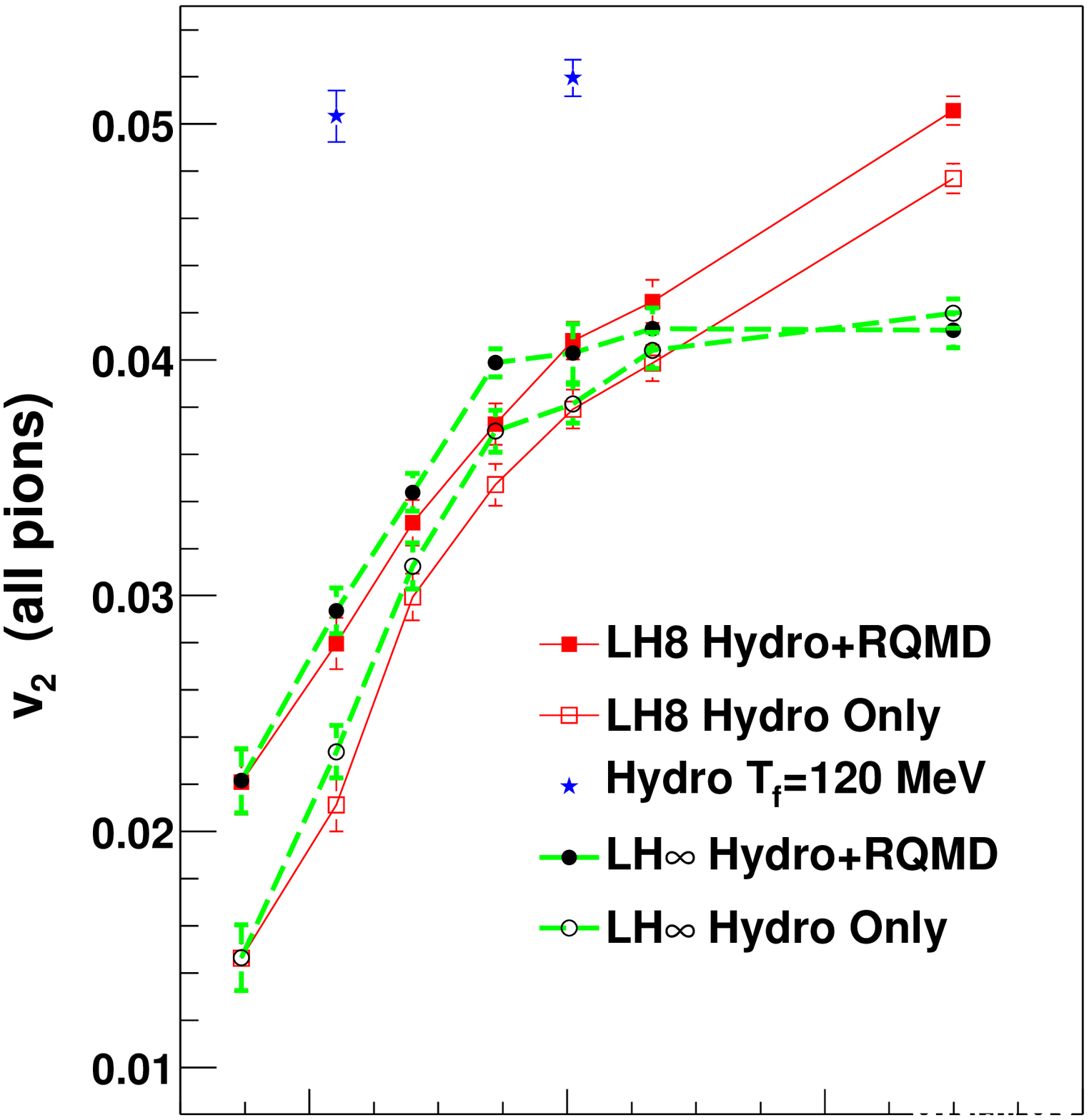}
   \includegraphics[height=2.8in, width=2.8in] {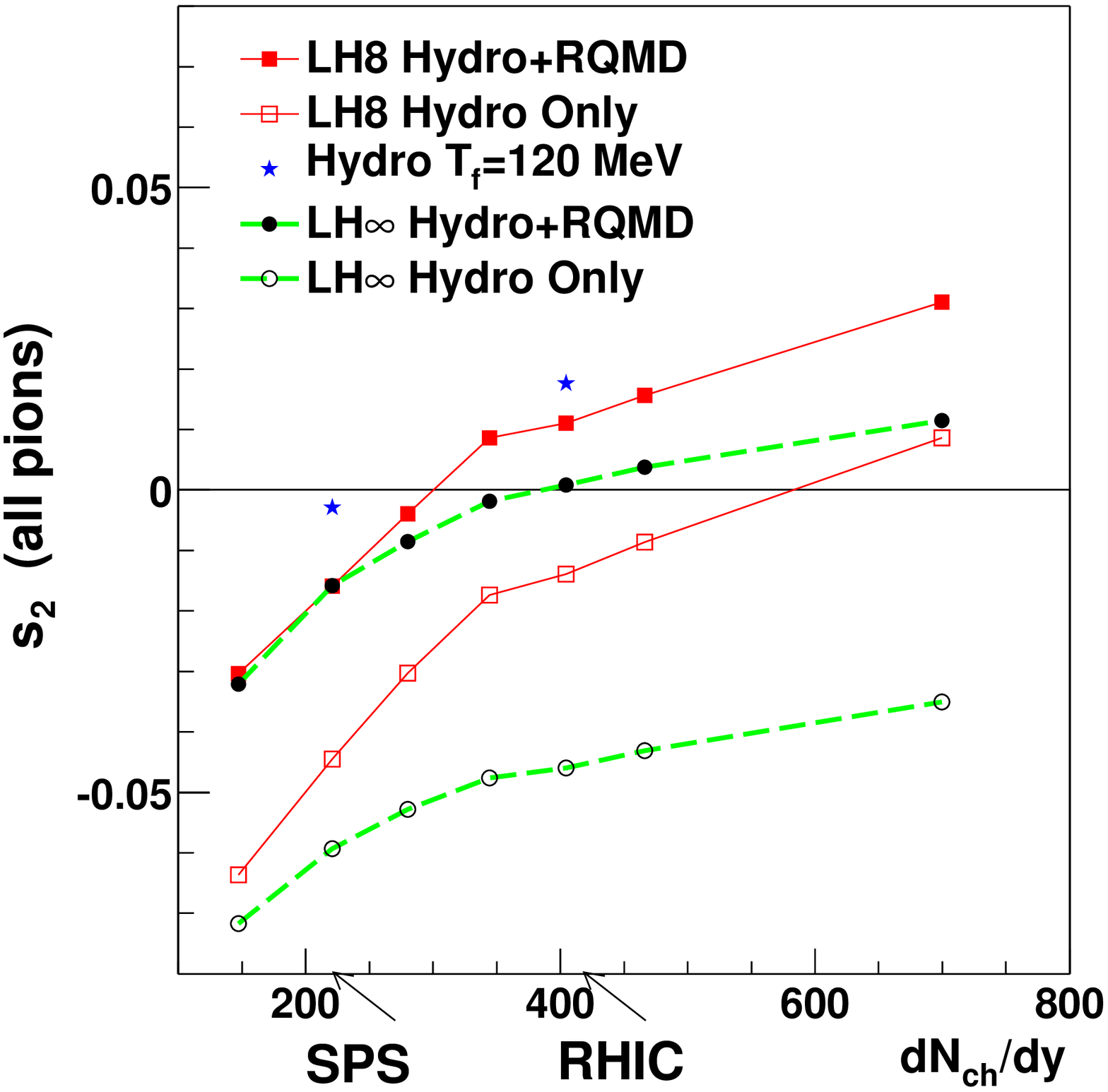}
   \end{center}
   \caption[The elliptic flow and spatial anisotropy 
   as a  function of the total multiplicity in a PbPb collision
   ]{
   \label{v2dndy}
       Panels (a)-(c) show three related 
       quantities as a function of
       the total multiplicity in a PbPb collision at $\mbox{b}=6\,\mbox{fm}$.
       (a) shows the integrated elliptic flow $v_2$ of pions for
       different EOS and freezeout conditions;
       (b) shows the integrated elliptic flow $v_2$ with and without
       hadronic rescattering;
       (c) shows the spatial anisotropy $s_{2}$ with and without 
       hadronic rescattering.
      At the SPS, the NA49 $v_2$  
      data point is extrapolated to b=6\,fm using
      Fig.~3 in  \cite{NA49-BDependence}.  
      At RHIC, the STAR $v_2$ data point is extrapolated to 
      $N_{ch}/N_{ch}^{max}=0.545$ (b=6\,fm in AuAu) using  Fig.~3  
      in \cite{STAR-EllipticPRL}.  
   }
\end{figure}

Fig.\,\ref{v2dndy}(a) shows integrated elliptic flow of pions 
as a function of the total
multiplicity for different EOS.  Fig.~\ref{v2dndy}(b)
shows the relative contribution of RQMD to the
integrated elliptic flow.
Note that elliptic flow increases for all EOS and 
dramatically so for LH8.
Assume momentarily that elliptic flow for Hydro+RQMD stops developing at a 
temperature of $T\approx T_{c}\approx 165\,\mbox{MeV}$. 
(Note however that the radial flow develops well below this temperature). 
The
dramatic increase of elliptic flow in 
Fig.\,\ref{v2dndy}(a) can be understood  as 
the dynamic response of the QGP pressure. 
Recall Fig.~\ref{TimeVt}(b), which plots anisotropy of the hydrodynamic
stress tensor versus time and pay particular attention to the 
$T\approx T_c$ points (the solid symbols) 
on the LH8 and RG curves (ignore LH$\infty$ for now).
At the SPS, the anisotropy  of
the stress tensor increases rapidly at first and then stalls. 
The final stress tensor anisotropy is small at the end of the mixed phase. 
At RHIC, elliptic flow develops more rapidly  and stalls
only when the anisotropy is large.  Thus, provided the
elliptic flow stops developing at $T_{c}$,  the elliptic
flow increases dramatically as the QGP pressure appears.
For a  RG EOS at the SPS, there is no mixed phase and no stall
and consequently the RG elliptic flow is significantly larger than
LH8 and the data. However with RHIC collision energies, LH8
begins to behave as an ideal QGP. Consequently, at RHIC 
the RG elliptic flow is only 20\% larger than that of LH8  
and of the data.   


To understand when elliptic flow stops developing, it is important
to track the spatial geometry of the underlying source. 
When the spatial anisotropy, $s_{2}$, is negative, the
pressure drives elliptic flow. However, as $s_{2}$ approaches
zero, the pressure gradients drive radial motion rather than
elliptic motion.  The elliptic development then stops. 
Fig.~\ref{v2dndy}(c) shows the spatial anisotropy, $s_{2}$, 
for pions as a function of multiplicity from the
SPS to RHIC.  Compare the LH8 curves (the stars, the solid squares,
and the open squares) seen in Fig.\,\ref{v2dndy}(c).
The open squares (LH8 Hydro Only) show the spatial anisotropy, 
$s_{2}$, at the end of the
mixed phase, the closed squares (LH8 Hydro+RQMD) show 
$s_{2}$ after the cascade, and the
stars show $s_{2}$ when the hydrodynamic evolution is continued to
$T_{f}=120\,\mbox{MeV}$.
After the mixed phase (LH8 Hydro Only), 
the matter retains some of its initial almond shape.
Continuing the hydrodynamics 
destroys the initial almond shape completely and
increases  $v_{2}$ by
a factor of $\approx2-3$ (see Fig.~\ref{v2dndy}(c)). 
Cascading also changes the almond shape
but  increases $v_{2}$ by only a factor of $\approx1.5$. 
In either case, $s_{2}$ crosses zero between the 
SPS and RHIC and therefore elliptic development in the 
hadronic stage ceases to be significant between the
SPS and RHIC.

\begin{figure}[!tbp]
\begin{center}
   \includegraphics[height=3.0in,width=3.0in]{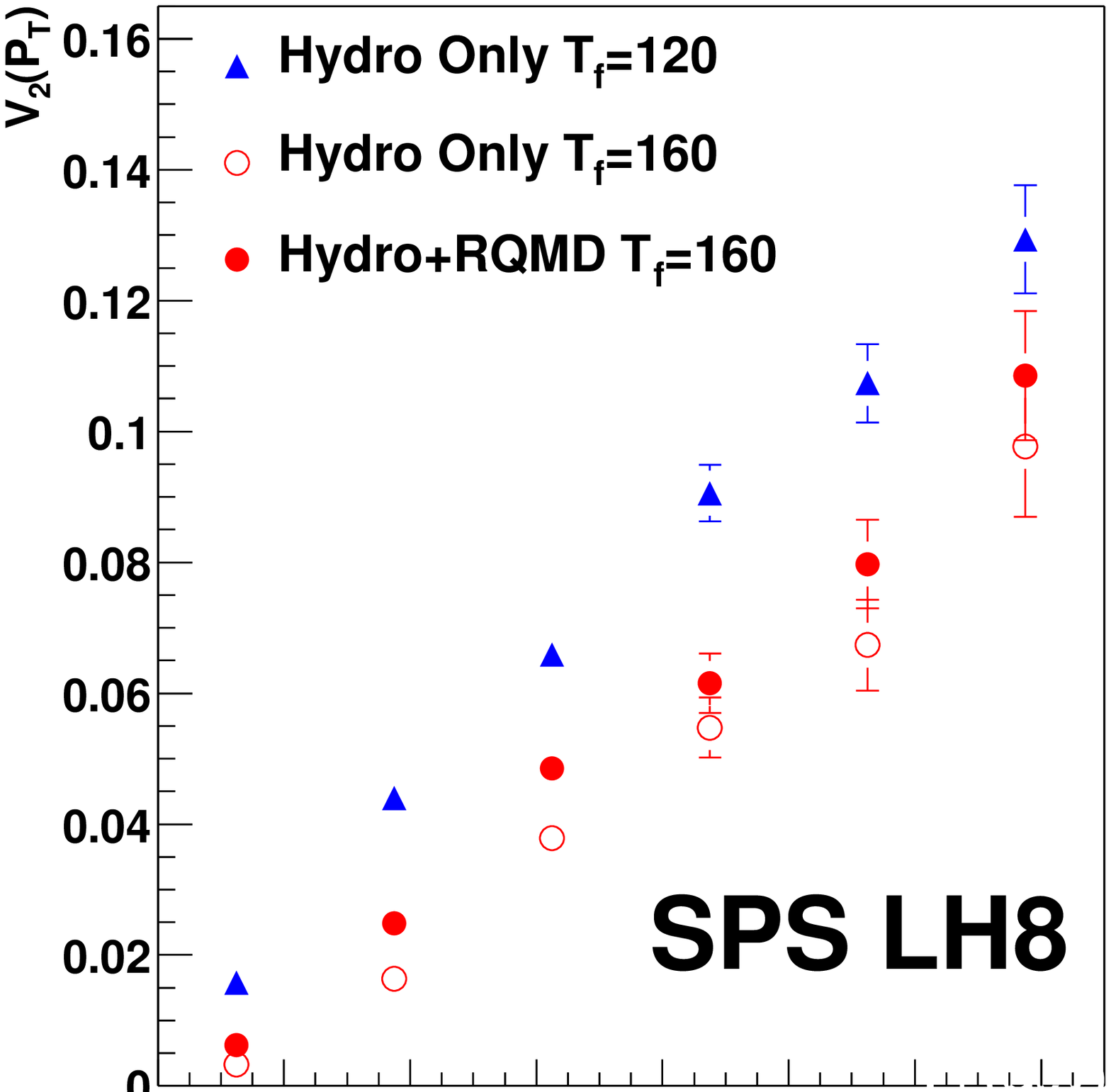}
   \includegraphics[height=3.0in,width=3.0in]{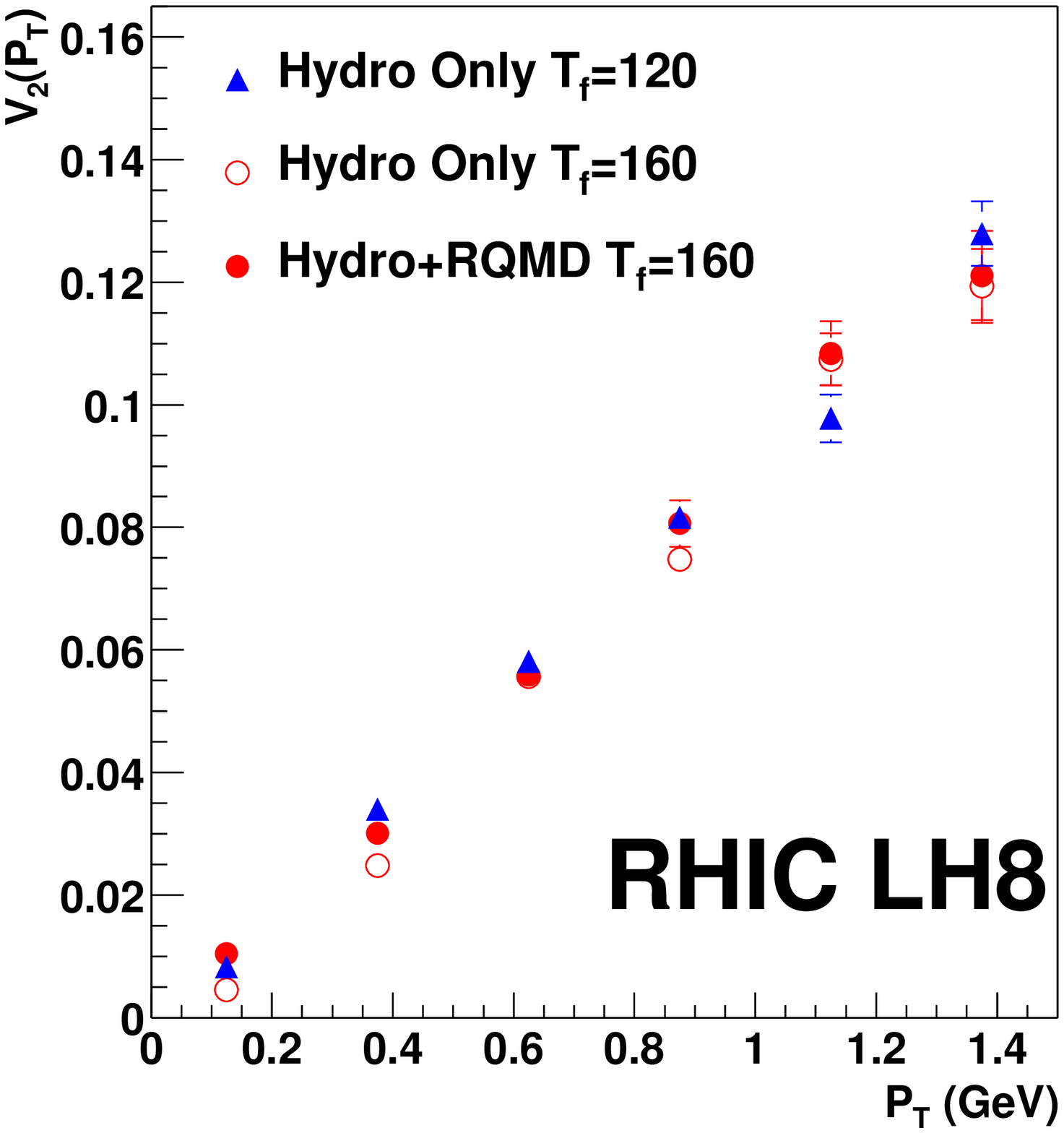}
\end{center}
\caption[The dependence of $v_{2}(p_{T})$ on the switching 
temperature at the SPS and RHIC] {
\label{v2ptcascade}
   The dependence of $v_{2}(p_{T})$ on the switching temperature 
   $T_{switch}$ at (a) the SPS and (b) RHIC for 
   an impact parameter of $\mbox{b}=6\,\mbox{fm}$.
}
\end{figure}

This fact is illustrated with Fig.\,\ref{v2ptcascade}
which contrasts the RQMD contribution to $v_2(p_T)$ 
with the contribution of the hadronic phase in a
pure hydrodynamic calculation at the SPS and RHIC.
At the SPS, 
$v_2(p_{T})$ increases by a factor of two when the
the hydrodynamics is continued to $T_{f} =120\,\mbox{MeV}$.
When the hydrodynamics is replaced with RQMD, 
$v_2(p_T)$ also develops but only by approximately $20\%$.
At RHIC, the spatial asymmetry is completely destroyed 
 by the end of the mixed phase and the elliptic 
development is frozen for all $p_T$.   
Continuing  with the cascade or the hydrodynamics increases
the elliptic flow marginally.

\subsection{The Impact Parameter Dependence of Elliptic Flow}
\label{FLEllipticFlow-Bdepend}

\begin{figure}[!tbp]
\begin{center}
   \includegraphics[height=3.0in,width=3.0in]{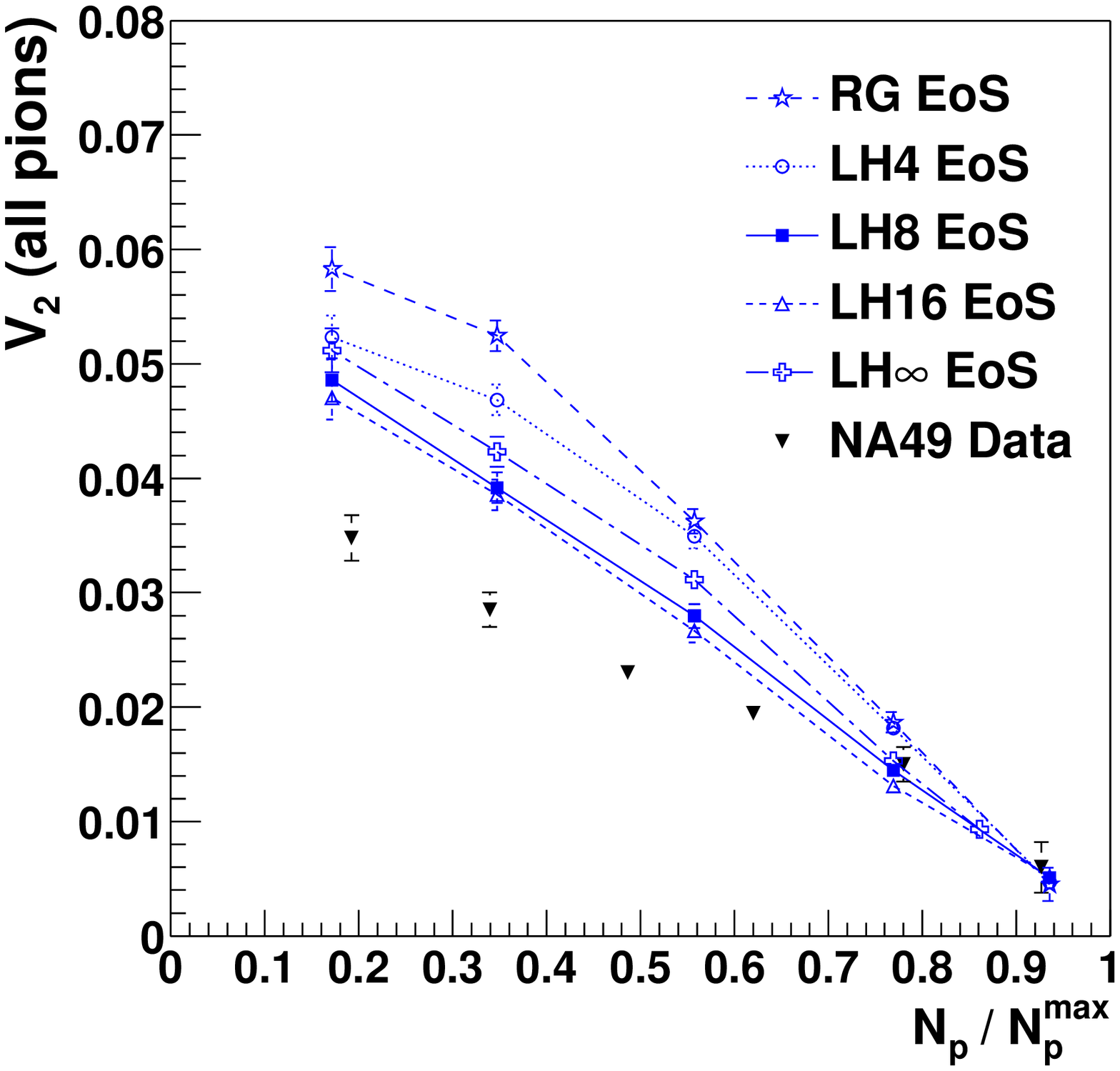}
   \includegraphics[height=3.0in,width=3.0in]{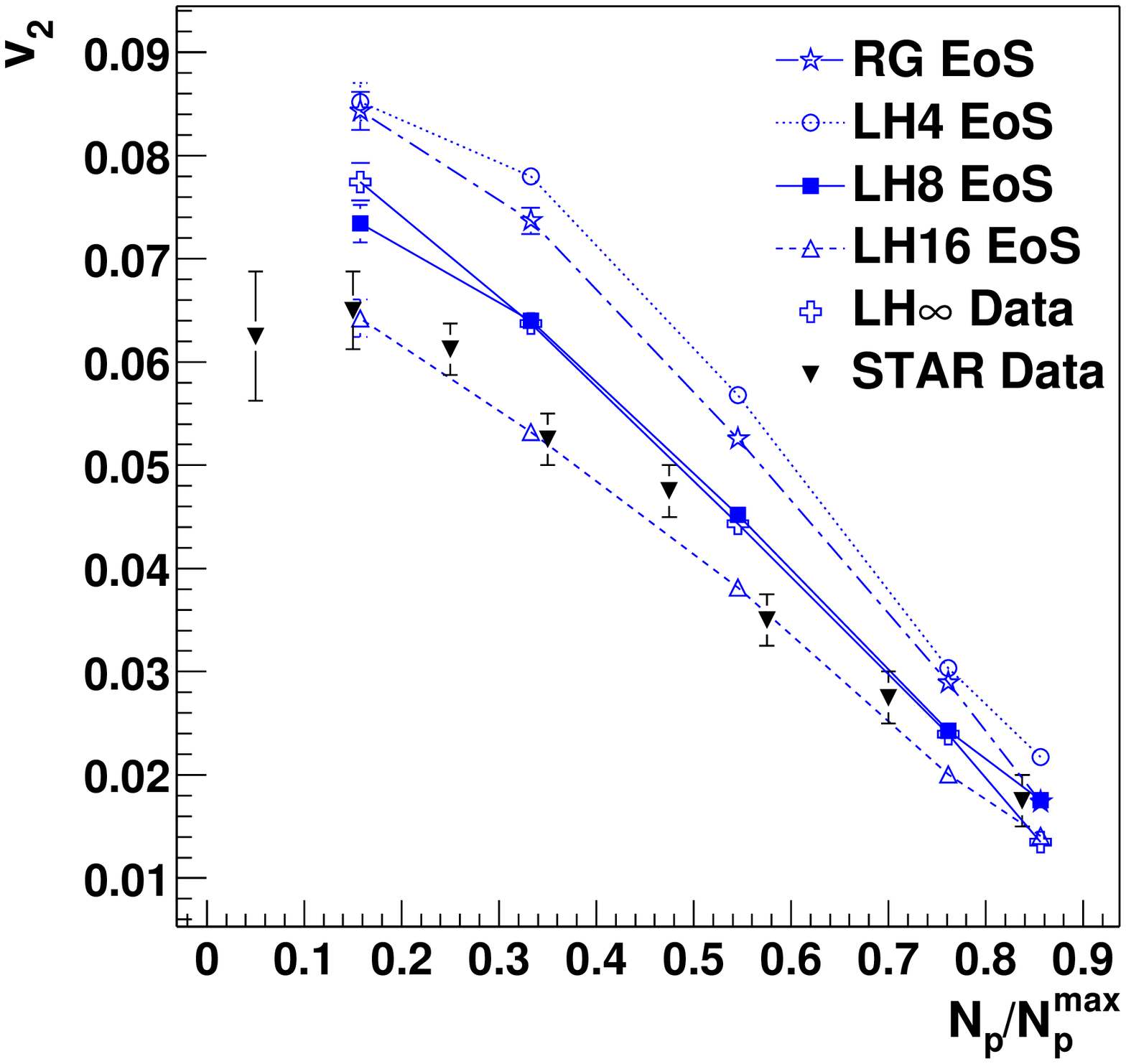}
\end{center}
\caption[Integrated elliptic flow as a function of impact parameter 
at the SPS and RHIC for different EOS]{
\label{v2bsps}
(a) $v_{2}$ for  pions at the SPS as a function
of participants (relative to the maximum) compared to NA49 data \cite{NA49-BDependence}.
(b) $v_{2}$ for charged particles at RHIC
as a function of participants (relative to the maximum) compared to
STAR data \cite{STAR-EllipticPRL}. 
}
\end{figure}

Fig.\,\ref{v2bsps} shows the b or  $N_{p}$ dependence of
integrated pion elliptic flow at the SPS and RHIC.

The
data restrict the underlying EOS.
At the SPS,
the data favor a soft EOS -- LH $\ge$ $0.8\,\mbox{GeV/fm}^{3}$.
LH4 and RG EOS generate too much elliptic flow. For LH8-LH16,
the model is about $20\%$ above the data.
However, the model to data
comparison is not completely fair -- the data points are 
integrated over rapidity, while the model
points are only strictly valid for mid-rapidity.  
This probably accounts for the 
residual model/data discrepancy.
As the latent heat is increased beyond LH32 to LH$\infty$ ,
the elliptic flow begins to rise. The
origin of this elliptic flow  was described in \cite{Htoh} and results
from
the slow evaporation of particles in an asymmetrical fashion
over a long time. This elliptic flow is generated without
radial flow \cite{Kolb-NoFlow} 
and the $p_T$ dependence of $v_{2}$  for nucleons (see below)
is modified accordingly \cite{Kolb-NoFlow}.
 At RHIC (Fig. \ref{v2bsps}(b)),
the comparison is fair and the data again  favor a relatively
soft EOS, LH8-LH16. Thus the elliptic flow data at the SPS and
RHIC are consistent with a single underlying EOS.

Note that the ordering of the EOS in Fig\,\ref{v2bsps}
 differs at the SPS and RHIC. At the SPS, LH8 and
LH16 generate approximately the same elliptic flow. At RHIC,
the hard QGP phase lives for  
substantially longer with LH8 than with LH16 and
therefore generates more elliptic flow. Additionally at RHIC,
LH4 generates more elliptic flow than
a RG EOS. Thus, the elliptic flow 
indicates that at high energy densities 
LH4 (with $c_{s}^{2}\approx1/3$)  
has a larger speed of sound than a RG EOS (with $c_{s}^2\approx 1/5$).
At asymptotically, high energy densities all EOS in the
LH(x) family approach the massless ideal gas limit.

\subsection{The $p_{T}$ Dependence of Elliptic Flow}
\label{FLEllipticFlow-PtDepend}

Having discussed qualitative changes from the SPS to RHIC,
we explore the $p_{T}$ dependence of elliptic flow. 
Experimental measurements are performed over a range of 
impact parameters. To find $v_{2}(p_{T})$  in a
specific impact parameter range, $b_{min} < b < b_{max}$, the
following integrals must be performed,
\begin{eqnarray}
\label{v2minbi}
v_{2} (p_{T},y)_{b_{min}}^{b_{max}} &\equiv&
\frac
   {
      \int_{b_{min}}^{b_{max}} v_{2}(p_{T},y;b)\,
      \frac
      {  dN  }
      {  dy\,dp_{T}  }(b)
      \, 2\pi b\,db
   }
   {
     \int \frac{ dN}{dy\,dp_{T}}(b) \, 2\pi b\,db
   }. \nonumber \\
\end{eqnarray}
Again, we drop the $y$, $b_{min}$ and $b_{max}$ labels below when
it is not confusing.
$v_{2}(p_T)^{min-bias}$ denotes the elliptic flow integrated 
over all events, or $v_{2}(p_{T})_{0}^{\infty}$.

\begin{figure}[!tbp]
   \begin{center}
      \includegraphics[height=2.7in,width=2.7in]{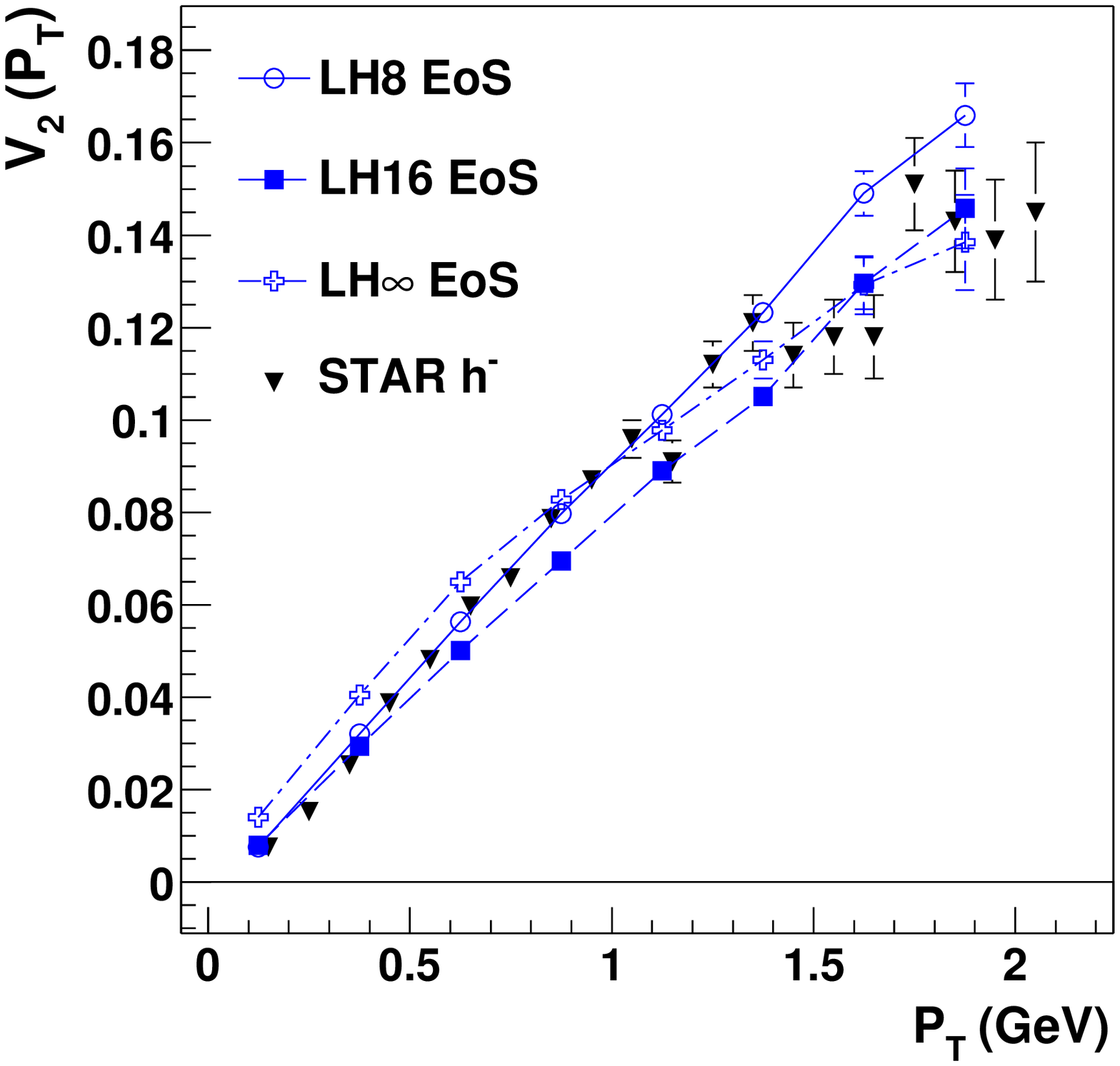}
      \includegraphics[height=2.7in,width=2.7in]{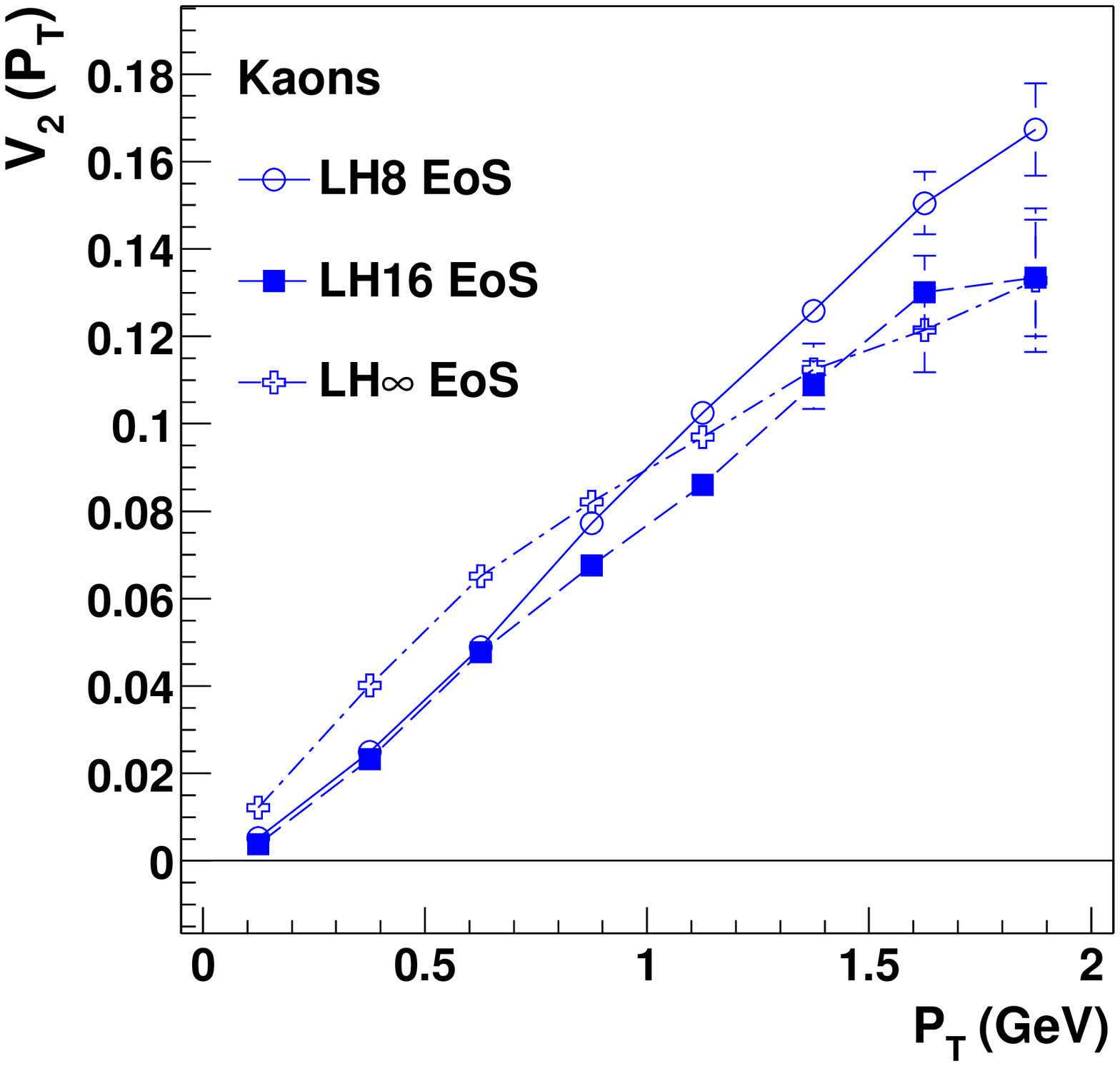}
      \includegraphics[height=2.7in,width=2.7in]{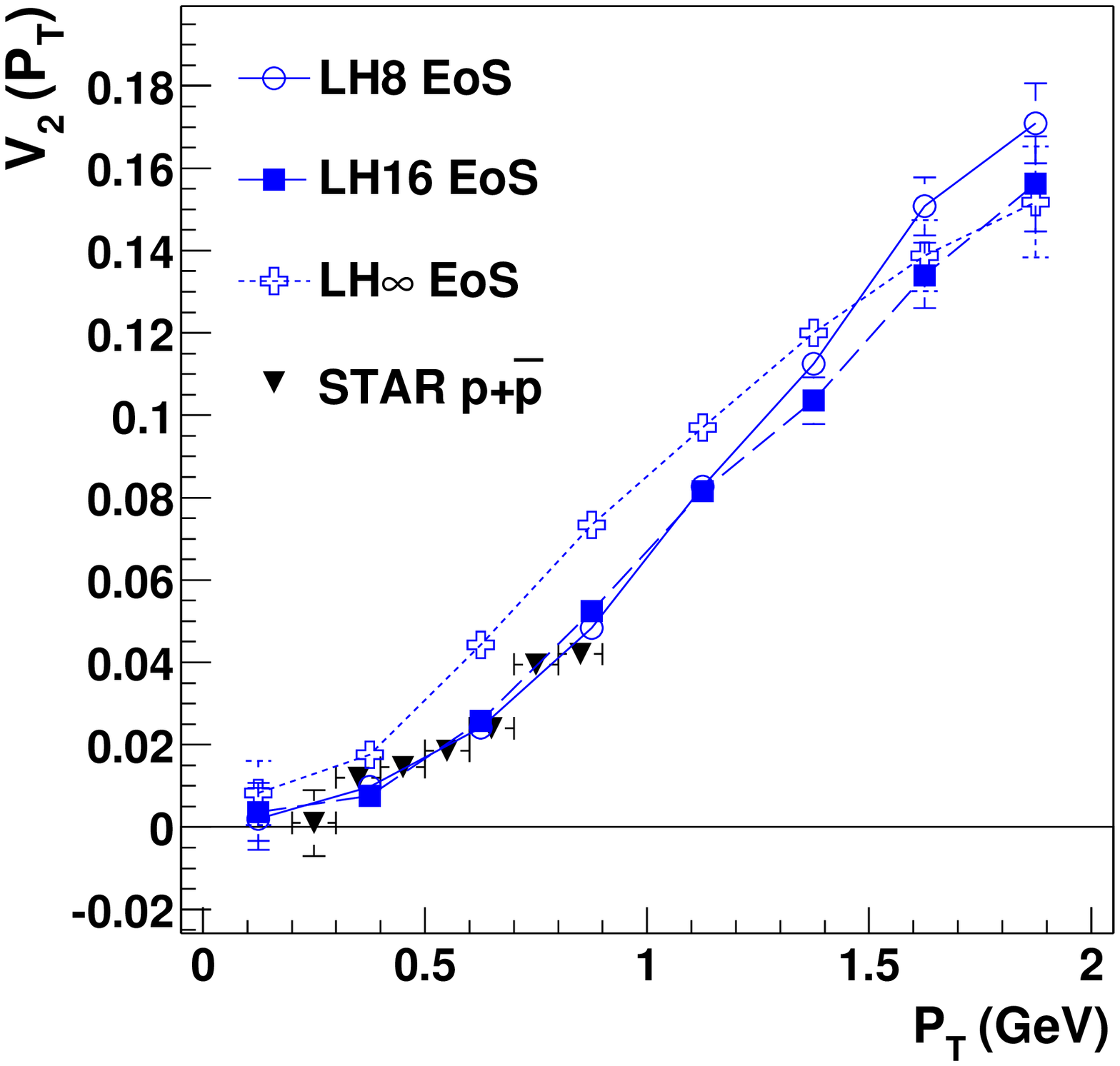}
   \end{center}
   \caption[A comparison of model results for different EOS 
   to STAR data on 
   elliptic flow as a function of $p_T$ for pions and kaons
   ]{
   \label{Eosdependence}
   $v_{2}(p_T)^{min-bias}$ (see Eq.~\ref{v2minbi})  for
	three different EOS compared to data. Panels (a), (b) and (c) 
	are for negative hadrons, all kaons, and p+$\bar{\mbox{p}}$, respectively.
   The data are from \cite{STAR-EllipticPRL,STAR-EllipticParticle}.
   }
\end{figure}

Fig.~\ref{Eosdependence}(a), (b) and (c) show $v_{2}(p_{T})^{min-bias}$ 
for negative hadrons, kaons, and  nucleons at RHIC.
Look first at  the negative hadrons (a):
Although LH8 and LH16 both show a strong linear rise,
the slope is smaller for LH16.
For LH$\infty$,  $v_{2}(p_{T})$ is
curved and bends over. For small $p_{T}$,
LH$\infty$ is above LH8, but by $p_{T}\approx 2.0\,\mbox{GeV}$,
LH$\infty$ is substantially below LH8.
The data show a strong linear rise and agree remarkably well with
the slope of LH8.  $v_{2}(p_{T})$  slightly
favors LH8 over LH16. The kaon $v_{2}(p_{T})$ curve has the
same shape and magnitude as the $h^{-}$ spectrum.
At the
SPS the kaon elliptic flow is slightly negative \cite{SPS-NegKaV2}. 
This
anti-elliptic flow  is most likely a remnant of the repulsive 
mean field observed at higher baryon densities.  
At RHIC, the baryon density is
lower than at the SPS and kaons should flow along with the pions
if the space time picture of the model is correct.

For nucleons, the $v_{2}$ spectral shape is different and
is initially curved upwards.  
A useful thermal model has been given to explain
the shape of $v_{2}(p_T)$ \cite{Kolb-Radial,SR-FlowProfile}. 
For nucleons, LH8 and LH16 are
concave up, indicating a strong radial expansion. 
By contrast, LH$\infty$ 
shows a linear rise in $v_{2} (p_{T})$, indicating a weak
transverse  expansion. As discussed in Sect. \ref{FLSpaceTime}, 
LH$\infty$ slowly evaporates particles into 
RQMD and generates elliptic flow
only at small $p_{T}$.
The curvature of $v_{2}(p_{T})$ for LH$\infty$ resembles 
the $p_{T}$ dependence expected if only surface
evaporation were present \cite{Kolb-NoFlow}.
However, LH$\infty$ does develop a substantial radial
flow over its long lifetime which gives the 
LH$\infty$ $v_{2}(p_T)$ some shape. The data favor 
the strong transverse expansion of LH8 over the weak expansion
of LH$\infty$. 

\begin{figure}[!tbp]
   \begin{center}
      \includegraphics[height=3.0in,width=3.0in]{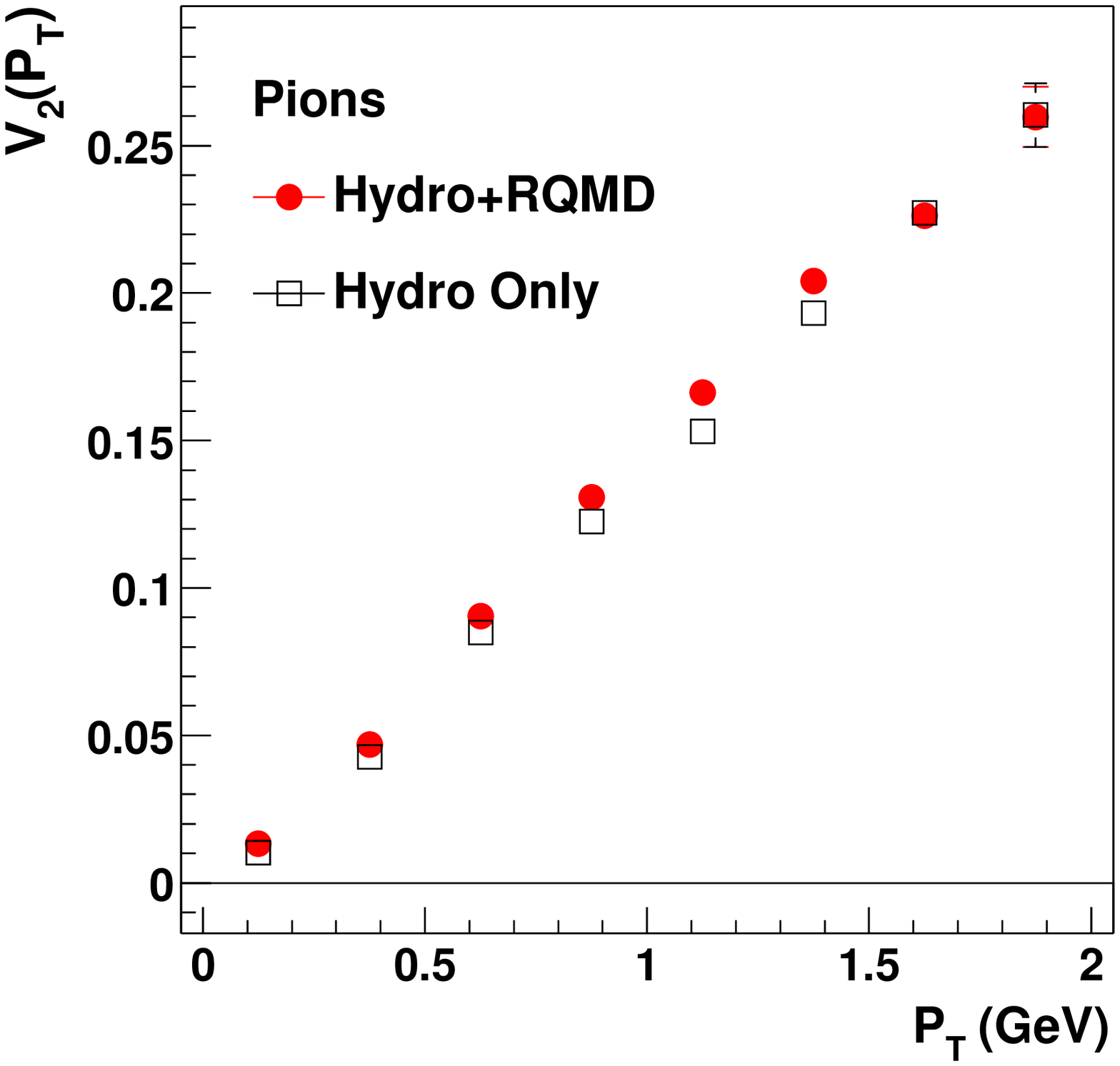}
      \includegraphics[height=3.0in,width=3.0in]{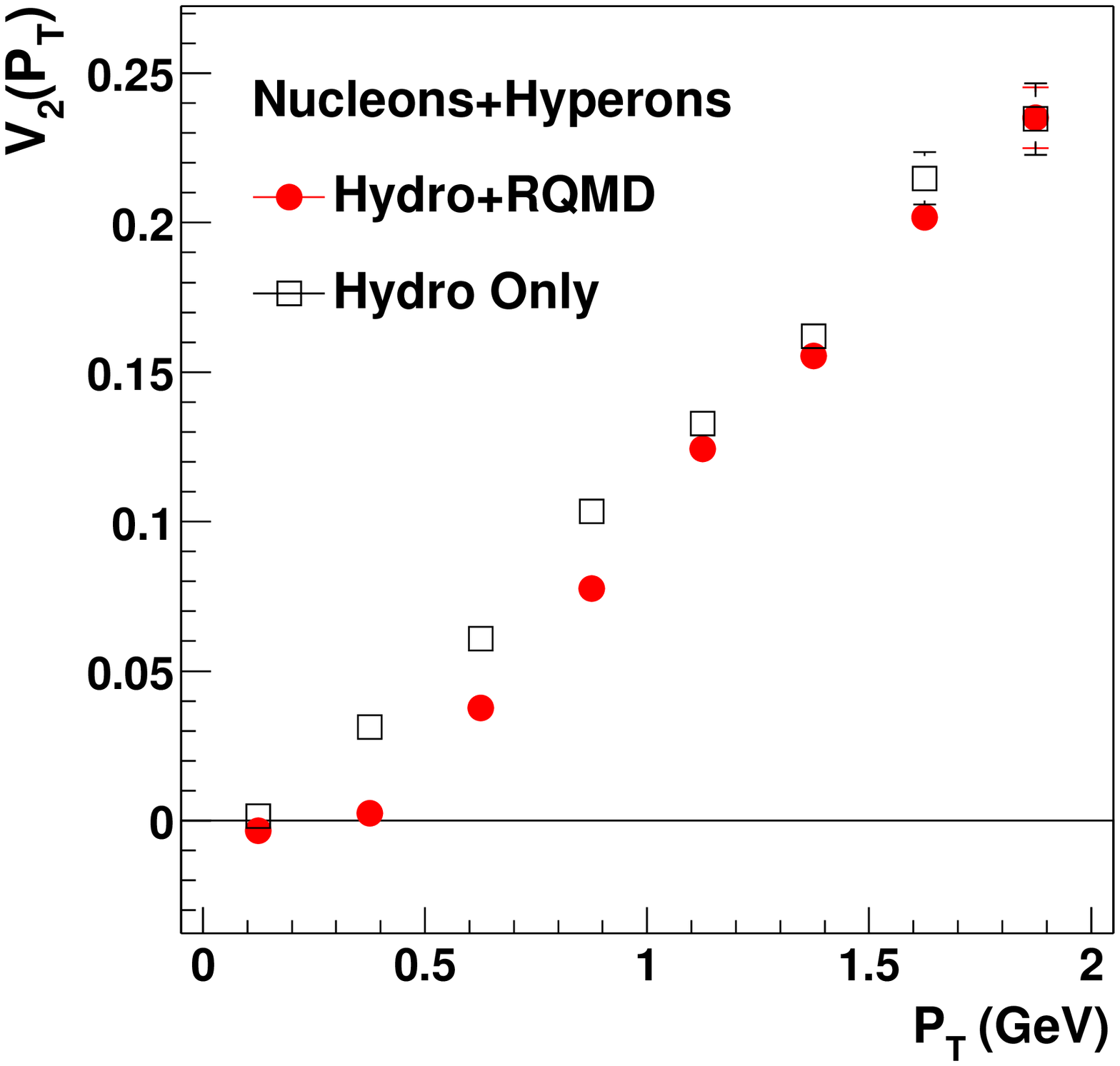}
   \end{center}
   \caption[The effect of hadronic rescattering on elliptic flow
   as a function of $p_{T}$ for pions and baryons at RHIC]{
   \label{v2ptRQMD}
   $v_{2}$ as a function of $p_{T}$ for (a) all pions and (b) 
	nucleons+hyperons with and without the RQMD after-burner for AuAu collisions at 
	b=6\,fm, for LH8 EOS. 
   }
\end{figure}

We now demonstrate that pion nucleon scattering on top of a baseline
elliptic flow is responsible for the curvature of $v_{2}(p_{T})$
seen in data. Fig.~\ref{v2ptRQMD} shows $v_{2}(p_{T})$ with and without
hadronic rescattering. Here the discussion parallels the
discussion of the radial flow. Pion nucleon scattering 
increases the radial flow of the nucleon spectrum and cools the
pion spectrum.
Consequently the pion $v_{2}$ spectrum with the RQMD after-burner
is slightly above the ``Hydro Only'' spectrum. Similarly, the nucleon
$v_{2}$ spectrum with the after-burner is curved upward by $\pi-N$
scattering within RQMD. Similar 
features were found in all the EOS studied above.

\begin{figure}[!tbp]
   \begin{center}
      \includegraphics[height=3.0in,width=3.0in]{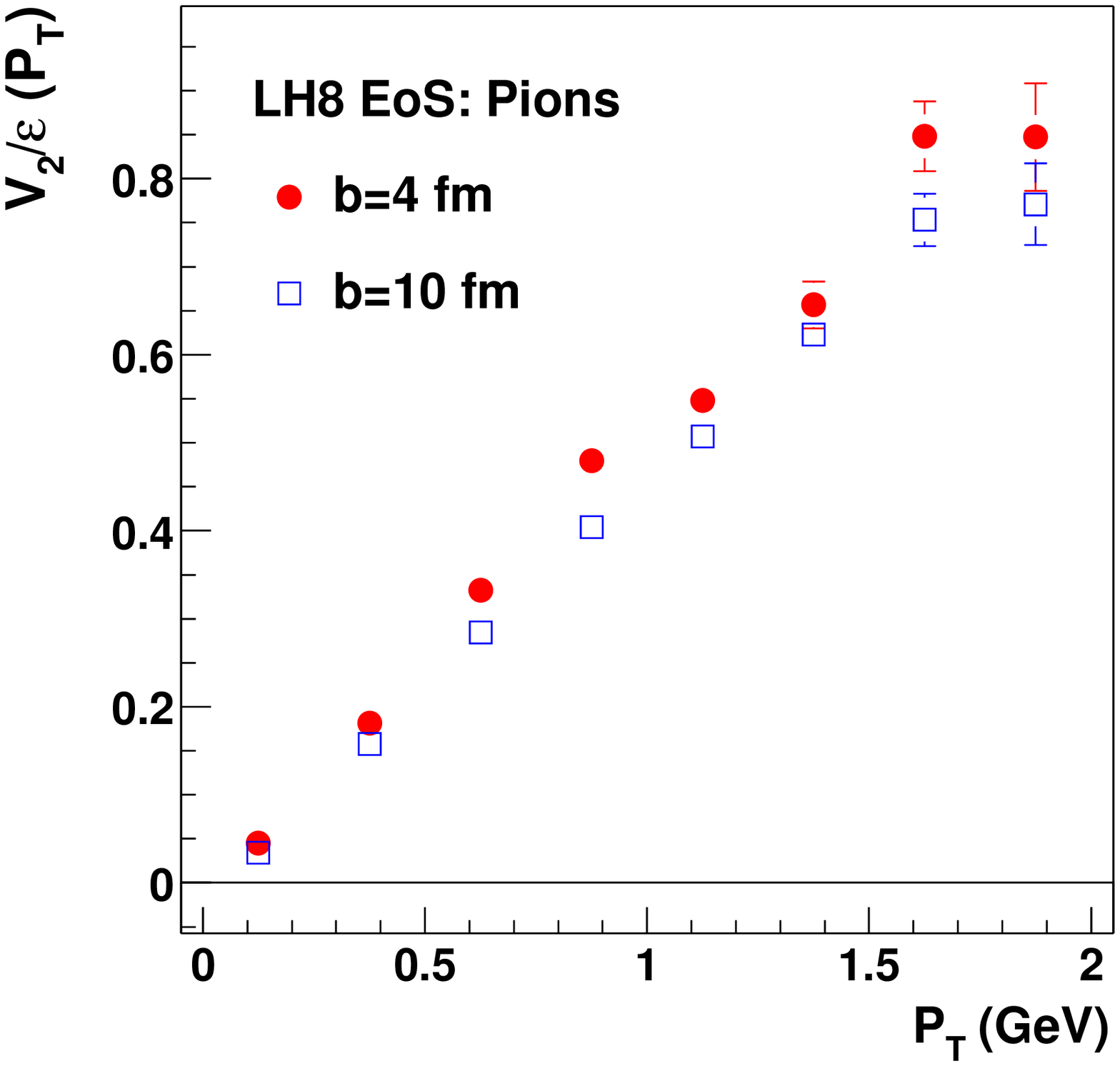}
      \includegraphics[height=3.0in,width=3.0in]{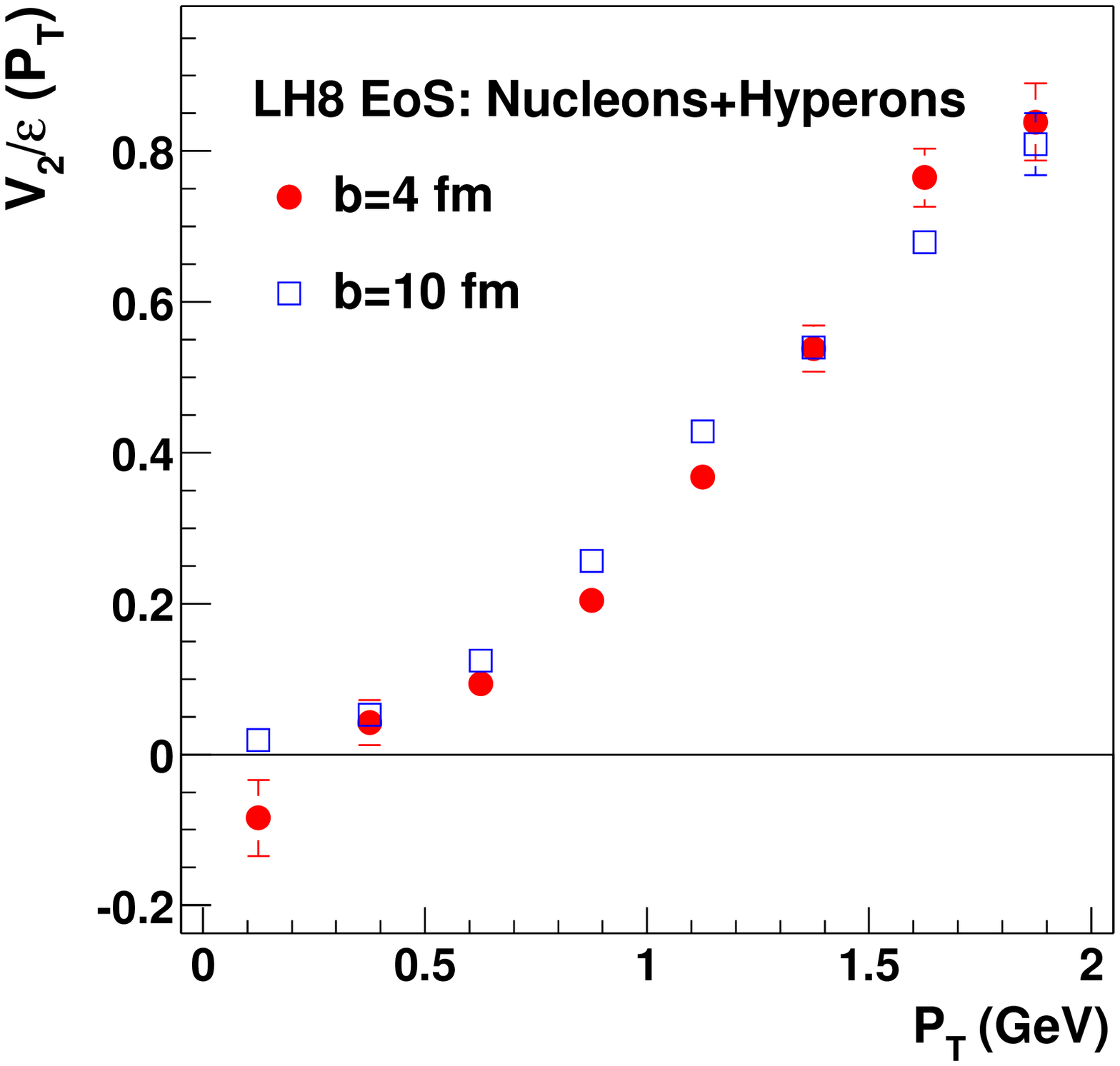}
   \end{center}
   \caption[Elliptic flow as a function of $p_{T}$ for pions 
   and protons in peripheral and semi-central collisions]{
   \label{V2Bdepend}
   $v_{2}(p_{T})/\epsilon$ for semi-central (b=4\,fm) and 
   semi-peripheral (b=10\,fm) collisions with LH8 EOS. Panel (a) is for pions and (b)
   is for nucleons+hyperons.
   }
\end{figure}

Now to illustrate the  impact parameter dependence, 
Fig.~\ref{V2Bdepend} shows the b-dependence of $v_{2}(p_{T})$. 
To compare different impact parameters,
$v_{2}(p_{T})$ for pions and nucleons
is divided by the initial space anisotropy $\epsilon$, for peripheral
(b=10\,fm) and semi-central (b=4\,fm) collisions.
The model response basically follows naive geometric considerations.
However, closer inspection reveals that the model captures 
some finite size effects during the late hadronic stages.
As the impact parameter is scanned, the total number of pions
decreases roughly as the number of participants, and pion-nucleon 
scattering decreases similarly. Consequently,
for central collisions
pions show slightly larger elliptic flow at small $p_{T}$
while nucleons show a smaller  (more curved)
elliptic flow at small $p_{T}$.
Thus, together these curves
indicate a slightly stronger hadronic expansion in central
collisions.

\begin{figure}[!tbp]
  \begin{center}
      \includegraphics[height=3.0in,width=3.0in]{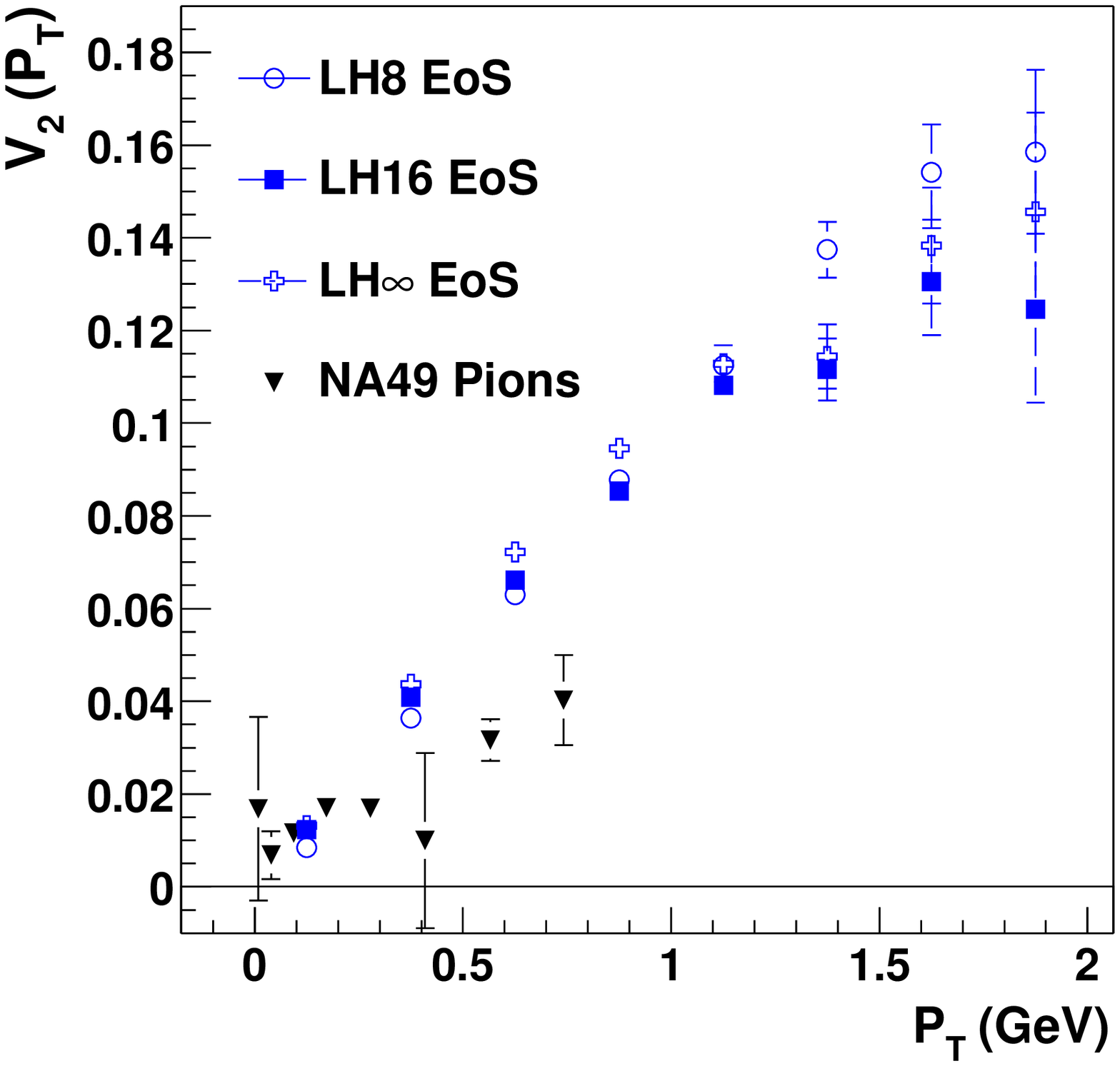}
      \includegraphics[height=3.0in,width=3.0in]{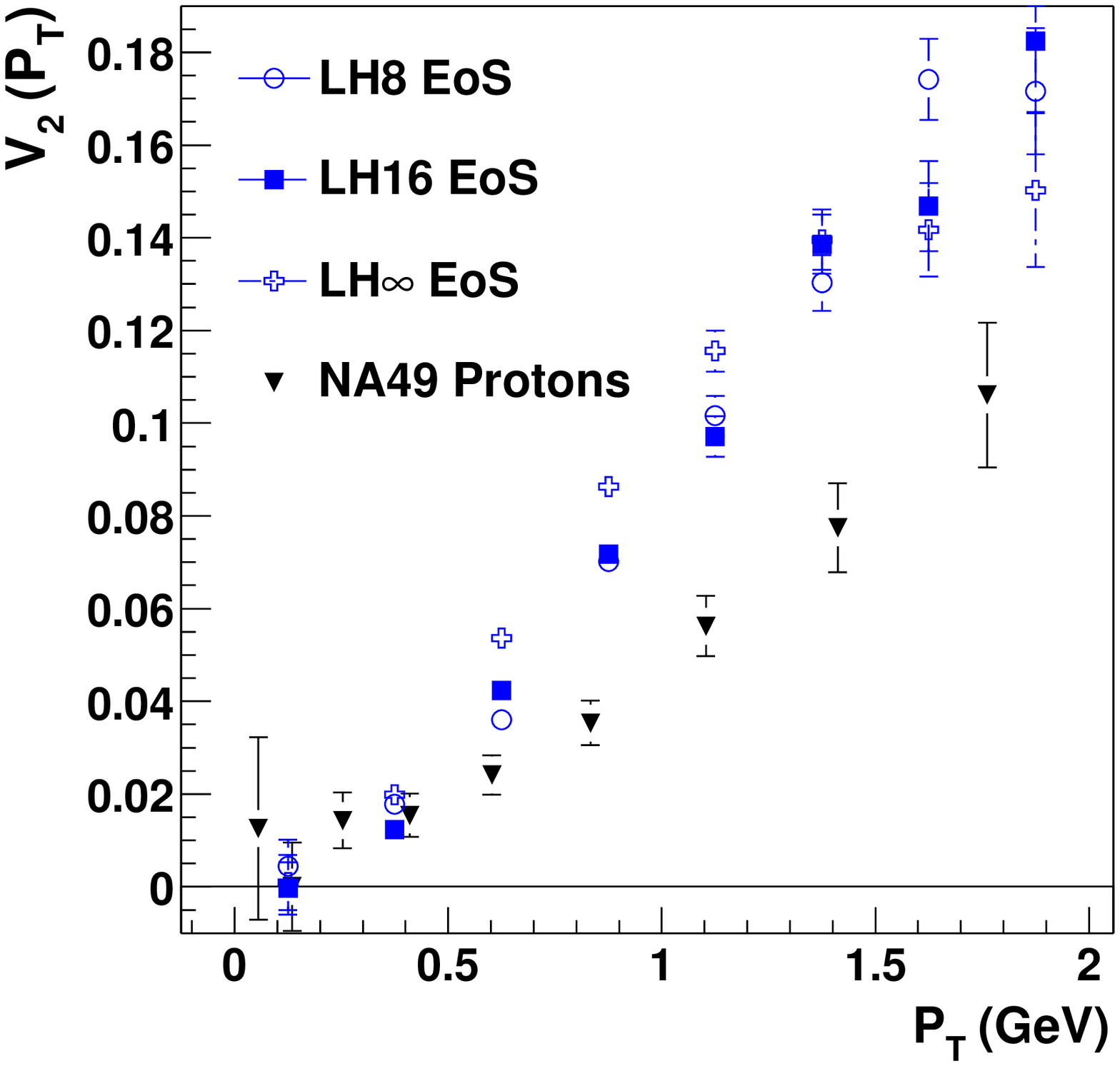}
   \end{center}
   \caption[A comparison of model results to NA49 data 
   on  elliptic flow as a function of $p_T$ for pions and protons.]{
   \label{V2ptsps}
   $v_{2}$ as a function of $p_{T}$ for pions and protons 
in PbPb collisions at the SPS.
The data points are for events with   
$b\approx6.5-8\,\mbox{fm}$  or more precisely for events
with 45-60\% of the beam energy in the NA49 zero-degree  
calorimeter \cite{NA49-EllipticPRL}. The model points
are the results of integrals between $\mbox{b}=6.5-8\,\mbox{fm}$ 
or $v_{2}(p_{T})_{6.5}^{8.0}$ as described by Eq.~\ref{v2minbi}.
The model data comparison
is not completely fair -- the model points are for mid-rapidity
while the data points have been integrated over rapidity (see text).
   }
\end{figure}

We now return to the SPS and compare the model to NA49 elliptic
flow data.  Fig.~\ref{V2ptsps} 
compares the model and data $v_2(p_{T})$ for pions and
nucleons. These data
are generally less well produced than at RHIC.
Some caveats should be mentioned. The data are
forward in rapidity, $4<y<5$, while strictly speaking, the model 
is only valid for mid-rapidity data (y=2.92). For pions, $v_{2}$  has a
fairly flat rapidity dependence but the total abundance changes
significantly from y=2.92 to y=5. For nucleons, $v_{2}$ has
strong rapidity dependence and is almost a factor of
two larger at mid-rapidity.  At mid-rapidity, the elliptic flow
is certainly stronger, which should improve the 
agreement with the model. Furthermore, in the 
forward rapidity region, the dynamics are complex; 
pions and nucleons have a significant directed flow  
indicating that details of stopping may play a significant
role. In conclusion,
much more data are needed to establish
the applicability of hydrodynamics at the SPS.

\label{FLconclusions}

\subsection{Summary and Comparison}

The principle motivation of this work was to demonstrate
that the body of heavy ion data from the SPS to RHIC 
can be described with thermodynamics and hydrodynamics.
To this end, we have compared of the hydro+cascade 
model of \cite{Htoh} to the radial and elliptic flow 
data from the SPS and RHIC.  A simultaneous analysis 
of available flow constrains the EOS. Only an EOS 
exhibiting the hard and soft features of the QCD
phase transition systematically reproduces the
observed radial and elliptic flow. 

The model incorporates strong radial and elliptic flow,
chemical freezeout at the phase boundary, subsequent
hadronic rescattering and differential freezeout. 
With these ingredients, the model explains a number of
features of the new RHIC data.
\begin{itemize}
\item
The ``anomalous'' $\bar{p}/\pi^{-}$ ratio
(which exceeds one for $p_{T} \sim 2\,\mbox{GeV}$)
is a simple consequence of the increase in the
radial flow and chemical freezeout.
In a thermal picture,
anti-protons are enhanced
relative to proton-proton collisions.
Then, the strong radial flow drives these anti-protons
to large $p_{T}$. Subsequent hadronic rescattering
makes the spectra cross.  

\item
The $M_{T}$ spectra (which are ``curved'' as opposed to simply exponential) 
are  readily explained in a hydro+cascade
model.  The curvature is due to a combination of 
the flow profile expected from hydrodynamics 
\cite{SR-FlowProfile,SSH-FlowProfile} and
hadronic rescattering.  With these ingredients,  the
mass dependence of the spectra measured by the
STAR and PHENIX collaborations are naturally explained.  
The strong
b-dependence
of the STAR slope parameters for anti-protons \cite{STAR-Spectra} is 
a consequence pion-nucleon scattering. In contrast,
the slope parameter for the $\phi$, which suffers few hadronic collisions in the model, shows little b-dependence. 

\item
The observed elliptic flow rises
rapidly as a function of $p_{T}$ and favors a strong
transverse expansion.
Unlike the radial spectra,
the elliptic spectra are less sensitive to hadronic
rescattering and differential freezeout. 
Therefore, our results on the $p_{T}$ 
spectrum of $v_{2}$  are largely similar to
the hydrodynamic analysis in \cite{Kolb-LowDensity,Kolb-Radial}.
\end{itemize}

These features are generic to a radially and elliptically 
expanding thermal source and do not immediately implicate
hydrodynamics as the dynamic origin of the 
radial and elliptic flow. However,  running
hydrodynamic 
up to the phase boundary (with the same EOS) quantitatively reproduces
the necessary radial and elliptic flow velocities both at the SPS and RHIC.
In particular, hydrodynamics reproduces the observed 
changes from the SPS to RHIC:

\begin{itemize}
\item
In a hydrodynamic framework, 
the radial flow  velocity increases at high 
energy densities for an EOS with a phase transition
to the QGP \cite{Ollitrault-MixedPhase, Kataja-MixedPhase}. 
At the SPS, LH8 gave the best agreement  with 
spectra  and predicted a $\approx 20\%$
increase in the radial flow velocity from
the SPS to RHIC \cite{Htoh}. 
The first
RHIC spectra confirm the predictions 
of LH8 and the predictions of other hydrodynamic
works \cite{Kolb-Radial,Kolb-UU}. Generally, LH8  has 
a smaller latent heat than that used in other
works and therefore LH8 predicts a larger increase in
the radial flow. In particular, already at RHIC, the $m_{T}$ spectrum of the
$\Omega^{-}$ is significantly curved by the radial flow \cite{HydroUrqmd}. 

\item
At RHIC and to a lesser extent at the SPS,  
the magnitude of the integrated elliptic flow 
is reproduced. 
During the early stage of this work,
a $\approx 40\%$ change in elliptic flow 
the SPS to RHIC was predicted and subsequently observed 
\cite{STAR-EllipticPRL}.
This increase is a direct consequence
of the QGP pressure \cite{Kolb-UU,Shuryak-QM99} and the early freezeout of elliptic
flow at the SPS \cite{Htoh}.   

\end{itemize}

\subsection{Fixing the EOS}

Taking the radial and elliptic flow together, we argue that
the momentum correlations in the data
reflect the hydrodynamic response of excited
matter exhibiting the soft and hard features of the
QCD phase transition. For an EOS without softness,
e.g. a resonance gas EOS, the flow of multi-strange baryons is
dramatically missed (see Fig.~\ref{slopessps}). 
In addition, the elliptic flow is
significantly too large both at the SPS and at RHIC 
(see Fig.~\ref{v2bsps}(a) and (b)). 
These
observations indicate that without softness the initial
hydrodynamic response of the fireball is too strong.

For an EOS without a hard component, e.g. LH$\infty$, the
spectra are significantly too soft both at the SPS and
RHIC (see Fig.~\ref{lhfig} and Fig.~\ref{rhic-sp2}). 
The flow of the multi-strange baryons is 
even too small (see Fig.~\ref{slopessps}).
Although LH$\infty$ generates a large $v_{2}$  by
evaporating particles anisotropically
through the freezeout 
surface, the $p_{T}$ dependence of this elliptic flow is
qualitatively wrong (see Fig.~\ref{Eosdependence}).  For LH$\infty$,
hadronic rescattering does generate
some transverse flow, but this transverse flow
is insufficient to explain the strong $p_{T}$ dependence of
the elliptic flow.
The strong curvature for $v_{2}(p_T)$ seen in the
nucleon data
implicates a violent explosion and this violent
explosion
is not generated by LH$\infty$.
Out of all the EOS considered, the best agreement is
found with LH8. The same EOS was deduced prior to 
the analysis of the first RHIC data \cite{Htoh}. 
LH8 has a latent heat of $800 ~\mbox{MeV/fm}^{3}$ and
represents a balance between soft and hard.  

\subsection{Outlook}

Finally, we turn to open problems.  
Hanbury-Brown Twiss (HBT) correlations
provide information about the spatial and temporal
extent of the freezeout region.
Such measurements
have been performed
at the AGS, the SPS \cite{Uli-Review} and recently at RHIC
\cite{STAR-Pion}. 
Although the HBT radii fall with the pion pair momentum $K_T$
providing additional evidence for transverse flow \cite{Uli-Review}, 
the measured radii at RHIC are approximately only
50\% percent of our preliminary 
radii \cite{Teaney-HBT}.

The dynamic origin of these small HBT radii is not understood
and much more work is needed \cite{Hirano-HBT}. The small radii indicate that although the
final velocities are correctly reproduced within the model, 
the model expansion time is too long. 
Future experiments 
will measure HBT radii and deuteron production  
as a function of impact parameter and azimuthal angle. 
Such detailed experimental information is essential 
to a complete understanding of the excited matter produced 
in ultra-relativistic heavy ion reactions.

{\bf Acknowledgments}
The continued support of the nuclear chemistry group at Stony Brook
is gratefully acknowledged.
This work was supported in parts by the US-DOE grants
DE-FG-88ER40388 and by DE-FG02-87ER 40331.

\end{document}